\documentclass[aps, prl, twocolumn, superscriptaddress, amsmath,  tightenlines, longbibliography]{revtex4-1}

\usepackage{dcolumn}
\usepackage{graphicx}
\usepackage{mathrsfs}
\usepackage{subfigure}
\usepackage{booktabs}
\usepackage{amsmath}
\usepackage{physics}
\usepackage{dsfont}
\usepackage{amstext}
\usepackage{amssymb}
\usepackage{amsbsy}
\usepackage{bbm}
\usepackage{amsthm}
\usepackage{graphicx}
\usepackage{color}
\usepackage[colorlinks,citecolor=blue]{hyperref}

\usepackage{url}
\usepackage[colorlinks]{hyperref}
\hypersetup{%
	plainpages=true,
	breaklinks=true,       
	hypertexnames=false,  
	pageanchor=true,
	colorlinks=true,
	linkcolor={blue},
	citecolor={red},
	urlcolor={blue},
	anchorcolor={black}
}

\newcommand{\bs}[1]{{\boldsymbol#1}}

 \makeatletter

\newcommand{\Rmnum}[1]{\expandafter\@slowromancap\romannumeral #1@}
\makeatother

\begin{document}

\title{Chiral-Extended Photon-Emitter Dressed States in  Non-Hermitian Topological Baths}
\author{Zhao-Fan Cai}
\thanks{These authors contributed equally}
\affiliation{School of Physics and Optoelectronics, South China University of Technology,  Guangzhou 510640, China}
\author{Xin Wang}
\thanks{These authors contributed equally}
\affiliation{Institute of Theoretical Physics, School of Physics, Xi'an Jiaotong University, Xi'an 710049,  China}
\author{Zi-Xuan Liang}
\affiliation{School of Physics and Optoelectronics, South China University of Technology,  Guangzhou 510640, China}
\author{Tao Liu}
\email[E-mail: ]{liutao0716@scut.edu.cn}
\affiliation{School of Physics and Optoelectronics, South China University of Technology,  Guangzhou 510640, China}
\author{Franco Nori}
\affiliation{Center for Quantum Computing, RIKEN, Wakoshi, Saitama 351-0198, Japan}
\affiliation{Department of Physics, University of Michigan, Ann Arbor, Michigan 48109-1040, USA}

\date{{\small \today}}


\begin{abstract}
	The interplay of quantum emitters and non-Hermitian structured baths has received increasing attention in recent years.  Here, we predict unconventional quantum optical behaviors of quantum emitters coupled to a non-Hermitian topological bath, which is realized in a 1D Su-Schrieffer-Heeger photonic chain subjected to nonlocal dissipation. In addition to the Hermitian-like  chiral bound states in the middle line gap and skin-mode-like hidden bound states inside the point gap, we identify peculiar in-gap chiral and extended photon-emitter dressed states. This is due to the competition of topological-edge localization and non-Hermitian skin-mode localization in combination with the non-Bloch bulk-boundary correspondence. Strikingly, dissipation can shape  the wavefunction profile  of the dressed state. Furthermore, when two emitters are coupled to the same bath,  such in-gap dressed states can mediate the nonreciprocal long-range emitter-emitter interactions, with the interaction range limited only by the dissipation of the bath.  Our work opens the door to further study  rich quantum optical phenomena and exotic many-body physics utilizing quantum emitters coupled to non-Hermitian baths.
\end{abstract}

\maketitle

\textit{Introduction}.---Recent years have witnessed considerable interest in controlling photon-emitter interactions utilizing structured nanophotonic environments due to their potential applications in quantum networks and quantum simulation of many-body physics \cite{PhysRevLett.119.143602,PhysRevA.96.043811,PhysRevLett.128.013601,PhysRevLett.120.140404, PhysRevLett.125.163602,PhysRevLett.126.043602,PhysRevLett.126.063601, PhysRevLett.126.103603, PhysRevLett.119.023603,PhysRevA.96.041603, Barik2018, Bello2019,PhysRevX.11.011015,PhysRevA.104.053522,PhysRevResearch.5.023031, PRXQuantum.3.010336,PhysRevLett.127.250402, RevModPhys.90.031002,PhysRevLett.128.203602, RevModPhys.95.015002,PhysRevResearch.6.013279,PhysRevA.110.053706,arXiv:2405.03675,PhysRevResearch.6.043226, GonzlezTudela2024,arXiv:2404.09829}.  Among them, one of the promising strategies is to couple quantum emitters with topological waveguides \cite{PhysRevLett.126.063601,PhysRevLett.126.103603,PhysRevLett.119.023603,PhysRevA.96.041603, Barik2018,Bello2019,PhysRevX.11.011015,PhysRevA.104.053522, PhysRevResearch.5.023031,PRXQuantum.3.010336}, where the topological nature of the bath can give rise to unconventional quantum optical phenomena robustness against disorder, e.g., chiral photon-emitter bound states, band topology-dependent super/subradiant states, and exotic many-body phases resulting from the tunable emitter-emitter interactions mediated by the bound states \cite{Bello2019}.

A photonic structure is unavoidably coupled to the external reservoir, which can be effectively described by non-Hermitian Hamiltonians \cite{Ashida2020}. Non-Hermitian physics is currently a burgeoning field due to the unique physical phenomenon without Hermitian counterparts \cite{Gao2015,Monifi2016,ZhangJ2018,Ashida2020,PhysRevLett.116.133903, PhysRevLett.118.040401, PhysRevLett.118.045701,arXiv:1802.07964,Peng2014b,El-Ganainy2018,ShunyuYao2018,PhysRevLett.125.126402,PhysRevLett.123.066404,YaoarXiv:1804.04672,PhysRevLett.121.026808,PhysRevLett.122.076801,PhysRevLett.123.170401, PhysRevLett.123.206404,PhysRevLett.123.066405,PhysRevLett.123.206404,PhysRevLett.131.103602, PhysRevB.100.054105,PhysRevB.99.235112,Zhao2019,PhysRevX.9.041015,PhysRevLett.124.056802,PhysRevB.102.235151,Bliokh2019, PhysRevB.104.165117,     PhysRevLett.124.086801, PhysRevLett.125.186802,PhysRevLett.127.196801, RevModPhys.93.015005,PhysRevLett.128.223903, Leefmans2022,  Zhang2022,Parto2023, Ren2022,PhysRevX.13.021007,PhysRevLett.131.036402,PhysRevLett.131.116601,arXiv:2311.03777,PhysRevA.109.063329,arXiv:2403.07459,PhysRevX.14.021011,PhysRevLett.132.050402,Leefmans2024}. An intriguing  physical phenomenon is the non-Hermitian skin effect (NHSE), with the emergence of  localized bulk modes at boundaries \cite{ ShunyuYao2018, YaoarXiv:1804.04672,PhysRevLett.122.076801,PhysRevLett.123.066404,PhysRevLett.125.126402,  PhysRevLett.121.026808}, which has the intrinsic topological origin associated to the point gap \cite{PhysRevX.9.041015,PhysRevLett.124.086801}.  In recent years, the interplay of quantum emitters and non-Hermitian structured baths has attracted much attention \cite{Roccati2022,PhysRevA.106.053517,PhysRevLett.129.223601,PhysRevResearch.5.L042040,Roccati2024,arXiv:1902.00967}, leading to exotic  quantum optical behaviors, e.g.  skin-mode-like bound state inside the point-gap loop  and anomalous quantum emitter dynamics without Hermitian counterparts \cite{PhysRevLett.129.223601}. 

In this work, we predict the unique photon-emitter dressed states and long-range emitter-emitter interaction by studying a paradigm of photon-emitter interactions in a nonreciprocal Su-Schrieffer-Heeger (SSH)  photonic chain. In addition to the existence  of conventional chiral bound states and hidden bound states inside the line and point gaps, respectively,  we unveil  unusual chiral and extended photon-emitter dressed states without Hermitian counterparts.  Moreover, we demonstrate the directional long-range emitter-emitter interaction mediated by dressed states, where the interaction range is  limited only by the bath dissipation.
 
\begin{figure}[!b]
	\centering
	\includegraphics[width=8.6cm]{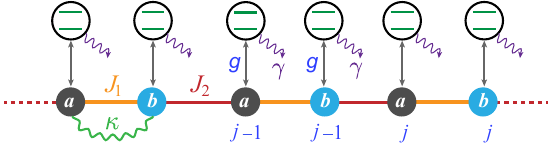}
	\caption{Schematic showing a set of $N$ identical two-level  atoms (acting as quantum emitters) coupled to a 1D SSH photonic bath. The bath consists of coupled cavities, subject to correlated photon decay (with loss rate $\kappa$) between two cavities in each unit cell. $J_1$ and $J_2$ denote the intracell and intercell hopping strength, $\gamma$ is the atomic decay rate, and $g$ is the atom-photon coupling strength.}\label{lattice}
\end{figure}

\textit{Model}.---We consider a set of $N$ identical  atoms, as quantum emitters, coupled to a 1D  SSH  photonic chain with $L$ unit cells, as shown in Fig.~\ref{lattice}. Each  two-level atom, with ground state $\ket{g}$ and excited state  $\ket{e}$, is coupled to each cavity in the lattice, and its decay rate is denoted by $\gamma$. The SSH photonic chain consists of coupled cavities  subject to  an engineered nonlocal photon dissipation between two sublattices  $a$ and $b$ in each unit cell with  loss rate $\kappa$. In the single-excitation subspace,   the system dynamics is governed by the effective non-Hermitian Hamiltonian (see details for the nonlocal dissipation and effective non-Hermitian Hamiltonian   in the Sec.~\Rmnum{1}   of the Supplemental Material (SM) in Ref.~\cite{NonHermitianBathS2023}) 
\begin{align}\label{Heffr}
	\hat{\mathcal{H}}_\textrm{eff} = &   \sum_{n=1}^N \Delta \hat{\sigma}_n^+ \hat{\sigma}_n^-
	+ \sum_{j=1}^{L-1}  J_2 \left(   \hat{a}_{j+1}^\dagger \hat{b}_{j} +   \hat{b}_{j}^\dagger  \hat{a}_{j+1} \right)
	    \nonumber \\ 
	&   + \sum_{j=1}^{L}  \left[ \left(J_1 + \frac{\kappa}{2} \right) \hat{b}_{j}^\dagger \hat{a}_{j} + \left(J_1 - \frac{\kappa}{2} \right)\hat{a}_{j}^\dagger  \hat{b}_{j}   \right]  \nonumber \\ 
	&  + \sum_{n=1}^N \sum_{\alpha \in \{a,b\}} \left[ -i\frac{\kappa}{2}\hat{\alpha}_{ n}^\dagger \hat{\alpha}_{ n}  + g\left(\hat{\alpha}_{j_n}^\dagger \hat{\sigma}_n^- +  \textrm{H.c.}\right) \right]  ,
\end{align}
where $\hat{\sigma}_n^- = (\hat{\sigma}_n^+)^\dagger =  |g_n  \rangle \langle e_n| $ is the pseudospin ladder operator of the $n$th atom, $\Delta=\Delta_0 - i\gamma/2$ with  frequency detuning $\Delta_0$, $\hat{a}_j$ and $\hat{b}_j$ annihilate  photons at sublattices $a$ and $b$ of the $j$th unit cell  (see Fig.~\ref{lattice}), $g$ is the photon-emitter interacting strength, and $j_n$ labels the unit cell at which the $n$th atom is located.  Unless otherwise specified, we assume $\gamma=\kappa$.

\begin{figure}[!tb]
	\centering
	\includegraphics[width=8.7cm]{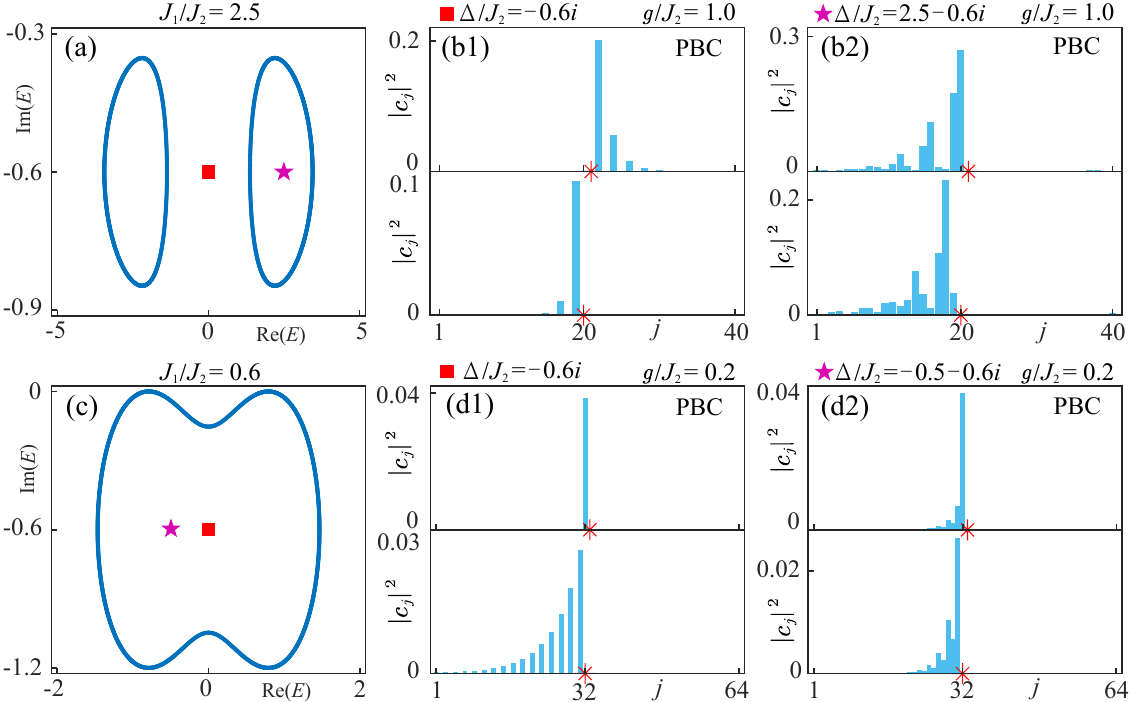}
	\caption{Single-excitation spectrum (blue loops) under PBCs  (a) with the coexistence of point and line gaps for  $J_1/J_2=2.5$, and (c) with only a point gap for  $J_1/J_2=0.6$. The markers denote the eigenenergies of the bound states of a single emitter coupled to the bath for different  $\Delta/J_2$. The corresponding site-resolved  photon weights $\abs{c_j}^2$ are shown in (b1,b2) and (d1,d2), where the emitter is coupled to the sublattice $a$ ($b$), denoted by the red asterisk, for the top (bottom) plot. The other parameters  used are $\kappa/J_2=1.2$.}\label{Fig2}
\end{figure}

\textit{Chiral and Hidden Bound States}.---We couple a single emitter to the sublattice $\alpha \in \{a,b\}$  within the unit cell $j_0$ of the bath, and study the bound states lying within the regimes of both line and point gaps of the SSH bath. In the single-excitation subspace,  the bound state using periodic boundary conditions (PBCs) can be written as $\ket{\psi_{\textrm{b}}} = [L^{-1/2}\sum_{k} (c_{k,a} \hat{a}_k^\dagger+ c_{k,b} \hat{b}_k^\dagger) +  c_{e }\hat{\sigma}^+_{j_0} ]\ket{g} \otimes \ket{\textrm{vac}}$, with $\hat{\alpha}_{k} = L^{-1/2} \sum_{j} e^{-ikj} \hat{\alpha}_{j}$ ($\alpha=a,b$), which satisfies $\hat{\mathcal{H}}_\textrm{eff}(k) \ket{\psi_{\textrm{b}}} = E_\textrm{b} \ket{\psi_{\textrm{b}}}$. For the photon-emitter bound states, we require $\textbf{\emph{c}}_e\neq0$. This yields \cite{NonHermitianBathS2023}
\begin{align}\label{E_b}
	\text{det}\left[E_\textrm{b} - \Delta - \Sigma\left(E_\textrm{b}\right) \right]  =  0,
\end{align}
where  $\Sigma\left(z\right)$  is the atomic self-energy, given by
\begin{align}\label{selfenergy}
	\Sigma\left(z\right)  =  \frac{1}{L} \sum_{k} \textbf{\emph{g}}_k^\dagger \left(z-H_k\right)^{-1}  \textbf{\emph{g}}_k,
\end{align}
with the bath's Bloch Hamiltonian $H_k = -i\frac{\kappa}{2} \tau_0 + \left(J_1+J_2\cos k\right)\tau_x + \left(J_2\sin k-i \kappa/2 \right) \tau_y$, and $\textbf{\emph{g}}_{k} =  [g_a e^{-i k j_0},~g_b e^{-i k j_0}]^T$ ($g_a, g_b \in \{0, g\}$).

In the presence of line gap for $\abs{J_2} < \abs{J_1 -\kappa/2}$, we can analytically solve the real-space wavefunction of the bound state  with $E_\textrm{b}=-i\kappa/2$ for $\Delta=-i\kappa/2$  \cite{NonHermitianBathS2023}). For the emitter coupled to the sublattice $a$,  we have $c_{j,a}=0$,    $c_{j,b} = -g c_{e}(-J_2)^{j-j_0} (J_1-\kappa/2)^{-j+j_0-1} $ if $j \ge j_0$, and $c_{j,b}=0$ if $j<j_0$. For the emitter  coupled to the sublattice $b$,  we have $c_{j,b}=0$,    $c_{j,a} = g c_{e}J_2^{j_0-j} (-J_1-\kappa/2)^{j-j_0-1} $ if $j \le j_0$, and $c_{j,a}=0$ if $j>j_0$. These indicate that the bound state lying within the line gap [see red filled square marker  in Fig.~\ref{Fig2}(a)] shows perfect chiral photon weight $\abs{c_j}^2$  for $\Delta = -i\kappa/2$, as shown in Fig.~\ref{Fig2}(b1). Such a chiral bound state  can be interpreted as a boundary between two semi-infinite chains with different topology \cite{Bello2019},  its chirality thus  depends on the sublattice $a$ or $b$ which the emitter is coupled to, and  is insensitive to the NHSE. Note that the chirality of the bound state is sensitive to $\Delta$  (see details in the Sec.~\Rmnum{3} of SM in Ref.~\cite{NonHermitianBathS2023}).

As a comparison, we calculate the bound state lying inside the point gap, which can be analytically solved  out  for $J_1 =  \kappa/2$  \cite{NonHermitianBathS2023}. The self-energy of the bound state is obtained as 
\begin{align}\label{PointgapSelfEnergy}
	\Sigma(E_\textrm{b})  = \begin{cases}
		-\frac{g^2(E_\textrm{b}+\frac{i\kappa}{2})}{J_2^2-(E_\textrm{b}+\frac{i\kappa}{2})^2}, & \text{ $\abs{\kappa J_2} < \abs{J_2^2-(E_\textrm{b}+i\kappa/2)^2 }$} \\
		0, & \text{ $\abs{\kappa J_2} > \abs{J_2^2-(E_\textrm{b}+i\kappa/2)^2 }$}
	\end{cases}.
\end{align}
The analytical results for the real-space wavefunctions are provided in SM \cite{NonHermitianBathS2023}. The self-energy in Eq.~(\ref{PointgapSelfEnergy}) vanishes for $E_\textrm{b}$ lying inside the loop of the point gap, dubbed hidden bound state \cite{PhysRevLett.129.223601}. In contrast to conventional bound states, such a bound state exhibits skin-mode-like localization independent of $\Delta$  [see Fig.~\ref{Fig2}(c,d1,d2) and also (b2)], which is determined by the NHSE associated with the point-gap topology of the bath. Note that the emergence of hidden bound states  does not rely on the coupling strength $g$ (see details in the Sec.~\Rmnum{3} of SM in Ref.~\cite{NonHermitianBathS2023}).
 
\begin{figure*}[!tb]
	\centering
	\includegraphics[width=18cm]{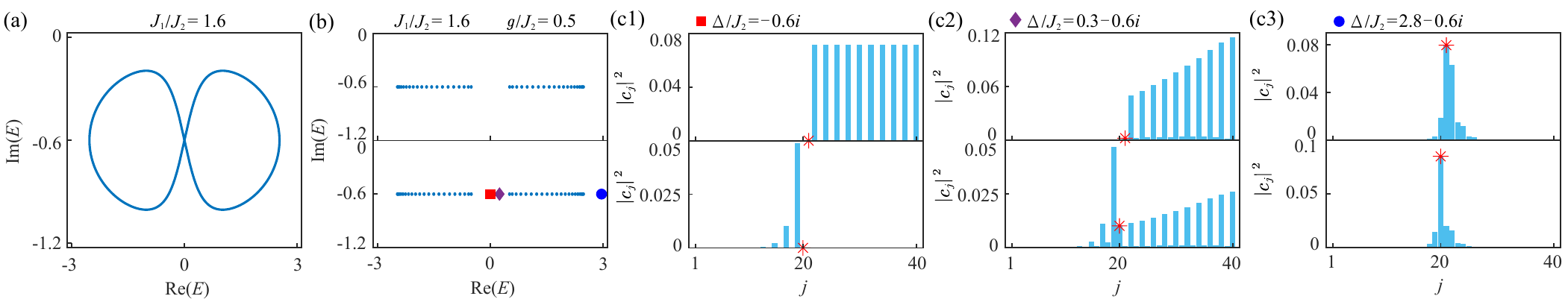}
	\caption{Single-excitation spectrum of the SSH bath at transition point $J_2=J_1 - \kappa/2$ under (a) PBCs, and (b) OBCs (top plot). The markers, shown in bottom plot of (b), denote the eigenenergies of the dressed states of a single emitter coupled to the chain for different values of $\Delta/J_2$. The corresponding site-resolved  photon weights $\abs{c_j}^2$ are shown in (c1-c3) under OBCs, where the emitter is coupled to the sublattice $a$ ($b$), denoted by the red asterisk, for the top (bottom) plot.  	The parameters used are    $g/J_2=0.5$, $\kappa/J_2=1.2$, $J_1/J_2 = 1.6$,  and $L=20$. }\label{Exceptionalpoint}
\end{figure*}

\textit{Chiral-Extended Dressed States}.---In addition to  localized chiral and hidden bound states, we identify an \textit{unique  in-gap photon-emitter dressed state}, exhibiting the chiral and extended mode distribution  under OBCs. We consider the system parameter satisfying $J_2=J_1 \pm \kappa/2$, where the line band-gap closes (with the appearance of an exceptional point) under PBCs [see the PBC spectrum in Fig.~\ref{Exceptionalpoint}(a)]. According to the non-Bloch bulk-boundary correspondence in a generalized Brillouin zone \cite{NonHermitianBathS2023}, the true topological-phase transition point of band topology is determined by $J_1 = \pm\sqrt{J_2^2+(\kappa/2)^2}$. It is thus topologically trivial for $J_2=J_1 -\kappa/2$ with {\color{red} $J_1    > \kappa/2 $}  [see OBC spectrum in the top plot of Fig.~\ref{Exceptionalpoint}(b)]. Unless otherwise specified, we consider this condition for system parameters below. 

We first study a single emitter coupled to the sublattice $\alpha \in \{a,b\}$ of the unit cell  $j_{0}$. In the single-excitation subspace under OBC,  the photon-emitter dressed state is written as $\ket{\psi_{\textrm{d}}} = (\sum_{j,\alpha \in \{a,b\}} c_{j,\alpha} \hat{\alpha}_{j}^\dagger+ c_{e}\hat{\sigma}^+ )\ket{g} \otimes \ket{\textrm{0}}$, which satisfies $\hat{\mathcal{H}}_\textrm{eff} \ket{\psi_{\textrm{d}}} = E_\textrm{d} \ket{\psi_{\textrm{d}}}$. Then, we achieve 
\begin{align}\label{eq1}
	\Delta c_e + g c_{j_0, \alpha} = E_\textrm{d} c_e,
\end{align}
\begin{align}\label{eq2}
 	g c_e \delta_{j,j_0}\delta_{\alpha,a}  +  J_2 \left( c_{j, b} + c_{j-1, b} \right)  = \left(E_\textrm{d} + i \kappa/2 \right) c_{j, a},
\end{align}
\begin{align}\label{eq3}
 	g c_e \delta_{j,j_0}\delta_{\alpha,b}  +\left(J_2 + \kappa \right) c_{j, a} + J_2 c_{j+1, a} = \left(E_\textrm{d} + i \kappa/2 \right) c_{j, b}.
\end{align}

For $\Delta = -i\kappa/2$, we can find the analytical solution for the dressed state with its eigenenergy $E_\textrm{d} = \Delta$. In this case, when the emitter is coupled to the sublattice $a$ ($\alpha = a$)  in Eqs.~(\ref{eq1}-\ref{eq3}), we obtain $c_{j,a}=0$, $c_{j,b}=0$ for $j < j_0$,  $c_e = -J_2 c_{j,b}/g$ for $j = j_0$, and $c_{j,b}=-c_{j-1,b}$ for $j > j_0$. The analytical results indicate that the in-gap  photon-emitter dressed state  exhibits an unconventional feature different from the one of the bound state when the emitter is coupled to the sublattice $a$. In addition to the chiral property with its eigenstate only distributed on the right side of the emitter, the dressed state is uniformly distributed along the $b$ sites under OBC, as shown in the top plot of Fig.~\ref{Exceptionalpoint}(c1) [Its eigenenergy is indicated by the red square marker in the bottom plot of Fig.~\ref{Exceptionalpoint}(b)]. Noticeably, the chiral and extended photon-emitter dressed states are quite robust against the disordered-distributed cavity frequencies and disordered photonic hopping between cavities, as explained in the Sec.~\Rmnum{4} of SM \cite{NonHermitianBathS2023}. 

In contrast, when the emitter is coupled to the sublattice $b$ ($\alpha = b$) in Eqs.~(\ref{eq1}-\ref{eq3}), we obtain $c_{j,b}=0$, $c_{j-1,a}=-J_2  c_{j,a}/(J_2+\kappa)$ for $j < j_0$,  $c_e = -(J_2+\kappa) c_{j,a}/g$ for $j = j_0$, and $c_{j,a}=0$ for $j > j_0$. It turns out that the in-gap  photon-emitter dressed state is bounded, and its photonic profile [see the bottom plot of Fig.~\ref{Exceptionalpoint}(c1)] is localized at the left side of the emitter, i.e., showing the emergence of a chiral bound state.

\begin{figure}[!tb]
	\centering
	\includegraphics[width=8.6cm]{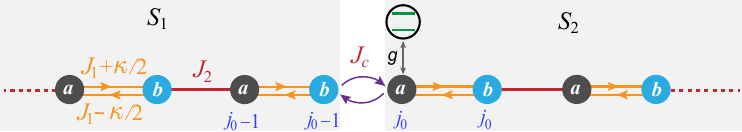}
	\caption{Schematic for understanding the chiral and extended dressed state. When the emitter is coupled to the sublattice $a$ under OBC,  the hybrid system is divided into  $S_1$ and $S_2$ subsystems by breaking the intercell coupling $J_c=J_2$ that exists on the left side of the sublattice lattice $a$ at the unit cell $j_0$.}\label{lattice2}
\end{figure}

The physical intuition of the appearance of the in-gap chiral and extended photon-emitter dressed states for $\Delta =-i\kappa/2$ with $J_2=J_1-\kappa/2$   can be attributed to the competition of topological-edge localization and non-Hermitian skin-mode localization with the combination of the non-Bloch bulk-boundary correspondence of a non-Hermitian topological bath. Namely, when the emitter is coupled to the sublattice $a$ under OBC, we divide the photon-emitter system into two subsystems $S_1$ and $S_2$, by breaking the intercell coupling $J_c=J_2$ that exists on the left side of the sublattice lattice $a$ at the unit cell $j_0$, as shown in Fig.~\ref{lattice2}. The subsystem $S_1$ is topologically trivial, while the subsystem $S_2$ hosts an in-gap topological edge mode  where the emitter acts as the effective boundary of $S_2$. Instead of topological-edge localization on the left side of the subsystem $S_2$, the competition from the opposite mode localization towards the right side induced by NHSE leads to  the extended mode distribution along the $S_2$ at  $J_2=J_1-\kappa/2$ \cite{PhysRevB.103.195414, Wang2022}. The coupling of the trivial subsystem $S_1$ to $S_2$  only has a minor effect on the dressed state due to its in-gap  topological protection and zero occupations on $a$ sublattices.  Here, the broken bulk-boundary correspondence of the topological bath due to NHSE excludes the coupling of the photon-emitter dressed state with the edge states of the SSH bath. However, when the emitter is coupled to the sublattice $b$, two subsystems $S_1$ and $S_2$ are constructed by breaking the intercell coupling $J_c=J_2$ that exists on the right side of the sublattice $b$ at the unit  cell $(j_0-1)$. In this  case, both topological edge-mode localization and NHSE lead  to the formation of the chiral localized in-gap bound state.

For  arbitrary $\Delta$, we can still  achieve the analytical solution  of the eigenenergy $E_d = E - i\kappa/2$  for the  dressed state \cite{NonHermitianBathS2023}, with $E$ satisfying  
\begin{align}\label{OBeffective2239}
	 E - \Delta_0 - g^2 \sum_{m=1}^{2L} \frac{\abs{\varphi_{m,\alpha}(j_0)}^2}{(E-\varepsilon_m) \mathcal{N}_m}  = 0,
\end{align} 
where $\varepsilon_{m}  =  (-1)^m \sqrt{2\bar{J}_1 J_2 \cos \theta_m + \bar{J}_1^2 + J_2^2}$, with $\bar{J}_1=\sqrt{(J_1- \kappa / 2 )(J_1+\kappa / 2)}$, and real number $\theta_m$,  is the analytical eigenvalue of the non-Hermitian bath, and $ \varphi_{m,\alpha}(j) $ ($\alpha=a,b$) is the element  of the analytical eigenvectors of the Hermitian SSH lattice  in the similarity-transformed basis with $\bar{\mathcal{H}}_{\alpha} = S_\alpha^{-1} \mathcal{H}_{\alpha} S_\alpha$ ($\mathcal{H}_{\alpha}$ is the Hamiltonian matrix of $\hat{\mathcal{H}}_\textrm{eff}$ for the emitter coupled to the sublattice $\alpha$, and $S_\alpha$ is the  diagonal matrix $\textrm{diag}[1, ~ r^{-(j_0-\delta_{\alpha,a})}, ~r^{1-(j_0-\delta_{\alpha,a})}, ~\cdots,   ~r^{L-(j_0-\delta_{\alpha,a})}]$ with $r=\sqrt{(J_1+\kappa/2)/(J_1-\kappa/2)}$, see details in the Sec.~\Rmnum{5} of Ref.~\cite{NonHermitianBathS2023}), and $\mathcal{N}_m$ is a normalization.  This analytical result provides us an additional  understanding of the chiral and extended dressed states for $\alpha=a$: in the similarity-transformed basis, the photon-emitter dressed state is bound with the photon weight power-law decaying towards the right side of the emitter \cite{Bello2019}. After employing the inverse of the similarity transformation, the bound state becomes extended due to the power-law increase  for each element of $S_a$ starting at the site $j_0$. 

Figure \ref{Exceptionalpoint}(c2) shows the photon weight  $\abs{c_j}^2$ with $\Delta \neq -i\kappa/2$ for the emitter coupled to the sublattice  $\alpha=a$   ($\alpha=b$) in the top (bottom)  plot. The extended dressed states remain chiral with the uniform site-resolved photon weight  for $\alpha=a$, while the bound state becomes extended distribution for $\alpha=b$.   Note that there exist only bound states  when the $\Delta$ is set to be outside the middle gap of the OBC spectrum [see Fig.~\ref{Exceptionalpoint}(c3)].

\begin{figure}[!tb]
	\centering
	\includegraphics[width=8.7cm]{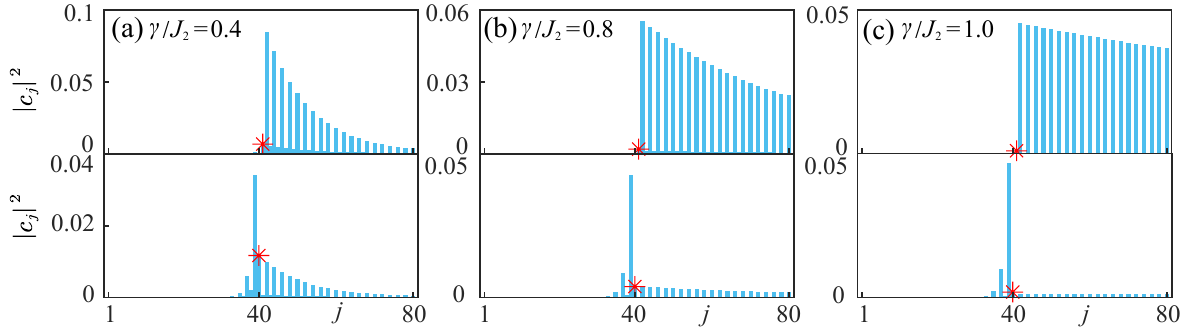}
	\caption{Site-resolved  photon weights $\abs{c_j}^2$ at   $J_2=J_1 - \kappa/2$    for (a) $\gamma/J_2=0.4$, (b) $\gamma/J_2=0.8$,  and (c) $\gamma/J_2=1.0$ under OBCs.  The emitter is coupled to the sublattice $a$ ($b$) for the top (bottom) plot.  	The other parameters used are    $g/J_2=0.5$, $\kappa/J_2=1.2$, $J_1/J_2 = 1.6$  and $L=40$. }\label{s4}
\end{figure}

\textit{Dissipation-Controlled State Profiles}.---We have discussed the chiral-extended photon-emitter dressed states for $\gamma=\kappa$. We now study the effects of  the emitter decay rate $\gamma$ on the mode distribution of the dressed state with $\Delta_0=0$ and $J_2=J_1 - \kappa/2$. As shown in Fig.~\ref{s4}, we show the site-resolved  photon weights $\abs{c_j}^2$ for the emitter   coupled to the sublattice $a$ ($b$) in the top (bottom) plot. 

As discussed above, when the emitter is coupled to the sublattice $a$ with $\gamma=\kappa$, an extended uniform distribution of chiral dressed states is achieved. This chiral-extended state distribution remains quite robust even when $\gamma$ deviates from $\kappa$. As shown in top plots of Fig.~\ref{s4}(a-c), the photon-emitter dressed states maintain chiral and extended distributions despite a significant deviation in emitter decay compared to cavity loss. Furthermore, as $\gamma$ deviates from $\kappa$, the state distribution becomes non-uniform, with the photon weight gradually diminishing across the lattice sites . These indicate that the emitter dissipation can be  utilized to control the wavefunction profiles of dressed states, and can also be employed to modulate  interaction dynamics between two quantum emitters. In addition, when the emitter is coupled to the sublattice $b$, the wavefunction profiles of  bound states are great changed as $\gamma$ deviates from $\kappa$ [see bottom plots of Fig.~\ref{s4}(a-c)].

\textit{Two Emitters}.---We now consider the consequences of such dressed states when two quantum emitters are coupled to the bath with $J_2=J_1-\kappa/2$.  The bound states can mediate the emitter-emitter interactions, giving rise to the exotic many-body phases \cite{Bello2019}. The distance of two interacting emitters is determined by the localization length of the bound state, leading to short-range interactions. In contrast, the extended in-gap dressed state can mediate long-range interactions, and its chiral character causes the directional interactions between emitters.

\begin{figure*}[!tb]
	\centering
	\includegraphics[width=17.8cm]{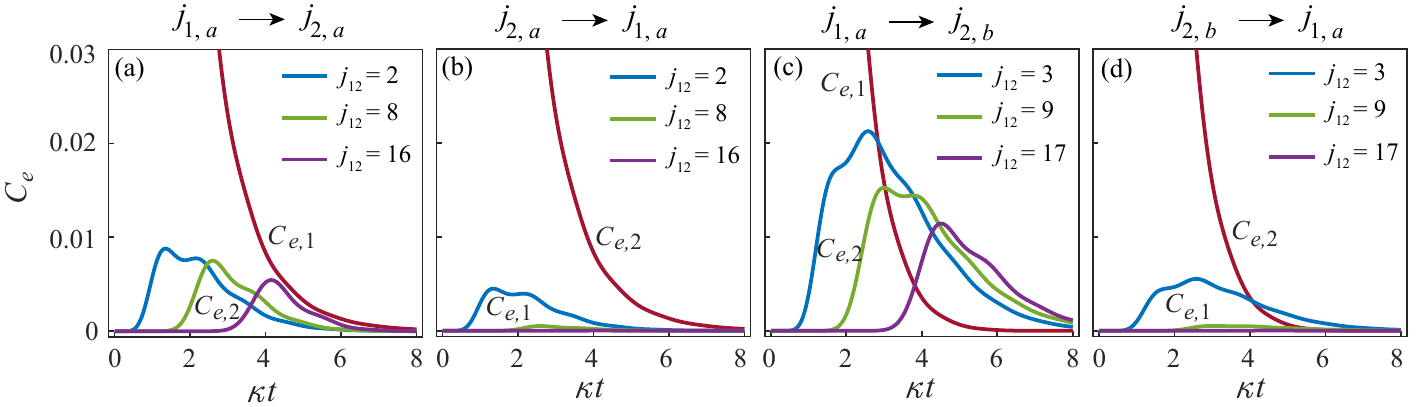}
	\caption{Excited-state probability $C_{e,i}=\abs{c_{e_i}(t)}^2$ ($i=1,2$) for two emitters coupled to sites $j_{1, \alpha_1}$ and $j_{2, \alpha_2}$ ($\alpha_1,\alpha_2=a ~\textrm{or} ~ b$, and $j_{2, \alpha_2} > j_{1, \alpha_1}$) of the bath, where the emitter, coupled to the site $j_{1, a}$, $j_{2, a}$, $j_{1, a}$, and $j_{2, b}$, is initially excited for (a-d), respectively. The parameters  used are $g/J_2=0.4$, $\kappa/J_2=0.4$, $J_1/J_2 = 1.2$,   $\Delta/J_2=-0.2i$, and $L=100$. }\label{Fig5}
\end{figure*}

In order to demonstrate such long-range interactions, we calculate the    non-unitary  real-time dynamics governed by $\ket{\psi_t} = e^{-i\hat{\mathcal{H}}_\textrm{eff} t} \ket{\psi_0}$ for two emitters (labeled as 1 and 2)   coupled to sites $j_{1, \alpha_1}$ and $j_{2, \alpha_2}$ ($\alpha_1,\alpha_2=a ~\textrm{or} ~ b$) of the bath with $j_{2, \alpha_2}>j_{1, \alpha_1}$, respectively.  The initial state  is chosen as one excited emitter $\ket{e_1}$ or $\ket{e_2}$ with $\ket{\psi_0} = \ket{e_n} \ket{\textrm{vac}}$ ($n=1 ~\textrm{or} ~ 2$), and  the time-evolved state can be expanded as $\ket{\psi_t} = \left(\sum_{m=1}^{2N}c_{m}(t) \ket{\varphi_m^R} \bra{\textrm{vac}}  + \sum_{n=1}^{2}c_{e_n}(t)\ket{e_n}\bra{g}  \right)\ket{gg} \otimes \ket{\textrm{vac}}$ ($\ket{\varphi_m^R}$ is the right eigenvector of the non-Hermitian bath). Using the  resolvent method \cite{CCohenTannoudji1Atom,Economou_2006}, we can express $\bs{c}_e(t) = [c_{e_1}(t), ~c_{e_2}(t)]^T$ as \cite{NonHermitianBathS2023}
\begin{align}\label{ce20}
	\bs{c}_e(t) = \frac{i}{2\pi}\int_{-\infty}^{+\infty} d E \: \mathcal{G}_p \left(E+i0^+\right) e^{-iE t} \bs{c}_e(0),
\end{align}
where the Green's function $\mathcal{G}_p \left(z\right)$ are given by

\begin{equation}\label{ce30}
 \mathcal{G}_p \left(E\right) = 
	\begingroup
	\setlength{\tabcolsep}{5pt}             
	\renewcommand{\arraystretch}{1.6}        
	\begin{pmatrix}
	\frac{1}{E-\Delta- \mathcal{T}(\alpha_1, \alpha_1)} & \frac{1}{E-  \mathcal{F}(\alpha_1, \alpha_2) \mathcal{T}(\alpha_1, \alpha_2) }    \\
\frac{1}{E-  \mathcal{F}(\alpha_2, \alpha_1) \mathcal{T}(\alpha_1, \alpha_2) }   & \frac{1}{E-\Delta- \mathcal{T}(\alpha_2, \alpha_2) }    \\	 
	\end{pmatrix},
	\endgroup
\end{equation}

with
\begin{align}\label{ce810}
	\mathcal{T}(\alpha_1, \alpha_2) = g^2\sum_{m=1}^{2L} \frac{\varphi_{m,\alpha_1}(j_{1,\alpha_1}) \varphi_{m,\alpha_2}(j_{2,\alpha_2})}{(E-\varepsilon_m +  i\kappa/2 )\mathcal{N}_m},
\end{align}
\begin{align}\label{ce80}
 	\mathcal{F}(\alpha_1, \alpha_2) = \frac{\left(J_1+\kappa/2\right)^{\frac{j_{1,\alpha_1} - j_{2,\alpha_2} + \delta_{\alpha_1,b} - \delta_{\alpha_2,b} }{2}}}{\left(J_1-\kappa/2\right)^{\frac{j_{1,\alpha_1} - j_{2,\alpha_2}+   \delta_{\alpha_1,b} - \delta_{\alpha_2,b}}{2}}}.
\end{align}

According to Eqs.~(\ref{ce20})-(\ref{ce80}), the main contribution from the diagonal elements of the Green function $\mathcal{G}_p \left(z\right)$ to the time evolution is the dressed state for small $g$ and $\Delta=-i\kappa/2$. The off-diagonal elements contribute to the state exchanges between two emitters. Remarkably, such state exchange is asymmetrical [see Eq.~(\ref{ce80})]. To be specific, when the emitter at the site $j_{2,\alpha_2}$ is initially excited, there is no excitation transferred to the emitter at the site $j_{1,\alpha_1}$ for the large distance $\abs{j_{1,\alpha_1} - j_{2,\alpha_2}}$ between them due to the power-law decay of  $\mathcal{F}(\alpha_1, \alpha_2)$. Figure   \ref{Fig5} shows the excited-state probability $C_{e,i}=\abs{c_{e_i}(t)}^2$ ($i=1,2$) for two emitters coupled to sites $j_{1, \alpha_1}$ and $j_{2, \alpha_2}$  of the bath, where the emitter, coupled to the site $j_{1, a}$, $j_{2, a}$, $j_{1, a}$, and $j_{2, b}$, is initially excited for (a-d), respectively. When the first emitter coupled to the site $j_{1, a}$ is initially in the excited state, this will excite the second emitter at the  $j_{2, \alpha}$, even in a very large separation away from the first emitter  [see Fig.~\ref{Fig5}(a,c)], which is limited by the intrinsic dissipation of the bath. In contrast, when the second emitter coupled to the site $j_{2, \alpha}$ is initially in the excited state, no excitation is transferred to the first emitter at the site $j_{1,a}$ for a slight separation between them.  The nonreciprocal long-range emitter-emitter interaction can induce exotic many-body phenomena, which is worth further investigation.

\textit{Conclusion and Outlook}.---In summary, we have studied the conventional chiral and hidden bound states lying inside the line and point gaps of the 1D non-Hermitian topological bath, to which a single emitter is coupled. Most remarkably, we found a unique photon-emitter dressed state without Hermitian counterparts, showing the chiral and extended distribution on just one side of the emitter along the bath. Moreover, dissipation can shape the wavefunction profile  of the dressed state. The unconventional dressed states mediate the nonreciprocal long-range emitter-emitter interactions with the range limited by the bath dissipation. Our study opens many possible directions for future studies, e.g.,   exploring novel many-body phases of emergent spin models with long-range interactions of many emitters,  peculiar extended dressed states in higher-dimensional non-Hermitian topological baths,  and  non-Markovian dynamics.

\begin{acknowledgments}
T.L. acknowledges the support from National Natural Science Foundation of China (Grant No.~12274142), the Key Program of the National Natural Science Foundation of China (Grant No.~62434009), Introduced Innovative Team Project of Guangdong Pearl River Talents Program (Grant
No. 2021ZT09Z109),  and the Startup Grant of South China University of Technology (Grant No.~20210012).  X.W. is supported by 	the National Natural Science	Foundation of China (NSFC; No.~12174303 and Grant No.~11804270), and the 	Fundamental 	Research Funds for the Central Universities (No.~xzy012023053). F.N. is supported in part by:  the Japan Science and Technology Agency (JST)[via the Quantum Leap Flagship Program (Q-LEAP), and the Moonshot R$\&$D Grant Number JPMJMS2061], the Asian Office of Aerospace Research and Development (AOARD) (via Grant No.~FA2386-20-1-4069), and the Office of Naval Research (ONR) Global (via Grant No.~N62909-23-1-2074).
\end{acknowledgments}

%

\clearpage \widetext
\begin{center}
	\section{Supplemental Material for ``Chiral-Extended Photon-Emitter Dressed States in  Non-Hermitian Topological Baths"}
\end{center}
\setcounter{equation}{0} \setcounter{figure}{0}
\setcounter{table}{0} \setcounter{page}{1} \setcounter{secnumdepth}{3} \makeatletter
\renewcommand{\theequation}{S\arabic{equation}}
\renewcommand{\thefigure}{S\arabic{figure}}
\renewcommand{\bibnumfmt}[1]{[S#1]}
\renewcommand{\citenumfont}[1]{S#1}

\makeatletter
\def\@hangfrom@section#1#2#3{\@hangfrom{#1#2#3}}
\makeatother

\maketitle

\section{Effective non-Hermitian bath in single-excitation subspace}

As shown in the main text, we consider a set of $N$ identical atoms, as quantum emitters, coupled to a 1D Su-Schrieffer–Heeger (SSH) photonic chain with $L$ unit cells. The photonic chain consists of coupled cavities  subjected to engineered nonlocal dissipation, as shown in Fig.~1 in the main text. Each  two-level atom, with ground state $\ket{g}$ and excited state  $\ket{e}$, is coupled to each cavity in the lattice.  Under the Markovian and rotating-wave approximations,  the dissipative dynamics of the system (in the rotating frame) is governed by the Lindblad master equation \cite{Scully1997SM, Breuer2007SM, Agarwal2012SM,arXiv:1902.00967SM}:
\begin{align}\label{mastereqSM}
	\frac{d \hat{\rho}}{dt} =   -i \left[\hat{\mathcal{H}}_\textrm{e}+\hat{\mathcal{H}}_\textrm{p}+\hat{\mathcal{H}}_\textrm{int}, ~\hat{\rho}\right] +  \gamma \sum_{n=1}^{N}\mathcal{D}[\hat{\sigma}_n^-]\hat{\rho}  
	+ \kappa \sum_j   \mathcal{D}[\hat{L}_{j}]\hat{\rho},
\end{align}
where the Hamiltonians of atoms $\hat{\mathcal{H}}_\textrm{e}$, photonic SSH bath $\hat{\mathcal{H}}_\textrm{p}$ and photon-emitter interaction $\hat{\mathcal{H}}_\textrm{int}$ are written as 
\begin{align}\label{atom}
	\hat{\mathcal{H}}_\textrm{e}=\sum_{n=1}^N \Delta_0 \hat{\sigma}_n^+ \hat{\sigma}_n^-,
\end{align}
\begin{align}\label{photon}
	\hat{\mathcal{H}}_\textrm{p}=   \sum_{j=1}^{L}  \left( J_1 \hat{b}_{j}^\dagger \hat{a}_{j} + J_2 \hat{a}_{j+1}^\dagger \hat{b}_{j} + \textrm{H.c.} \right),
\end{align}
\begin{align}\label{interaction}
	\hat{\mathcal{H}}_\textrm{int} = \sum_{n=1}^N \sum_{\alpha \in \{a,b\}} g\left(\hat{\alpha}_{j_n}^\dagger \hat{\sigma}_n^- +  \textrm{H.c.}\right).
\end{align}
Here, $\hat{\sigma}_n^- = (\hat{\sigma}_n^+)^\dagger =  |g_n  \rangle \langle e_n| $ is the pseudospin ladder operator of the $n$th atom, $\Delta_0$ is frequency detuning of the atom with respect to the cavity frequency, $\hat{a}_j$ and $\hat{b}_j$ annihilate  photons at sublattices $a$ and $b$ of the $j$th unit cell  (see Fig.~1 in the main text), $g$ is the photon-emitter interacting strength, and $j_n$ labels the unit cell at which the $n$th atom is located. Moreover, $\hat{\rho}$ is the system density matrix, the Lindblad superoperator $\mathcal{D}[\mathcal{L}]\hat{\rho} =\mathcal{L} \hat{\rho} \mathcal{L}^\dagger - \{\mathcal{L}^\dagger \mathcal{L}, ~\hat{\rho}\}/2 $ represents atomic and photonic dissipation,  $\gamma$ is the atomic decay  rate, and $\kappa$ denotes the photonic loss. In this work, we consider the nonlocal photon decay between two sublattices $a$ and $b$ in each unit cell with $\hat{L}_{j} = \hat{a}_{j} - i \hat{b}_{j}$.  This type of nonlocal dissipation has been extensively studied in theoretical frameworks \cite{Gong2018SM,PhysRevLett.122.076801SM,PhysRevX.13.031009SM}, and demonstrated in experimental settings \cite{PhysRevLett.129.070401SM}. The nonlocal dissipation of the photonic waveguide can be realized by coupling it to an auxiliary bath, When the auxiliary bath operates under conditions of large detuning or strong dissipation, it can be adiabatically eliminated, effectively implementing the desired nonlocal dissipation $\hat{L}_{j} = \hat{a}_{j} - i \hat{b}_{j}$ \cite{Gong2018SM}. 

We consider the single-excitation subspace with an initial state $\ket{\psi_0} = \hat{\sigma}^+_n \ket{g} \otimes \ket{\textrm{vac}}$, here $\ket{g} \equiv \ket{g_1g_2 \dots g_N}$ and $\ket{\textrm{vac}}$ is the photon vacuum state, and so as the initial density matrix $\hat{\rho}_0= |\psi_0  \rangle \langle \psi_0|$. Then, the master equation in Eq.~(\ref{mastereqSM}) can be solved as \cite{PhysRevA.96.043811SM,PhysRevLett.118.200401SM, PhysRevLett.129.223601SM,	PhysRevA.106.053517SM} 
\begin{align}\label{mastereqSM1}
	\hat{\rho}_t  =  e^{-i \hat{\mathcal{H}}_\textrm{eff} t} \hat{\rho}_0 e^{i \hat{\mathcal{H}}_\textrm{eff}^\dagger t } + p_t |g  \rangle \langle g| \otimes |\textrm{vac}  \rangle \langle \textrm{vac}|, 
\end{align}
with 
\begin{align}\label{mastereqSM2}
	p_t  =  1-\textrm{Tr}[e^{-i \hat{\mathcal{H}}_\textrm{eff} t} \hat{\rho}_0 e^{i \hat{\mathcal{H}}_\textrm{eff}^\dagger t }]. 
\end{align}
Therefore, when focusing only on the single-excitation subspace,  the system's dynamics is governed by the effective non-Hermitian Hamiltonian 
\begin{align}\label{heff}
	\hat{\mathcal{H}}_\textrm{eff} = \hat{\mathcal{H}}_\textrm{e}+\hat{\mathcal{H}}_\textrm{p}+\hat{\mathcal{H}}_\textrm{int} - i \gamma/2 \sum_{n} \hat{\sigma}_n^+ \hat{\sigma}_n^- - i \kappa/2 \sum_{j} \hat{L}_j^\dagger \hat{L}_j. 
\end{align}

Under periodic boundary conditions (PBCs), and using $\hat{\alpha}_{k} = L^{-1/2} \sum_{j} e^{-ikj} \hat{\alpha}_{j}$ ($\alpha=a,b$),  the momentum-space Hamiltonian becomes
\begin{align}\label{HamileffkSM}
	\hat{\mathcal{H}}_\textrm{eff}(k)  =   \Delta \sum_{n=1}^N  \hat{\sigma}_n^+ \hat{\sigma}_n^- + \sum_{k} \hat{\textbf{\emph{a}}}^\dagger_k H_k \hat{\textbf{\emph{a}}}_k    + \frac{1}{\sqrt{L}} \sum_{n=1}^{N} \sum_{k} \left( \hat{\sigma}_n^- \hat{\textbf{\emph{a}}}^\dagger_k \textbf{\emph{g}}_{kn} + \text{H.c.}\right),
\end{align}
where $\Delta=\Delta_0 - i\gamma/2$, $\hat{\textbf{\emph{a}}}_k \equiv [\hat{a}_k, ~\hat{b}_k ]^T$,  $\textbf{\emph{g}}_{kn} =  [g_a e^{-i k j_n},~g_b e^{-i k j_n}]^T$ with $g_a, g_b \in \{0, g\}$, and the Bloch Hamiltonian of the  non-Hermitian SSH bath $H_p(k)$ is
\begin{align}\label{hpk} 
	H_k = -i\frac{\kappa}{2} \tau_0 + \left(J_1+J_2\cos k\right)\tau_x + \left(J_2\sin k-i \kappa/2 \right) \tau_y,
\end{align}
with Pauli matrices $\tau_i$ $(i=x,y,z)$ and identity matrix $\tau_0$.

\section{Bulk-boundary correspondence of a non-Hermitian SSH bath}

As shown in Eq.~(\ref{hpk}), the  Bloch Hamiltonian of the non-Hermitian SSH bath is rewritten as 
\begin{align}
	H_k  = \mathcal{H}_{k} -i(\kappa/2) \tau_0 , ~~\textrm{with} ~~ \mathcal{H}_{k} =  \left(J_1+J_2\cos k\right)\tau_x +\left(J_2\sin k-i\kappa/2\right) \tau_y,
\end{align}
where the Hamiltonians $H_k$ and $\mathcal{H}_{k}$ are topologically equivalent. The non-Hermitian skin effect leads to the breakdown of the conventional bulk-boundary correspondence. The topological-phase boundary can be recovered by the non-Bloch theory \cite{ShunyuYao2018SM}, where the non-Bloch Hamiltonian for $\mathcal{H}_{k}$ reads
\begin{equation}\label{similaritySSH}
	\mathcal{H}_{\beta} =  \begingroup
	\setlength{\tabcolsep}{5pt}             
	\renewcommand{\arraystretch}{1.5}        
	\begin{pmatrix}
		0  & J_1 - \frac{\kappa}{2} + J_2 \beta^{-1}\\
		J_1 + \frac{\kappa}{2} + J_2 \beta  & 0 \\
	\end{pmatrix}.
	\endgroup
\end{equation}
One can obtain the eigenvalue equation for $\beta$ as $\text{det}[\mathcal{H}_{\beta}-E]=0$. Therefore, we have
\begin{align}\label{eigenequation}
	\left[\left(J_1-\frac{\kappa}{2}\right)+J_2 \beta^{-1}\right] \left[\left(J_1+\frac{\kappa}{2}\right)+J_2 \beta\right]=E^2.
\end{align}
This leads to two solutions   
\begin{align}\label{beta}
	\beta_{1,2}(E) = \frac{[E^2 +\kappa^2/4 - J_1^2 -J_2^2 ] \pm \sqrt{[E^2 +\kappa^2/4 - J_1^2 -J_2^2 ]^2 -4 J_2^2(J_1^2-\kappa^2/4)} }{2J_2 (J_1-\kappa/2)},
\end{align}
where $+(-)$ corresponds to $\beta_1$ ($\beta_2$). Then, we obtain
\begin{align}\label{Relationbeta12}
	\beta_{1} \beta_2 = \frac{J_1+\kappa/2}{J_1-\kappa/2}.
\end{align}

According to the generalized Bloch band theory \cite{PhysRevLett.123.066404SM}, we require  $\abs{\beta_1}=\abs{\beta_2}$. This  leads to 
\begin{align}\label{Tranpoints}
	\abs{\beta_1} = \abs{\beta_2} =\sqrt{\abs{\frac{J_1+\kappa/2}{J_1-\kappa/2}}}.
\end{align}
Due to chiral symmetry, we can obtain a generalized $Q$ matrix \cite{ShunyuYao2018SM}, defined as 
\begin{align}\label{matrixQ}
	Q(\beta) = |\tilde{\psi}_R(\beta) \rangle \langle \tilde{\psi}_L(\beta)|-|\psi_R(\beta)\rangle \langle \psi_L(\beta)| = \left(\begin{matrix}
		0 & q \\
		q^{-1} & 0 
	\end{matrix}\right),
\end{align}
where $|\tilde{\psi}_R(\beta)\rangle = \sigma_z |\psi_R(\beta)\rangle$ and $|\tilde{\psi}_L(\beta)\rangle = \sigma_z |\psi_L(\beta)\rangle$, with the right and left eigenvectors given by the following eigenequations
\begin{align}\label{biorthVector}
	\mathcal{H}_\beta |\psi_R(\beta)\rangle = E(\beta) |\psi_R(\beta)\rangle,~~~ \mathcal{H}^\dagger_\beta |\psi_L(\beta) \rangle = E^\ast(\beta) |\psi_L(\beta)\rangle.
\end{align}
The non-Bloch winding number $\mathcal{W}$ is given by
\begin{align}\label{NonBlochWN}
	\mathcal{W} = \frac{i}{2\pi} \int_{\text{GBZ}} \frac{dq}{q},
\end{align}
where $\textrm{GBZ}$ denotes the generalized Brillouin zone.

According to Eqs.~(\ref{matrixQ}-\ref{NonBlochWN}), the true topological-phase transition points in the presence of non-Hermitian skin effects are given by
\begin{align}\label{trueTran}
	J_1 = \pm \sqrt{J_2^2+\kappa^2/4},
\end{align}
where the SSH bath is topologically nontrivial when $J_1 \in (-\sqrt{J_2^2+\kappa^2/4},~ \sqrt{J_2^2+\kappa^2/4})  $. 

This indicates that the bath is topologically trivial for $J_2 = J_1 - \kappa/2$  with  $J_1  > \kappa/2$ due to $\kappa\geq0$, where we explore this regime for chiral-extended photon-emitter dressed states.

\section{Chiral and hidden bound states}


Due to the particle number conservation of the system Hamiltonian $\hat{\mathcal{H}}_\textrm{eff}(k)$ in Eq.~(\ref{HamileffkSM}), in the single-excitation subspace,  the bound state can be written as 
\begin{align}\label{HamileffkS3}
	\ket{\psi_{\textrm{b}}} = \left(\frac{1}{\sqrt{L}}\sum_{k} \textbf{\emph{c}}_k \hat{\textbf{\emph{a}}}_{k}^\dagger+ \sum_{n=1}^{N}c_{e,n}\hat{\sigma}^+_n \right)\ket{g} \otimes \ket{\textrm{vac}},
\end{align} 
with $\textbf{\emph{c}}_k \equiv [c_{k,a},~c_{k,b}]^T$, which satisfies $\hat{\mathcal{H}}_\textrm{eff}(k) \ket{\psi_{\textrm{b}}} = E_\textrm{b} \ket{\psi_{\textrm{b}}}$. Then, we obtain 
\begin{align}\label{boundeq1}
	\Delta \textbf{\emph{c}}_e + \frac{1}{L}\sum_{k} \textbf{\emph{g}}_k^\dagger \textbf{\emph{c}}_k   = E_\textrm{b} \textbf{\emph{c}}_e, ~~~~ \textrm{and}~~ H_k \textbf{\emph{c}}_k + \textbf{\emph{g}}_k \textbf{\emph{c}}_e  = E_\textrm{b} \textbf{\emph{c}}_k,
\end{align}
for $\forall~ k$, where $\textbf{\emph{c}}_e \equiv [c_{e,1},~c_{e,2},\dots,~c_{e,N}]^T$, and $\textbf{\emph{g}}_k \equiv [\textbf{\emph{g}}_{k1},~\textbf{\emph{g}}_{k2},\dots,~\textbf{\emph{g}}_{kN}]$. 

According to Eqs.~(\ref{boundeq1}), we have
\begin{align}\label{E_bCe}
	\left[E_\textrm{b} - \Delta - \Sigma\left(E_\textrm{b}\right) \right]\textbf{\emph{c}}_e  =  0,
\end{align}
and for photon-emitter bound states, we require $\textbf{\emph{c}}_e\neq0$. This yields
\begin{align}\label{E_b}
	\text{det}\left[E_\textrm{b} - \Delta - \Sigma\left(E_\textrm{b}\right) \right]  =  0,
\end{align}
where  $\Sigma\left(z\right)$  is the self-energy of the emitters, given by
\begin{align}\label{selfenergy}
	\Sigma\left(z\right)  =  \frac{1}{L} \sum_{k} \textbf{\emph{g}}_k^\dagger \left(z-H_k\right)^{-1}  \textbf{\emph{g}}_k.
\end{align}
We then can determine the atomic and photonic weights as
\begin{align}\label{HamileffkS8}
	\abs{\textbf{\emph{c}}_e}^2 = \frac{ 1}{1 +\frac{ 1}{L}\sum_{k} \textbf{\emph{g}}_k^\dagger  \left[(E_\textrm{b}-H_k)(E_\textrm{b}^\ast-H_k^\dagger) \right]^{-1} \textbf{\emph{g}}_k}, ~~~~ \textrm{and}~~ \textbf{\emph{c}}_k = \frac{\textbf{\emph{g}}_k \textbf{\emph{c}}_e }{E_\textrm{b} - H_k}.
\end{align}

In this section, we are interest in a single emitter coupled to the sublattice $a$ or $b$ within the unit cell $j_0$ of the bath, and study the bound states with its eigenenergy lying within the regimes of both    line and point gaps of the non-Hermitian SSH bath. In this work, unless otherwise specified, we assume $\gamma=\kappa$.

\subsection{Line gap and chiral bound state}

In the presence of the line gap for the bath Hamiltonian $H_k$, to have a simple form of the analytical solution for the bound-state wavefunction in Eq.~(\ref{HamileffkS8}), we solve the bound state for $E_\textrm{b} = -i\kappa/2$.

According to Eqs.~(\ref{E_b}) and (\ref{selfenergy}), for $E_\textrm{b} = -i\kappa/2$, we immediately have
\begin{align}\label{boundchiral}
	\Sigma\left(E_\textrm{b}\right) = 0 , ~~~\textrm{and} ~~~\Delta = -i\kappa/2.  
\end{align}
Then,  according to Eq.~(\ref{HamileffkS8}), the atomic  weight  $\abs{\textbf{\emph{c}}_e}^2$ for the emitter coupled to the sublattice $a$ or $b$ is, respectively, derived as
\begin{align}\label{Linegapboundchiralatomicprof1}
	\abs{c_{e,a}}^2 = \frac{1}{1 + \frac{g^2 }{L} \sum_{k} \abs{J_1+J_2 e^{- ik} - \kappa/2}^{-2}},
\end{align}
or
\begin{align}\label{Linegapboundchiralatomicprof2}
	\abs{c_{e,b}}^2 = \frac{1}{1 + \frac{g^2 }{L} \sum_{k} \abs{J_1+J_2 e^{ ik} + \kappa/2}^{-2}}.
\end{align}

(i) When the emitter is coupled to the sublattice $a$, the photonic weight $c_{k,\alpha}$ ($\alpha = a, b$) for $\Delta = -i\kappa/2$ in momentum space is obtained as 
\begin{align}\label{LinegapckA}
	c_{k,a} = 0, ~~~ \textrm{and} ~~ c_{k,b} = -\frac{g_k c_{e,a} }{J_1+J_2 e^{- ik} - \kappa/2}, 
\end{align}
where $g_k = g e^{-i k j_0}$. The real-space photonic profile can be obtained by the inverse Fourier transformation of Eq.~(\ref{LinegapckA}). This leads to $c_{j,a}=0$, and 
\begin{align}\label{LinegaprealspacePhotoncjb}
	c_{j,b} &= -\frac{g c_{e,a}}{L} \sum_{k} \frac{e^{ik(j-j_0)}}{J_1+J_2 e^{-ik}-\frac{\kappa}{2} }  = -\frac{g c_{e,a}}{2\pi i} \oint_{\abs{y}=1} dy \frac{y^{j-j_0}}{J_2 + (J_1-\kappa/2)y}.
\end{align}

According to Eq.~(\ref{LinegaprealspacePhotoncjb}), for $\abs{J_2} < \abs{J_1 -\kappa/2}$, where the PBC spectrum of the SSH bath Hamiltonian $H_k$ exhibits a line gap,  we obtain
\begin{align}\label{LinegaprealspacePhotoncjbFinal}
	c_{j,b} = \begin{cases}
		-\frac{g c_{e,a}}{J_1-\kappa/2} \left(-\frac{J_2}{J_1-\kappa/2}\right)^{j-j_0}, & \text{ $j \ge j_0$}, \\
		0, & \text{ $j<j_0$}.
	\end{cases}
\end{align} 

(ii) When the emitter is coupled to the sublattice $b$, the photonic weight $c_{\alpha,k}$ ($\alpha = a, b$) for $\Delta = -i\kappa/2$ in momentum space is obtained as 
\begin{align}\label{LinegapckB}
	c_{k,a} = -\frac{g_k c_{e,b}}{J_1+J_2 e^{ik} + \kappa/2}, ~~~ \textrm{and} ~~ c_{k,b} = 0. 
\end{align}
The real-space photonic profile is obtained by the inverse Fourier transformation of Eq.~(\ref{LinegapckB}). This leads to $c_{j,b}=0$, and 
\begin{align}\label{LinegaprealspacePhotoncja}
	c_{j,a} &= -\frac{g c_{e,b}}{L} \sum_{k} \frac{e^{ik(j-j_0)}}{J_1+J_2 e^{ik}+\frac{\kappa}{2} }  = -\frac{g c_{e,b}}{2\pi i} \oint_{\abs{y}=1} dy \frac{y^{j-j_0-1}}{J_2 y +  J_1+\kappa/2 }.
\end{align}

According to Eq.~(\ref{LinegaprealspacePhotoncja}), for $\abs{J_2} < \abs{J_1+\kappa/2}$,  where the PBC spectrum exhibits a line gap of the SSH bath Hamiltonian $H_k$,  we obtain
\begin{align}\label{LinegaprealspacePhotoncjaFinal}
	c_{j,a} = \begin{cases}
		0, & \text{ $j > j_0$}, \\
		\frac{g c_{e,b}}{J_2} \left(-\frac{J_1+\kappa/2}{J_2}\right)^{j-j_0-1}, & \text{ $j \le j_0$}.
	\end{cases}
\end{align}

\begin{figure*}[!tb]
	\centering
	\includegraphics[width=18cm]{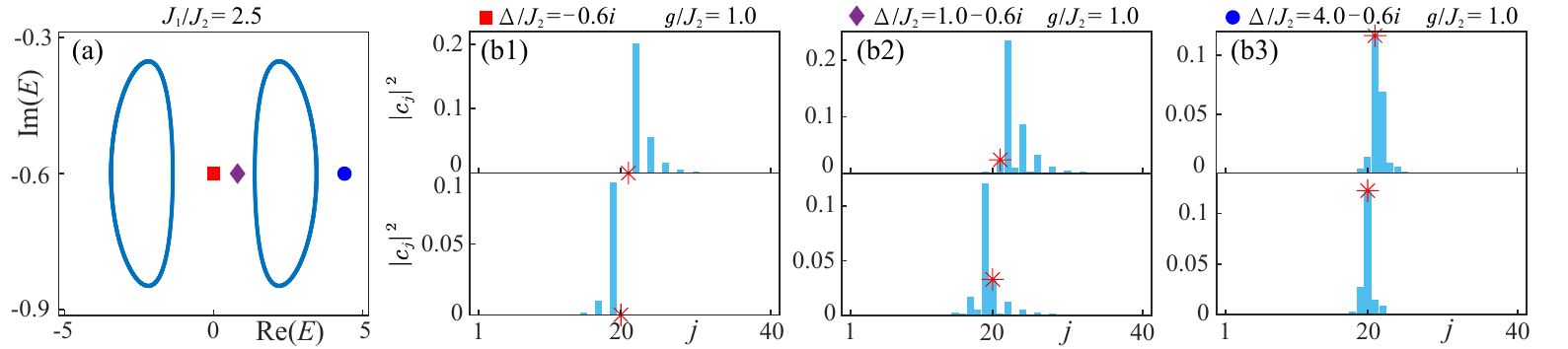}
	\caption{Single-excitation line-gap spectrum (blue loops) of the  bath $H_k$  under PBCs   for  $J_1/J_2=2.5$, The markers denote the eigenenergies of the bound states of a single emitter coupled to the bath for different  $\Delta$. The corresponding site-resolved  photon weights $\abs{c_j}^2$ are shown in (b1-b3), where the emitter is coupled to the sublattice $a$ ($b$), denoted by the red asterisk, for the top (bottom) plot. The other parameters  used  are $\kappa/J_2=1.2$.}\label{FigS0}
\end{figure*}

The above analytical results show that the bound state, with its eigenenergy lying inside the line gap,  has its eigenstate located on just the left or right side of the emitter, depending on the sublattice $a$ or $b$ to which the emitter is coupled, for $\Delta = -i\kappa/2$, as shown in Fig.~\ref{FigS0}(a,b1). As explained in the main text, such a chiral bound state has a topological origin, which can be interpreted as the effective domain-wall state between two semi-infinite chains with different topology. 

In spite of the nonreciprocal hopping, the bound state inside the line gap behaves more like conventional Hermitian bound states for $\Delta = -i\kappa/2$ \cite{Bello2019SM}. When the detuning $\Delta$ is deviated  from the $ -i\kappa/2$, the  chirality of the bound states  inside the line gap  decreases due to coupling with the bulk bands of the bath, as shown in Fig.~\ref{FigS0}(b2). Outside the eigenenergy-spectrum range (i.e., above or below the two bands), the bound states do not exhibit  chirality due to the absence of topological protection [see Fig.~\ref{FigS0}(b3)].

\subsection{Point gap and hidden bound state}

We now consider the bound state with its eigenenergy enclosed by the point gap of the bath Hamiltonian $H_k$. To  have a simple form of the analytical solution for the bound-state wavefunction in Eq.~(\ref{HamileffkS8}), we solve the bound state for $J_1 =  \kappa/2$. In this case, the eigenenergy of the  bath Hamiltonian $H_k$ in Eq.~(\ref{hpk}) reads
\begin{align}
	E_0 = -i(\kappa/2) \pm \left[J_2 (J_2 + \kappa e^{-ik})\right]^{-1/2}. 
\end{align}

According to Eq.~(\ref{selfenergy}), the self energy of the bound state is written as 
\begin{align}\label{PointgapSelfEnergy}
	\Sigma(E_\textrm{b}) = & ~\frac{g^2}{2\pi i} \oint_{\abs{y}=1} dy \frac{E_\textrm{b}+\frac{i\kappa}{2}}{\left[  \left(E_\textrm{b} + i\kappa /2\right)^2 -J_2^2\right] y -J_2 \kappa}= \begin{cases}
		-\frac{g^2(E_\textrm{b}+\frac{i\kappa}{2})}{J_2^2-(E_\textrm{b}+\frac{i\kappa}{2})^2}, & \text{ $\abs{\kappa J_2} < \abs{J_2^2-(E_\textrm{b}+i\kappa/2)^2 }$}, \\
		0, & \text{ $\abs{\kappa J_2} > \abs{J_2^2-(E_\textrm{b}+i\kappa/2)^2 }$}.
	\end{cases}
\end{align}
It turns out that the self-energy vanishes when the $E_\textrm{b}$ lies inside the loop of the point-gap spectrum. We will reveal that such a bound state shows a skin-mode-like photonic profile resulting from the nontrivial point-gap topology, \textit{dubbed hidden bound state } \cite{PhysRevA.106.053517SM,PhysRevLett.129.223601SM}.

(i) When the emitter is coupled to the sublattice $a$, the photonic weight $c_{k,\alpha}$ ($\alpha = a, b$) in momentum space is obtained as 
\begin{align}\label{PointgapckAB_A}
	c_{k,a} = c_{e,a} g_k \left[E_{b}+\frac{i\kappa}{2} - \frac{J_2^2 + \kappa J_2 e^{-ik}}{E_{b}+i\kappa/2}\right]^{-1}, ~~~ \textrm{and} ~~ c_{k,b} = c_{e,a} g_k \left[\frac{(E_{b}+i\kappa/2)^2}{\kappa+J_2 e^{ik}}-J_2 e^{-ik}\right]^{-1}, 
\end{align}
where $g_k = g e^{-i k j_0}$. The real-space photonic profile can be obtained by the inverse Fourier transformation of Eq.~(\ref{PointgapckAB_A}). This leads to  
\begin{align}\label{PointgaprealspacePhotoncja_A}
	c_{j,a}  = \frac{g c_{e,a} (E_{b}+\frac{i\kappa}{2})}{L} \sum_{k} \frac{e^{ik(j-j_0)}}{[(E_{b}+\frac{i\kappa}{2})^2-J_2^2] - \kappa J_2 e^{-ik}}  = \frac{g c_{e,a}(E_{b}+\frac{i\kappa}{2}) y_A}{2\pi i \kappa J_2 } \oint_{\abs{y}=1} dy \frac{y^{j-j_0}}{y-\eta},
\end{align}
and
\begin{align}\label{PointgaprealspacePhotoncjb_A}
	c_{j,b} &= \frac{g c_{e,a}}{L} \sum_{k} \frac{(\kappa+J_2 e^{ik}) e^{ik(j-j_0)}}{[(E_{b}+\frac{i\kappa}{2})^2-J_2^2] - \kappa J_2 e^{-ik}}  = \frac{g c_{e,a} \eta}{2\pi i  \kappa J_2} \oint_{\abs{y}=1} dy \frac{\kappa y^{j-j_0} +J_2 y^{j-j_0+1}}{y-\eta},
\end{align}
where $\eta = (\kappa J_2)/[(E_{b}+i\kappa/2)^2-J_2^2]$.

We consider the $E_\textrm{b}$ lies inside the loop of the point-gap spectrum with  $\abs{\kappa J_2} > \abs{J_2^2 - (E_{b}+i\kappa/2)^2}$. According to Eqs.~(\ref{PointgaprealspacePhotoncja_A}) and (\ref{PointgaprealspacePhotoncjb_A}), we obtain the real-space photonic profile as
\begin{align}\label{PointgaprealspacePhotoncja_AFinal}
	c_{j,a} = \begin{cases}
		0, & \text{ $j \ge j_0$}, \\
		\frac{-gc_{e,a}(E_{b}+i\kappa/2)\eta^{j-j_0+1}}{\kappa J_2}, & \text{ $j<j_0$},
	\end{cases}
\end{align} 
and
\begin{align}\label{PointgaprealspacePhotoncjb_AFinal}
	c_{j,b} = \begin{cases}
		0, & \text{ $j \ge j_0$}, \\
		-\frac{gc_{e,a}}{J_2}, & \text{ $j = j_0-1$}, \\
		-\frac{gc_{e,a}\eta^{j-j_0+1}}{J_2} -\frac{gc_{c,a}\eta^{j-j_0+2}}{\kappa}, & \text{ $j<j_0-1$}.
	\end{cases}
\end{align} 

(ii) When the emitter is coupled to the sublattice $b$, the photonic weight $c_{k,\alpha}$ ($\alpha = a, b$) in momentum space is obtained as 
\begin{align}\label{PointgapckAB_B}
	c_{k,a} = \frac{gc_{e,b} J_2 e^{-ik}}{(E_{b}+\frac{i\kappa}{2})^2 - J_2^2 -\kappa J_2}, ~~~ \textrm{and} ~~ c_{k,b} = \frac{gc_{e,b} (E_{b}+\frac{i\kappa}{2})}{(E_{b}+\frac{i\kappa}{2})^2 - J_2^2 -\kappa J_2}. 
\end{align}
The real-space photonic profile is obtained by the inverse Fourier transformation of Eq.~(\ref{PointgapckAB_B}). This leads to
\begin{align}\label{PointgaprealspacePhotoncja_B}
	c_{j,a} &= \frac{gc_{e,b}J_2}{L} \sum_{k} \frac{e^{ik(j-j_0-1)}}{(E_{b}+\frac{i\kappa}{2})^2 -J_2^2 -\kappa J_2 e^{-ik}}  = \frac{g c_{e,b} \eta}{2\pi i \kappa} \oint_{\abs{y}=1} dy \frac{y^{j-j_0-1}}{y-\eta},
\end{align}
and
\begin{align}\label{PointgaprealspacePhotoncjb_B}
	c_{j,b} &= \frac{gc_{e,b}(E_{b}+\frac{i\kappa}{2})}{L} \sum_{k} \frac{e^{ik(j-j_0)}}{(E_{b}+\frac{i\kappa}{2})^2 -J_2^2 -\kappa J_2 e^{-ik}}  = \frac{g c_{e,b} (E_{b} +\frac{i\kappa}{2}) \eta}{2\pi i \kappa J_2 } \oint_{\abs{y}=1} dy \frac{y^{j-j_0}}{y-\eta}.
\end{align}

$E_\textrm{b}$ lies inside the loop of the point-gap spectrum for  $\abs{\kappa J_2} > \abs{J_2^2 - (E_{b}+i\kappa/2)^2}$. According to Eqs.~(\ref{PointgaprealspacePhotoncja_B}) and (\ref{PointgaprealspacePhotoncjb_B}),   we obtain the real-space photonic profile as
\begin{align}\label{PointgaprealspacePhotoncja_BFinal}
	c_{j,a} = \begin{cases}
		0, & \text{ $j > j_0$}, \\
		-\frac{g c_{e,b} \eta^{(j-j_0)}}{\kappa}, & \text{ $j \le j_0$},
	\end{cases}
\end{align} 
and
\begin{align}\label{PointgaprealspacePhotoncjb_BFinal}
	c_{j,b} = \begin{cases}
		0, & \text{ $j \ge j_0$}, \\
		-\frac{g c_{e,b} (E_{b}+\frac{i\kappa}{2}) \eta^{(j-j_0+1)}}{\kappa J_2}, & \text{ $j < j_0$}.
	\end{cases}
\end{align} 

\begin{figure*}[!tb]
	\centering
	\includegraphics[width=17cm]{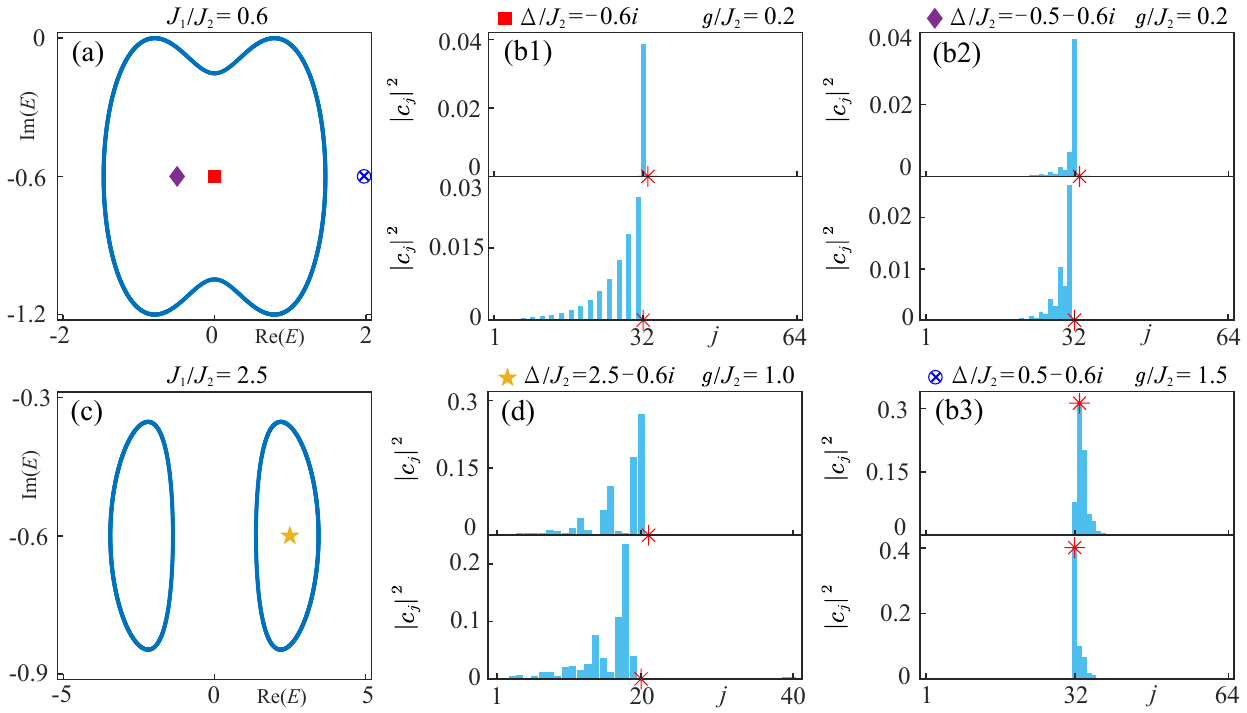}
	\caption{Single-excitation point-gap spectrum (blue loops) of the bath under PBCs  (a)  for  $J_1/J_2=2.5$, and (c)   for  $J_1/J_2=0.6$. The markers denote the eigenenergies of the bound states of a single emitter coupled to the bath for different  $\Delta$. The corresponding site-resolved  photon weights $\abs{c_j}^2$ are shown in (b1-b3) and (d), where the emitter is coupled to the sublattice $a$ ($b$), denoted by the red asterisk, for the top (bottom) plot. The other parameters  used are $\kappa/J_2=1.2$.}\label{figS1}
\end{figure*}

The above analytical results show that the bound state, with its eigenenergy lying inside the point gap,  has its eigenstate located on  only the left side of the emitter, no matter if the emitter is coupled to the sublattice $a$ or $b$. Such a bound state behaves like the skin modes. Figure \ref{figS1}(a,b1,b2) plot the bound states and site-resolved photon weight for $J_1=\kappa/2$. In spite of the detuning $\Delta$, the eigenstates are located on  only the left side of the emitter due to the NHSE. While, outside the loop of the point gap, the bound state behaves like the conventional Hermitian bound states [see Figure \ref{figS1}(b3)]. In addition, in spite of the coexistence of point gap and line gap, bound states inside the point gap behave like  skin modes, as shown in   Fig.~\ref{figS1}(c,d)

\subsection{Atomic weight}

The atomic weight can be directly solved out using Eq.~(\ref{HamileffkS8}). To  have a simple form of the analytical solution for the bound-state wavefunction in Eq.~(\ref{HamileffkS8}), we solve out the bound state for $J_1 =  \kappa/2$. When the emitter is coupled to the sublattice $a$, the photonic weight reads 
\begin{align}
	|c_{e,a}|^2 =\left[1 + \frac{4g^2}{w} + \frac{g^2(4J_2\kappa z_+^2 + u_a z_+ + 4J_2 \kappa)}{4 J_2 \kappa v z_+(z_+ - z_-)} \theta(1-\abs{z_+}) + \frac{g^2(4J_2^2\kappa z_-^2 + u_a z_- + 4J_2 \kappa)}{4 J_2 \kappa v z_-(z_--z_+)} \theta(1-\abs{z_-}) \right]^{-1}, \label{ceA4}
\end{align}
where 
\begin{align}\label{u_A}
	u_a =  4J_2^2-2iE_b\kappa + 5\kappa^2 +4\abs{E_\textrm{b}}^2 + 2i\kappa E_\textrm{b}^\ast,
\end{align}
\begin{align}\label{w}
	w=4J_2^2 + \kappa^2 + 4i\kappa E_\textrm{b}^\ast - 4(E_\textrm{b}^\ast)^2, ~~~ \textrm{and} ~~~v=J_2^2- \left(E_\textrm{b} + \frac{i\kappa}{2}\right)^2,
\end{align}
\begin{align}\label{v}
	p=4J_2^2 \kappa^2 +vw, ~~~ \textrm{and} ~~~ z_\pm = \frac{-p\pm \sqrt{p^2-16J_2^2\kappa^2vw}}{8J_2\kappa v}.
\end{align}

When the emitter is coupled to the sublattice $b$, the photonic weight reads
\begin{align}
	|c_{e,b}|^2 =\left[1  + \frac{g^2 u_b}{4J_2 \kappa v(z_+-z_-)} \theta(1-\abs{z_+}) + \frac{g^2 u_b }{4J_2 \kappa v(z_- - z_+)} \theta(1-\abs{z_-}) \right]^{-1}, \label{ceB4}
\end{align}
where 
\begin{align}\label{u}
	u_b = 4J_2^2-2iE_b\kappa + \kappa^2 +4\abs{E_\textrm{b}}^2 + 2i\kappa E_\textrm{b}^\ast.
\end{align}

\begin{figure*}[!tb]
	\centering
	\includegraphics[width=11cm]{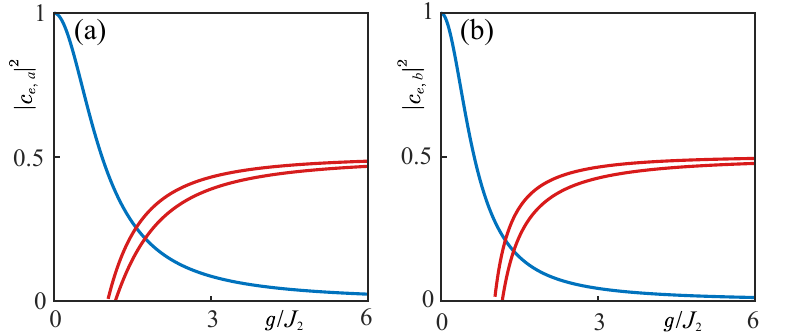}
	\caption{Dependence of the atomic weights $\abs{c_{e,a}}$ and $\abs{c_{e,b}}$ of the hidden
		(blue curves) and Hermitian-like (red curves) bound states on the coupling strength $g$   under PBCs   for  $J_1=\kappa/2$, where the emitter is coupled to the sublattice $a$ or $b$. The parameters used are $\Delta/J_2 = 0.2-0.4i$.}\label{figS2}
\end{figure*}

In Fig.~\ref{figS2}, we calculate the dependence of the atomic weights $\abs{c_{e,a}}$ and $\abs{c_{e,b}}$ of the hidden
(blue curves) and Hermitian-like (red curves) bound states on the coupling strength $g$   under PBCs   for  $J_1=\kappa/2$, where the emitter is coupled to the sublattice $a$ or $b$. The emergence of hidden bound states with eigenenergies inside the point gap does not rely on the coupling strength $g$. In contrast, the conventional bound states with energies outside the point gap only appear for sufficiently large $g$ [see also Fig.~\ref{figS1}(a,b1-b3)].


\section{Effects of disorder on chiral and extended photon-emitter dressed states}

The chiral-extended photon-emitter dressed state has the topological origin, and it is thus robust against the disorder. To illustrate this, we investigate the effect of two types of disorders: (a) the random  cavity frequencies with the addition of the diagonal terms to the original Hamiltonian $\hat{\mathcal{H}}_\textrm{eff} \to \hat{\mathcal{H}}_\textrm{eff} + \hat{\mathcal{H}}_\textrm{diag} $, and (b)  the random hopping between cavities with the addition of the off-diagonal terms to the original Hamiltonian $\hat{\mathcal{H}}_\textrm{eff} \to \hat{\mathcal{H}}_\textrm{eff} + \hat{\mathcal{H}}_\textrm{off} $.  The Hamiltonians $\hat{\mathcal{H}}_\textrm{diag}$ and $\hat{\mathcal{H}}_\textrm{off}$ are written as 
\begin{align}\label{diag}
	\hat{\mathcal{H}}_\textrm{diag} = \sum_{j=1}^{L}  \left( \varepsilon_{a,j} \hat{a}_{j}^\dagger  \hat{a}_{j}   + \varepsilon_{b,j} \hat{b}_{j}^\dagger  \hat{b}_{j}  \right),
\end{align}
and
\begin{align}\label{off}
	\hat{\mathcal{H}}_\textrm{off} = \sum_{j=1}^{L}  \left( \varepsilon_{1,j}   \hat{b}_{j}^\dagger \hat{a}_{j} + \textrm{H.c.}\right) +	\sum_{j=1}^{L-1}  \left( \varepsilon_{2,j}   \hat{b}_{j}^\dagger \hat{a}_{j+1} + \textrm{H.c.}\right),
\end{align}
where $\varepsilon_{a,j}$, $\varepsilon_{b,j}$, $\varepsilon_{1,j}$ and $\varepsilon_{2,j}$ are taken from a uniform distribution within the range $[-V/2, ~V/2]$ with the disorder strength $V$.

\begin{figure*}[!b]
	\centering
	\includegraphics[width=18cm]{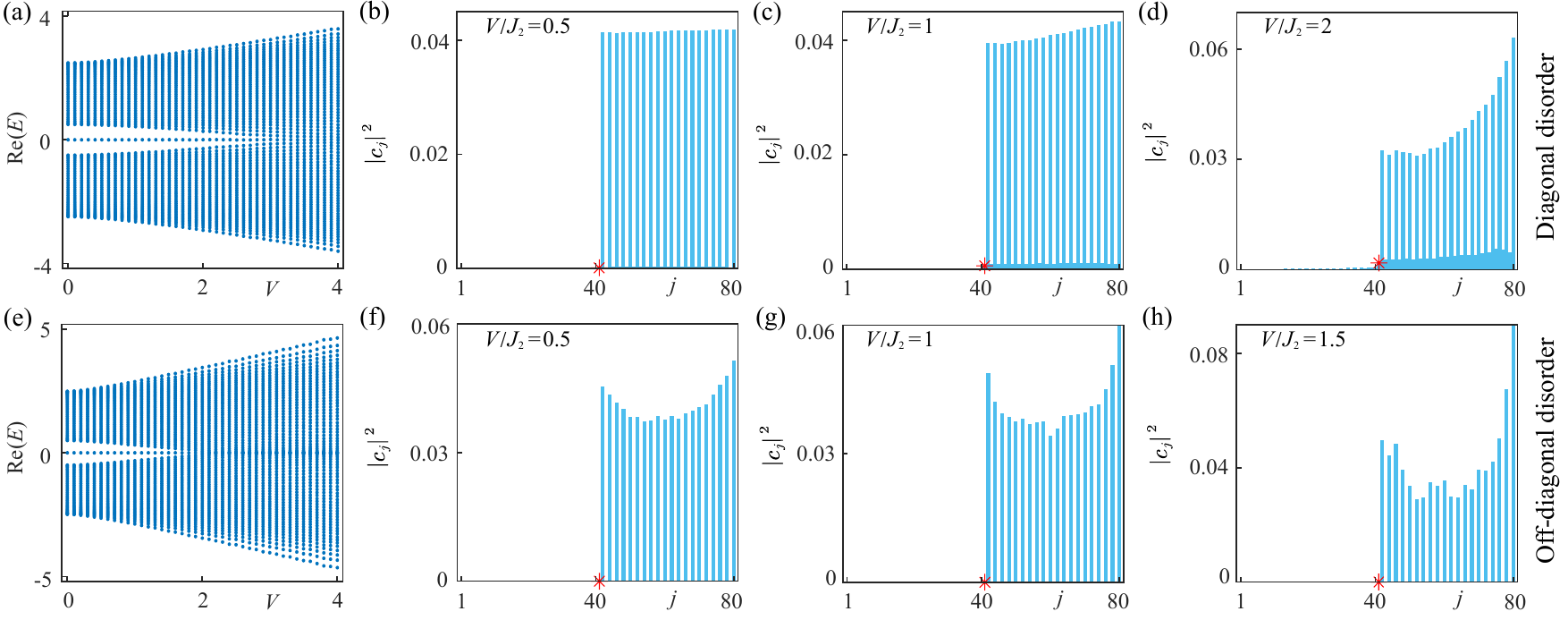}
	\caption{Real part of the complex eigenspectrum $E$ and the corresponding site-resolved photon weights $\abs{c_j}^2$     at the transition point $J_2=J_1 - \kappa/2$ under OBCs as a function of the disorder strength $V$, where   a single emitter is coupled to the sublattice $a$ of the  disordered bath. The in-gap modes are the dressed photon-emitter state. (a-d) Disorder is applied to the cavity frequencies (diagonal disorder), and (e-h) the disorder is applied to the intercell couplings between cavities (off-diagonal disorder).  The results are averaged over $1000$ random realizations. The other parameters  used are  $\Delta/J_2=-0.6i$, $g/J_2=0.5$, $\kappa/J_2=1.2$, $J_1/J_2 = 1.6$ and $L=40$.}\label{figS3}
\end{figure*}

We consider a single emitter coupled to the sublattice $a$ of the disordered SSH bath. Figure \ref{figS3}(a-d) plots the real part of the complex eigenspectrum $E$ and the corresponding site-resolved photon weights $\abs{c_j}^2$ for  randomly disordered cavity frequencies (diagonal disorder). The chirality of the in-gap dressed photon-emitter state, along with its extended photon profile, remains remarkably robust even under strong disorder.

Figure \ref{figS3}(e-h) shows the real part of the complex eigenspectrum $E$ and the corresponding site-resolved photon weights $\abs{c_j}^2$ for the random hopping between cavities (off-diagonal disorder). The chirality and extended photon profile of the dressed state remain remarkably robust  even in the presence of strong off-diagonal disorder caused by random hopping between cavities.

\section{Analytical solution of chiral-extended photon-emitter dressed states }

\subsection{Single Emitter}

We now consider a single emitter coupled to the sublattice $\alpha \in \{a,b\}$ of the unit cell  $j_{0}$. In the single-excitation subspace, spanned by $\{\ket{e}\ket{\textrm{vac}},~\ket{g}\ket{j,a},\ket{g}\ket{j,b}\}$ with $j\in[1, ~L]$, and under the open boundary condition (OBC),  the Hamiltonian of the photon-emitter hybrid system is written as
\begin{equation}\label{halpa}
	\mathcal{H}_{\alpha}= 
	\begingroup
	\setlength{\tabcolsep}{5pt}             
	\renewcommand{\arraystretch}{1.5}        
	\begin{pmatrix}
		\Delta  & V_{\alpha} \\
		V_{\alpha}^\dagger & H_p \\
	\end{pmatrix},
	\endgroup
\end{equation}
where  $\alpha=a,b$, indicating the  sublattice $a$ or $b$ to which the emitter is coupled, the coupling vector
\begin{align}\label{Valpha}
	V_\alpha = (0,~0,~0,~0,~\cdots, ~g \delta_{\alpha,a}, ~g \delta_{\alpha,a},  ~0,~0, ~\cdots, ~0),
\end{align}
and the Hamiltonian matrix of the SSH chain $H_p$ becomes 
\begin{equation}\label{Hp}
	H_p= 
	\begingroup
	\setlength{\tabcolsep}{10pt}             
	\renewcommand{\arraystretch}{2.0}        
	\begin{pmatrix}
		-i \frac{\kappa}{2}    & J_1-\frac{\kappa }{2}   & 0 & 0 & \cdots & 0 & 0  \\
		J_1+\frac{\kappa }{2} & -i \frac{\kappa}{2}    & J_2 & 0 & \cdots & 0 & 0 \\	
		0 & J_2  & -i \frac{\kappa}{2}  & J_1-\frac{\kappa }{2} & \cdots & 0 & 0  \\
		0 & 0   & J_1+\frac{\kappa }{2} & -i \frac{\kappa}{2}  & \cdots & 0 & 0  \\
		\vdots & \vdots & \vdots & \vdots & \vdots & \vdots & \vdots  \\
		0 & 0  & 0 & 0 & \cdots & -i \frac{\kappa}{2}  & J_1-\frac{\kappa }{2}  \\
		0 & 0 & 0 & 0 & \cdots  & J_1+\frac{\kappa }{2} & -i \frac{\kappa}{2}   \\	  
	\end{pmatrix}.
	\endgroup
\end{equation}

We first discuss the emitter  coupled to the sublattice $\alpha = a$. We implement a similarity transformation to the Hamiltonian $\mathcal{H}_{a}$ in Eq.~(\ref{halpa}) with
\begin{align}\label{similarity}
	\bar{\mathcal{H}}_{a} = S_a^{-1} \mathcal{H}_{a} S_a,
\end{align}
where  $S_a$ is a diagonal matrix whose diagonal elements are 
\begin{align}\label{S}
	\{1, ~ r^{-(j_0-1)}, ~r^{1-(j_0-1)}, ~r^{1-(j_0-1)}, ~r^{2-(j_0-1)},~\cdots,  ~r^{L-1-(j_0-1)}, ~r^{L-1-(j_0-1)}, ~r^{L-(j_0-1)}\}, ~~~\textrm{with}~r=\sqrt{\frac{J_1+\frac{\kappa }{2}}{J_1-\frac{\kappa }{2}}}. 
\end{align} 
Then, the $\bar{\mathcal{H}}_{a}$ is written as
\begin{equation}\label{H_a}
	\bar{\mathcal{H}}_{a}= 
	\begingroup
	\setlength{\tabcolsep}{5pt}             
	\renewcommand{\arraystretch}{1.5}        
	\begin{pmatrix}
		\Delta  & V_{a} \\
		V_{a}^\dagger & \bar{H}_{p} \\
	\end{pmatrix},
	\endgroup
\end{equation}
where $\bar{H}_p$ reads
\begin{equation}
	\bar{H}_p= 
	\begingroup
	\setlength{\tabcolsep}{10pt}             
	\renewcommand{\arraystretch}{2.0}        
	\begin{pmatrix}
		-i \frac{\kappa}{2}    & \bar{J}_1   & 0 & 0 & \cdots & 0 & 0  \\
		\bar{J}_1 & -i \frac{\kappa}{2}    & J_2 & 0 & \cdots & 0 & 0 \\	
		0 & J_2  & -i \frac{\kappa}{2}  & \bar{J}_1 & \cdots & 0 & 0  \\
		0 & 0   & \bar{J}_1 & -i \frac{\kappa}{2}  & \cdots & 0 & 0  \\
		\vdots & \vdots & \vdots & \vdots & \vdots & \vdots & \vdots  \\
		0 & 0  & 0 & 0 & \cdots & -i \frac{\kappa}{2}  & \bar{J}_1  \\
		0 & 0 & 0 & 0 & \cdots  & \bar{J}_1 & -i \frac{\kappa}{2}   \\	  
	\end{pmatrix},
	\endgroup
\end{equation}
with $\bar{J}_1 = \sqrt{(J_1- \kappa / 2 )(J_1+\kappa / 2)}$. After the similarity transformation, the photon-emitter Hamiltonian $\bar{\mathcal{H}}_{a}$ describes a Hermitian system subject to the uniform local dissipation with  rate $\kappa/2$.

When the emitter is coupled to the sublattice  $\alpha = b$, we can also implement a similarity transformation to the Hamiltonian $\mathcal{H}_{b}$ in Eq.~(\ref{halpa}) with 
\begin{align}\label{similarityb}
	\bar{\mathcal{H}}_{b} = S_b^{-1} \mathcal{H}_{a} S_b,
\end{align}
where  $S_b$ is a diagonal matrix whose diagonal elements are 
\begin{align}\label{Sb}
	\{1, ~ r^{-j_0}, ~r^{1-j_0}, ~r^{1-j_0}, ~r^{2-j_0},~\cdots,  ~r^{L-1-j_0}, ~r^{L-1-j_0}, ~r^{L-j_0}\}. 
\end{align} 
Then, $\bar{\mathcal{H}}_{b}$ is written as
\begin{equation}\label{H_b}
	\bar{\mathcal{H}}_{b}= 
	\begingroup
	\setlength{\tabcolsep}{5pt}             
	\renewcommand{\arraystretch}{1.5}        
	\begin{pmatrix}
		\Delta  & V_{b} \\
		V_{b}^\dagger & \bar{H}_{p} \\
	\end{pmatrix},
	\endgroup
\end{equation}
After the similarity transformation, the photon-emitter Hamiltonian $\bar{\mathcal{H}}_{b}$ also describes a Hermitian system subjected to the uniform local dissipation with  rate  $\kappa/2$.

According to the   Hamiltonian $\bar{\mathcal{H}}_{a}$ in Eq.~(\ref{H_a}) or $\bar{\mathcal{H}}_{b}$ in Eq.~(\ref{H_b}), in the single-excitation subspace, spanned by $\{\ket{e}\ket{\textrm{vac}},~\ket{g}\ket{j,\bar{a}},\ket{g}\ket{j,\bar{b}}\}$ with the similarity-transformed basis $\ket{g}\ket{j,\bar{\alpha}}$ ($\bar{\alpha}=\bar{a},\bar{b}$) and $j\in[1, ~L]$, the eigenequation for the photon-emitter dressed states reads 
\begin{align}\label{eigenvalues}
	\bar{\mathcal{H}} \ket{\bar{\psi}_\alpha} = \left(\bar{\mathcal{H}}_{0} + \bar{\mathcal{H}}_{e} + \bar{\mathcal{V}}_{\alpha}  \right) \ket{\bar{\psi}_\alpha} = E \ket{\bar{\psi}_\alpha},
\end{align} 
with the eigenenergy of the dressed state being $E_\textrm{d} = E - i\kappa/2$, and
\begin{align}\label{OBeffective21}
	\bar{\mathcal{H}}_{0}  =  \sum_{j=1}^{L} \left(\bar{J}_1 \ket{j,\bar{b}}\bra{j,\bar{a}} + J_2 \ket{j,\bar{b}}\bra{j+1,\bar{a}}+\textrm{H.c.}\right) , 
\end{align} 
\begin{align}\label{OBeffective22e}
	\bar{\mathcal{H}}_{e}   =  \Delta_0 \ket{e}\bra{e},
\end{align} 
\begin{align}\label{OBeffective22}
	\bar{\mathcal{V}}_{\alpha} =   g \left(\ket{e}\bra{j_0,\bar{\alpha}} + \ket{j_0,\bar{\alpha}}\bra{e} \right). 
\end{align} 

The Hermitian SSH Hamiltonian $\bar{\mathcal{H}}_{0}$ for $J_2 = J_1-\kappa/2$ (topological trivial phase) under OBCs can be solved out  as \cite{PhysRevLett.127.116801SM}  
\begin{align}\label{OBeffective211}
	\bar{\mathcal{H}}_{0}  =  \sum_{m=1}^{2L} \varepsilon_{m}  \ket{\varphi_m} \bra{\varphi_m},
\end{align} 
where the eigenvalue reads
\begin{align}\label{OBeffective212}
	\varepsilon_{m}  =  (-1)^m \sqrt{2\bar{J}_1 J_2 \cos \theta_m + \bar{J}_1^2 + J_2^2},
\end{align} 
with $\theta_m$ being a real number   satisfying
\begin{align}\label{OBeffective213}
	\bar{J}_1 \sin[(L+1)\theta_m] + J_2 \sin[L\theta_m] = 0, 
\end{align} 
and the eigenvectors are
\begin{align}\label{OBeffective214}
	\ket{\varphi_{m}}  =  \frac{1}{\sqrt{\mathcal{N}_m}} \sum_{j=1}^{L} \left(\varphi_{m,a}(j) \ket{j,\bar{a}} + \varphi_{m,b}(j) \ket{j,\bar{b}}\right).
\end{align} 
Here, $\mathcal{N}_m$ is the normalized constant for $\ket{\varphi_{m}}$,  and the amplitudes $\varphi_{m,a}(j)$ and $\varphi_{m,b}(j)$ are given by \cite{PhysRevLett.127.116801SM} 
\begin{align}\label{OBeffective215}
	\varphi_{m,a}(j) = \sin[j\theta_m] + \frac{J_2}{\bar{J}_1} \sin[(j-1)\theta_m],
\end{align} 
\begin{align}\label{OBeffective216}
	\varphi_{m,b}(j) =\frac{\varepsilon_{m}}{\bar{J}_1} \sin [j\theta_m].
\end{align} 

Utilizing the eigenvector $\{\ket{\varphi_m}\}$, we rewrite the Hamiltonian  $\bar{\mathcal{V}}_{\alpha}$ in Eq.~(\ref{OBeffective22}) as  
\begin{align}\label{OBeffective217}
	\bar{\mathcal{V}}_{\alpha}  &=  g  \left(\sum_{m=1}^{2L} \ket{e}\bra{j_0,\bar{\alpha}} \ket{\varphi_m} \bra{\varphi_m} + \textrm{H.c.} \right), \nonumber \notag \\
	&=g \left(\sum_{m=1}^{2L} \frac{\varphi_{m,\alpha}(j_0)}{\sqrt{\mathcal{N}_m}}  \ket{e} \bra{\varphi_m} + \text{H.c.}\right).
\end{align} 
Therefore, in the single-excitation subspace, spanned by $\{\ket{e}\ket{\textrm{vac}},~\ket{g}\ket{\varphi_m}\}$ with $m\in[1, ~2L]$, we would like to consider the dressed state 
\begin{align}\label{OBeffective218}
	\ket{\bar{\psi}_\alpha} = \sum_{m=1}^{2L} \bar{c}_m \ket{g}\ket{\varphi_m} + \bar{c}_e \ket{e}\ket{\textrm{vac}},
\end{align} 
that satisfies the eigenequation in Eq.~(\ref{eigenvalues}) with total Hamiltonian $\bar{\mathcal{H}}$ in the new basis $\{\ket{\varphi_m}\}$. Then we obtain
\begin{align}\label{OBeffective219}
	\varepsilon_m \bar{c}_m + \frac{g \varphi_{m,\alpha}^\ast(j_0) \bar{c}_e }{\sqrt{\mathcal{N}_m}} = E \bar{c}_m,~~~~  \forall m,
\end{align} 
\begin{align}\label{OBeffective220}
	g \sum_{m=1}^{2L} \frac{\varphi_{m,\alpha}(j_0) \bar{c}_m }{\sqrt{\mathcal{N}_m}} + \Delta_0 \bar{c}_e = E \bar{c}_e.
\end{align} 
According to Eq.~(\ref{OBeffective219}), the photon profile of the dressed state can be given by
\begin{align}\label{OBeffective221}
	\bar{c}_m = \frac{g\varphi_{m,\alpha}^\ast(j_0)}{\sqrt{\mathcal{N}_m} (E-\varepsilon_m)} \bar{c}_e,
\end{align} 
where the atom profile $\bar{c}_e$ can be determined by the normalization of the dressed state as
\begin{align}\label{OBeffective222}
	\abs{\bar{c}_e}^2 = \left[1 + \sum_{m=1}^{2L} \frac{g^2 \abs{\varphi_{m,\alpha}(j_0)}^2}{(E-\varepsilon_m)(E^\ast-\varepsilon_m^\ast)\mathcal{N}_m}\right]^{-1}.
\end{align} 
Inserting Eq.~(\ref{OBeffective221}) into Eq.~(\ref{OBeffective220}) to eliminate $\bar{c}_m$ yields 
\begin{align}\label{OBeffective223}
	\left[E - \Delta_0 - g^2 \sum_{m=1}^{2L} \frac{\abs{\varphi_{m,\alpha}(j_0)}^2}{(E-\varepsilon_m) \mathcal{N}_m}\right] \bar{c}_e = 0.
\end{align} 

According to Eqs.~(\ref{OBeffective221})-(\ref{OBeffective223}), we can solve $E$, $\bar{c}_m$, and $\bar{c}_e$. Then, at the basis $\{\ket{e}\ket{\textrm{vac}},~\ket{g}\ket{j,a},\ket{g}\ket{j,b}\}$, we obtain the   wavefunction of the dressed state $\psi_\alpha = S_\alpha \bar{\psi}_\alpha $, where
\begin{align}\label{OBeffective225}
	\bar{\psi}_\alpha = [\bar{c}_e, \phi_1, \phi_2, \cdots, \phi_m, \cdots, \phi_{2L}]^T,  ~~~\textrm{with} ~~\phi_m = \bra{g,\varphi_m}\ket{\bar{\psi}_\alpha}. 
\end{align} 

\subsection{Two Emitters}

We now consider two emitters ($\ket{g_1}, \ket{e_1}$) and ($\ket{g_2}, \ket{e_2}$) coupled to site  ($\alpha_1$, $j_1$) and ($\alpha_2$, $j_2$) of the  same SSH bath, respectively. In this subsection, we focus on the situation when the atom-photon interaction strength between both emitters and the SSH bath is set as $g_1=g_2=g$.  In the single-excitation subspace, spanned by $\{\ket{e_1}\ket{\textrm{vac}},~\ket{e_2}\ket{\textrm{vac}},  ~\ket{g}\ket{j,a},\ket{g}\ket{j,b}\}$ with $j\in[1, ~L]$, and under OBCs, the system Hamiltonian reads
\begin{equation}\label{halpatwo}
	\mathcal{H}_{\alpha_1 \alpha_2}= 
	\begingroup
	\setlength{\tabcolsep}{5pt}             
	\renewcommand{\arraystretch}{1.5}        
	\begin{pmatrix}
		\Delta & 0 & V_{\alpha_1}  \\
		0 & \Delta  & V_{\alpha_2} \\
		V_{\alpha_1}^\dagger & V_{\alpha_2}^\dagger & H_p \\
	\end{pmatrix}.
	\endgroup
\end{equation}
Then the eigenequation for the photon-emitter dressed states becomes
\begin{align}\label{OBeffective226}
	\mathcal{H} \ket{\Psi_{\alpha_1\alpha_2}} = \left( \mathcal{H}_p + \mathcal{H}_{e_1} + \mathcal{H}_{e_2} + \mathcal{V}_{\alpha_1\alpha_2}\right) \ket{\Psi_{\alpha_1\alpha_2}} = E_\textrm{d} \ket{\Psi_{\alpha_1\alpha_2}},
\end{align} 
with the eigenenergy of the dressed state being $E_\textrm{d}$, and
\begin{align}\label{OBeffective227}
	\mathcal{H}_{p}  &=  \sum_{j=1}^{L} \left[\left(J_1+\frac{\kappa}{2}\right) \ket{j,b}\bra{j,a} + \left(J_1-\frac{\kappa}{2}\right) \ket{j,a}\bra{j,b} \right] \nonumber \notag \\
	& + \sum_{j=1}^{L-1} \left(J_2 \ket{j,b} \bra{j+1,a} + J_2 \ket{j+1,a} \bra{j,b}\right) \nonumber \notag \\
	& - \sum_{j=1}^{L} \frac{i\kappa}{2} \left(\ket{j,a} \bra{j,a} + \ket{j,b} \bra{j,b}\right),
\end{align} 
\begin{align}\label{OBeffective228}
	\mathcal{H}_e = \mathcal{H}_{e_1} + \mathcal{H}_{e_2}   =  \Delta \left(\ket{e_1}\bra{e_1} + \ket{e_2}\bra{e_2}\right),
\end{align} 
\begin{align}\label{OBeffective229}
	\mathcal{V}_{\alpha_1\alpha_2} = g (\ket{e_1} \bra{j_1,\alpha_1} + \ket{j_1,\alpha_1} \bra{e_1}) + g (\ket{e_2} \bra{j_2,\alpha_2} + \ket{j_2,\alpha_2} \bra{e_2}).
\end{align} 

Due to the non-Hermiticity of the bath Hamiltonian $\mathcal{H}_p$ in Eq.~(\ref{OBeffective227}), the right and left eigenstates of $\mathcal{H}_p$ can be defined as 
\begin{align}\label{OBeffective230}
	\mathcal{H}_p \ket{\varphi_m^R} = E \ket{\varphi_m^R},~~~ \mathcal{H}^\dagger_p \ket{\varphi_m^L} = E^\ast \ket{\varphi_m^L},
\end{align} 
whose biorthogonal conditions and completeness conditions are given by $\bra{\varphi_m^R} \ket{\varphi_n^L}=\bra{\varphi_m^L} \ket{\varphi_n^R}=\delta_{mn} $ and $\sum_{m} \ket{\varphi_m^L} \bra{\varphi_m^R} = \sum_{m} \ket{\varphi_m^R} \bra{\varphi_m^L} = 1$, respectively. Using these relations, the bath Hamiltonian $\mathcal{H}_p$ can be expressed in terms of quasi-particle energy bands as
\begin{align}\label{OBeffective231}
	\mathcal{H}_p = \sum_{m=1}^{2L} \left(\varepsilon_m - \frac{i\kappa}{2} \right)  \ket{\varphi_m^R} \bra{\varphi_m^L},
\end{align} 
and the right and left eigenvectors are given by
\begin{align}\label{OBeffective234}
	\ket{\varphi_m^R} = \frac{1}{\sqrt{\bar{\mathcal{N}}_m}} \sum_{j=1}^{L} ( \varphi^R_{m,a}(j) \ket{j,a} + \varphi^R_{m,b}(j) \ket{j,b}),
\end{align} 
\begin{align}\label{OBeffective235}
	\ket{\varphi_m^L} = \frac{1}{\sqrt{\bar{\mathcal{N}}_m}} \sum_{j=1}^{L} ( \varphi^L_{m,a}(j) \ket{j,a} + \varphi^L_{m,b}(j) \ket{j,b}).
\end{align} 
Here, $\bar{\mathcal{N}}_m$   is the normalized constant in the biorthogonal condition. By utilizing the inverse of the similarity transformation to Eqs.~(\ref{OBeffective215}) and (\ref{OBeffective216}), the amplitudes of $\varphi^{R/L}_{m,a}(j)$ and $\varphi^{R/L}_{m,b}(j)$ are given by
\begin{align}\label{OBeffective236}
	\varphi^{R}_{m,a}(j) = \left(\frac{J_1+\kappa/2}{J_1-\kappa/2}\right)^{\frac{j}{2}} (\sin[j\theta_m] + \frac{J_2}{\bar{J_1}} \sin [(j-1)\theta_m] ),
\end{align} 
\begin{align}\label{OBeffective237}
	\varphi^{R}_{m,b}(j) = \frac{\varepsilon_m}{J_1-\kappa/2} \left(\frac{J_1+\kappa/2}{J_1-\kappa/2}\right)^{\frac{j}{2}} \sin[j\theta_m],
\end{align} 
\begin{align}\label{OBeffective238}
	\varphi^{L}_{m,a}(j) = \left(\frac{J_1-\kappa/2}{J_1+\kappa/2}\right)^{\frac{j}{2}} (\sin[j\theta_m] + \frac{J_2}{\bar{J_1}} \sin [(j-1)\theta_m] ),
\end{align} 
\begin{align}\label{OBeffective239}
	\varphi^{L}_{m,b}(j) = \frac{\varepsilon_m}{J_1+\kappa/2} \left(\frac{J_1-\kappa/2}{J_1+\kappa/2}\right)^{\frac{j}{2}} \sin[j\theta_m].
\end{align} 

Utilizing the completeness condition $\sum_{m=1}^{2L} \ket{\varphi_m^R} \bra{\varphi_m^L} = 1$, we rewrite the atom-photon interaction Hamiltonian $\mathcal{V}_{\alpha_1\alpha_2}$ in Eq.~(\ref{OBeffective229}) as
\begin{align}\label{OBeffective240}
	\mathcal{V}_{\alpha_1\alpha_2} = & ~g \left(\sum_{m=1}^{2L} \frac{\varphi^R_{m,\alpha_1}(j_1)}{\sqrt{\bar{\mathcal{N}}_m}} \ket{e_1} \bra{\varphi_m^L} + \sum_{m=1}^{2L} \frac{[\varphi^L_{m,\alpha_1}(j_1)]^\ast}{\sqrt{\bar{\mathcal{N}}_m}} \ket{\varphi_m^R}  \bra{e_1}\right) \nonumber \notag \\
	& + g \left(\sum_{m=1}^{2L} \frac{\varphi^R_{m,\alpha_2}(j_2)}{\sqrt{\bar{\mathcal{N}}_m}} \ket{e_2} \bra{\varphi_m^L} + \sum_{m=1}^{2L} \frac{[\varphi^L_{m,\alpha_2}(j_2)]^\ast}{\sqrt{\bar{\mathcal{N}}_m}} \ket{\varphi_m^R}  \bra{e_2}\right).
\end{align} 

We employ  the resolvent method to solve the evolution dynamics of two emitters coupled to the topological bath \cite{CCohenTannoudji1AtomSM,Economou_2006SM}. Using the Hamiltonian $\mathcal{H} = \mathcal{H}_p + \mathcal{H}_{e_1} + \mathcal{H}_{e_2} + \mathcal{V}_{\alpha_1\alpha_2}$  in Eqs.~(\ref{OBeffective228}), (\ref{OBeffective231}) and (\ref{OBeffective240}), the resolvent operator of the   whole system is defined as 
\begin{align}\label{gree125}
	\mathcal{G}(z) = \frac{1}{z-\mathcal{H}} = \frac{1}{z-\mathcal{H}_{pe}-\mathcal{V}_{\alpha_1\alpha_2}},
\end{align} 
where 
\begin{align}\label{gree1250}
	\mathcal{H}_{pe} = \mathcal{H}_p + \mathcal{H}_{e_1} + \mathcal{H}_{e_2}.
\end{align} 

We now consider   the single-excitation  spanned by   the emitter and bath Hamiltonian $\mathcal{H}_{pe}$, which consists of the atomic excitation $\{ \ket{e_1} \ket{\text{vac}},~ \ket{e_2} \ket{\text{vac}} \}$, and the quasi-particle excitation $\{ \ket{g} \ket{\varphi_m^R} \}$ with $m\in [1,~2L]$. The photon-emitter interaction term $\mathcal{V}_{\alpha_1\alpha_2}$ describes the coupling between the subspaces $\{ \ket{e_1} \ket{\text{vac}},~ \ket{e_2} \ket{\text{vac}} \}$ and $\{ \ket{g} \ket{\varphi_m^R} \}$. In the following, we use the following notations $\ket{e_1}:=\ket{e_1} \ket{\text{vac}}$, $\ket{e_2}:=\ket{e_2} \ket{\text{vac}}$ and $\ket{\varphi_m^R}:=\ket{g} \ket{\varphi_m^R}$ for convenience. Then, we define the projector operator
\begin{align}\label{OBeffective241}
	\mathcal{P} = \ket{e_1} \bra{e_1} + \ket{e_2} \bra{e_2},
\end{align} 
and its complementary
\begin{align}\label{OBeffective242}
	\mathcal{Q} = \sum_{m=1}^{2L} \ket{\varphi_m^R} \bra{\varphi_m^L}.
\end{align} 

Therefore, the constrained propagator $\mathcal{G}_p(z)$ is written as
\begin{align}\label{OBeffective243}
	\mathcal{G}_p(z) \equiv \mathcal{P} \mathcal{G}(z) \mathcal{P}.
\end{align} 

Starting from $(z-\mathcal{H}) \mathcal{G}(z)=\mathbf{1}$, and manipulating it on the right by $\mathcal{P}$ and on the left by $\mathcal{P}$ or $\mathcal{Q}$, the constrained propagator can be derived as
\begin{align}\label{OBeffective244}
	\mathcal{G}_p(z) = \frac{\mathcal{P}}{z - \mathcal{P} \mathcal{H}_{pe} \mathcal{P} - \mathcal{P} \Sigma(z) \mathcal{P} },
\end{align} 
where $\Sigma(z)$ is called the level-shift operator \cite{CCohenTannoudji1AtomSM}, defined as
\begin{align}\label{OBeffective245}
	\Sigma_(z) &= \mathcal{V}_{\alpha_1\alpha_2} + \mathcal{V}_{\alpha_1\alpha_2} \frac{\mathcal{Q}}{z - \mathcal{Q} \mathcal{H}_{pe} \mathcal{Q} - \mathcal{Q} \mathcal{V}_{\alpha_1\alpha_2} \mathcal{Q}} \mathcal{V}_{\alpha_1\alpha_2}, \nonumber \notag \\
	&= \mathcal{V}_{\alpha_1\alpha_2} + \mathcal{V}_{\alpha_1\alpha_2} \frac{\mathcal{Q}}{z-\mathcal{H}_p} \mathcal{V}_{\alpha_1\alpha_2},
\end{align} 
with 
\begin{align}\label{OBeffective246}
	\mathcal{V}_{\alpha_1\alpha_2} \frac{\mathcal{Q}}{z-\mathcal{H}_p} \mathcal{V}_{\alpha_1\alpha_2}  = &  \sum_{m=1}^{2L} \frac{g^2 \varphi^R_{m,\alpha_1}(j_1) [\varphi^L_{m,\alpha_1}(j_1)]^\ast / \bar{\mathcal{N}}_m  }{z-\varepsilon_m  +  i\kappa/2  } \ket{e_1} \bra{e_1} \nonumber \notag \\
	& + \sum_{m=1}^{2L} \frac{g^2 \varphi^R_{m,\alpha_1}(j_1) [\varphi^L_{m,\alpha_2}(j_2)] ^\ast / \bar{\mathcal{N}}_m  }{z-\varepsilon_m +  i\kappa/2 } \ket{e_1} \bra{e_2} \nonumber \notag \\
	&+ \sum_{m=1}^{2L} \frac{g^2 \varphi^R_{m,\alpha_2}(j_2) [\varphi^L_{m,\alpha_1}(j_1)] ^\ast / \bar{\mathcal{N}}_m  }{z-\varepsilon_m +  i\kappa/2 } \ket{e_2} \bra{e_1} \nonumber \notag \\
	& + \sum_{m=1}^{2L} \frac{g^2 \varphi^R_{m,\alpha_2}(j_2) [\varphi^L_{m,\alpha_2}(j_2)] ^\ast / \bar{\mathcal{N}}_m}{z-\varepsilon_m +  i\kappa/2 } \ket{e_2} \bra{e_2}.
\end{align} 

We now proceed to calculate the non-unitary  real-time dynamics governed by $\ket{\psi_t} = e^{-i\hat{\mathcal{H}}_\textrm{eff} t} \ket{\psi_0}$ for two emitters (labeled as 1 and 2)   coupled to sites $j_{1, \alpha_1}$ and $j_{2, \alpha_2}$ ($\alpha_1,\alpha_2=a ~\textrm{or} ~ b$) of the bath with $j_{2, \alpha_2}>j_{1, \alpha_1}$, respectively.  The initial state is chosen as one excited emitter $\ket{e_1}$ or $\ket{e_2}$ with $\ket{\psi_0} = \ket{e_n} \ket{\textrm{vac}}$ ($n=1 ~\textrm{or} ~ 2$), and  the time-evolved state can be expanded as
\begin{align}\label{phit2}
	\ket{\psi_t} = \left(\sum_{m=1}^{2N}c_{m}(t) \ket{\varphi_m^R} \bra{\textrm{vac}}  + \sum_{n=1}^{2}c_{e_n}(t)\ket{e_n}\bra{g}  \right)\ket{gg} \otimes \ket{\textrm{vac}}.
\end{align}
Then, the component $\mathcal{P}\ket{\psi_t}$ can be evaluated by the resolvent method  \cite{CCohenTannoudji1AtomSM} as
\begin{align}\label{ce}
	\mathcal{P}\ket{\psi_t} = \frac{i}{2\pi}\int_{-\infty}^{+\infty} d E \mathcal{G}_p \left(E+i0^+\right) e^{-iE t} \ket{\psi_0}.
\end{align}

Using Eq.~(\ref{ce}), we can express $\bs{c}_e(t) = [c_{e_1}(t), ~c_{e_2}(t)]^T$ as
\begin{align}\label{ce2}
	\bs{c}_e(t) = \frac{i}{2\pi}\int_{-\infty}^{+\infty} d E \mathcal{G}_p \left(E+i0^+\right) e^{-iE t} \bs{c}_e(0),
\end{align}
where, according to Eqs.~(\ref{OBeffective244})-(\ref{OBeffective246}), we explicitly write   $\mathcal{G}_p \left(z\right)$  as 
\begin{equation}\label{ce3}
	\mathcal{G}_p \left(E\right)= 
	\begingroup
	\setlength{\tabcolsep}{4pt}             
	\renewcommand{\arraystretch}{2}        
	\begin{pmatrix}
		\frac{1}{E-\Delta- \mathcal{T}(\alpha_1, \alpha_1)} & \frac{1}{E-  \mathcal{F}(\alpha_1, \alpha_2) \mathcal{T}(\alpha_1, \alpha_2) }    \\
		\frac{1}{E-  \mathcal{F}(\alpha_2, \alpha_1) \mathcal{T}(\alpha_1, \alpha_2) }   & \frac{1}{E-\Delta- \mathcal{T}(\alpha_2, \alpha_2) }    \\		 
	\end{pmatrix},
	\endgroup
\end{equation}
where
\begin{align}\label{ce81}
	\mathcal{T}(\alpha_1, \alpha_2) = g^2\sum_{m=1}^{2L} \frac{\varphi_{m,\alpha_1}(j_{1,\alpha_1}) \varphi_{m,\alpha_2}(j_{2,\alpha_2})}{(E-\varepsilon_m +  i\kappa/2 )\mathcal{N}_m},
\end{align}
and
\begin{align}\label{ce8}
	\mathcal{F}(\alpha_1, \alpha_2) = \left(\frac{J_1+\kappa/2}{J_1-\kappa/2}\right)^{\frac{\delta_{\alpha_1,b}}{2}} \left(\frac{J_1-\kappa/2}{J_1+\kappa/2}\right)^{\frac{\delta_{\alpha_2,b}}{2}}   \left(\frac{J_1+\kappa/2}{J_1-\kappa/2}\right)^{\frac{j_{1,\alpha_1} - j_{2,\alpha_2}}{2}}.
\end{align}

We assume a small $g$, a large band gap of the topological bath under OBCs and $\Delta = -i\kappa/2$. According to Eqs.~(\ref{ce2})-(\ref{ce8}) and Eq.~(\ref{OBeffective223}), the main contribution from the diagonal elements of the Green function $\mathcal{G}_p \left(z\right)$ to the time evolution is the dressed state for small $g$ and $\Delta=-i\kappa/2$. The off-diagonal elements contribute to the state exchanges between two emitters. Remarkably, such state exchange is asymmetric [see Eq.~(\ref{ce8})]. To be specific, when the emitter at the site $j_{2,\alpha_2}$ is initially excited, there is no excitation transferred to the emitter at   site $j_{1,\alpha_1}$ for the large distance $\abs{j_{1,\alpha_1} - j_{2,\alpha_2}}$ between them, due to the power-law decay of  $\mathcal{F}(\alpha_1, \alpha_2)$.  In principle, according to Eq.~(\ref{ce8}), as the $J_1$ approaches  $\kappa/2$ (while ensuring $J_1 \neq \kappa/2$),  the directional long-range emitter-emitter interaction is enhanced.  However, as $J_1$ gets closer to $\kappa/2$, the band gap of the open-boundary condition (OBC) spectrum of the SSH bath diminishes. A smaller band gap reduces the robustness of the system against disorder and increases the likelihood of coupling between the emitters and the bulk modes, which can undermine the desired directional transport properties.


\begin{thebibliography}{88}%
	\makeatletter
	\providecommand \@ifxundefined [1]{%
		\@ifx{#1\undefined}
	}%
	\providecommand \@ifnum [1]{%
		\ifnum #1\expandafter \@firstoftwo
		\else \expandafter \@secondoftwo
		\fi
	}%
	\providecommand \@ifx [1]{%
		\ifx #1\expandafter \@firstoftwo
		\else \expandafter \@secondoftwo
		\fi
	}%
	\providecommand \natexlab [1]{#1}%
	\providecommand \enquote  [1]{``#1''}%
	\providecommand \bibnamefont  [1]{#1}%
	\providecommand \bibfnamefont [1]{#1}%
	\providecommand \citenamefont [1]{#1}%
	\providecommand \href@noop [0]{\@secondoftwo}%
	\providecommand \href [0]{\begingroup \@sanitize@url \@href}%
	\providecommand \@href[1]{\@@startlink{#1}\@@href}%
	\providecommand \@@href[1]{\endgroup#1\@@endlink}%
	\providecommand \@sanitize@url [0]{\catcode `\\12\catcode `\$12\catcode
		`\&12\catcode `\#12\catcode `\^12\catcode `\_12\catcode `\%12\relax}%
	\providecommand \@@startlink[1]{}%
	\providecommand \@@endlink[0]{}%
	\providecommand \url  [0]{\begingroup\@sanitize@url \@url }%
	\providecommand \@url [1]{\endgroup\@href {#1}{\urlprefix }}%
	\providecommand \urlprefix  [0]{URL }%
	\providecommand \Eprint [0]{\href }%
	\providecommand \doibase [0]{http://dx.doi.org/}%
	\providecommand \selectlanguage [0]{\@gobble}%
	\providecommand \bibinfo  [0]{\@secondoftwo}%
	\providecommand \bibfield  [0]{\@secondoftwo}%
	\providecommand \translation [1]{[#1]}%
	\providecommand \BibitemOpen [0]{}%
	\providecommand \bibitemStop [0]{}%
	\providecommand \bibitemNoStop [0]{.\EOS\space}%
	\providecommand \EOS [0]{\spacefactor3000\relax}%
	\providecommand \BibitemShut  [1]{\csname bibitem#1\endcsname}%
	\let\auto@bib@innerbib\@empty
	\bibitem [{\citenamefont {Gonz\'alez-Tudela}\ and\ \citenamefont
		{Cirac}(2017{\natexlab{a}})}]{PhysRevLett.119.143602}%
	\BibitemOpen
	\bibfield  {author} {\bibinfo {author} {\bibfnamefont {A.}~\bibnamefont
			{Gonz\'alez-Tudela}}\ and\ \bibinfo {author} {\bibfnamefont {J.~I.}\
			\bibnamefont {Cirac}},\ }\bibfield  {title} {\enquote {\bibinfo {title}
			{Quantum emitters in two-dimensional structured reservoirs in the
				nonperturbative regime},}\ }\href {\doibase 10.1103/PhysRevLett.119.143602}
	{\bibfield  {journal} {\bibinfo  {journal} {Phys. Rev. Lett.}\ }\textbf
		{\bibinfo {volume} {119}},\ \bibinfo {pages} {143602} (\bibinfo {year}
		{2017}{\natexlab{a}})}\BibitemShut {NoStop}%
	\bibitem [{\citenamefont {Gonz\'alez-Tudela}\ and\ \citenamefont
		{Cirac}(2017{\natexlab{b}})}]{PhysRevA.96.043811}%
	\BibitemOpen
	\bibfield  {author} {\bibinfo {author} {\bibfnamefont {A.}~\bibnamefont
			{Gonz\'alez-Tudela}}\ and\ \bibinfo {author} {\bibfnamefont {J.~I.}\
			\bibnamefont {Cirac}},\ }\bibfield  {title} {\enquote {\bibinfo {title}
			{Markovian and non-{M}arkovian dynamics of quantum emitters coupled to
				two-dimensional structured reservoirs},}\ }\href {\doibase
		10.1103/PhysRevA.96.043811} {\bibfield  {journal} {\bibinfo  {journal} {Phys.
				Rev. A}\ }\textbf {\bibinfo {volume} {96}},\ \bibinfo {pages} {043811}
		(\bibinfo {year} {2017}{\natexlab{b}})}\BibitemShut {NoStop}%
	\bibitem [{\citenamefont {Bienias}\ \emph {et~al.}(2022)\citenamefont
		{Bienias}, \citenamefont {Boettcher}, \citenamefont {Belyansky},
		\citenamefont {Koll\'ar},\ and\ \citenamefont
		{Gorshkov}}]{PhysRevLett.128.013601}%
	\BibitemOpen
	\bibfield  {author} {\bibinfo {author} {\bibfnamefont {P.}~\bibnamefont
			{Bienias}}, \bibinfo {author} {\bibfnamefont {I.}~\bibnamefont {Boettcher}},
		\bibinfo {author} {\bibfnamefont {R.}~\bibnamefont {Belyansky}}, \bibinfo
		{author} {\bibfnamefont {A.~J.}\ \bibnamefont {Koll\'ar}}, \ and\ \bibinfo
		{author} {\bibfnamefont {A.~V.}\ \bibnamefont {Gorshkov}},\ }\bibfield
	{title} {\enquote {\bibinfo {title} {Circuit quantum electrodynamics in
				hyperbolic space: {F}rom photon bound states to frustrated spin models},}\
	}\href {\doibase 10.1103/PhysRevLett.128.013601} {\bibfield  {journal}
		{\bibinfo  {journal} {Phys. Rev. Lett.}\ }\textbf {\bibinfo {volume} {128}},\
		\bibinfo {pages} {013601} (\bibinfo {year} {2022})}\BibitemShut {NoStop}%
	\bibitem [{\citenamefont {Kockum}\ \emph {et~al.}(2018)\citenamefont {Kockum},
		\citenamefont {Johansson},\ and\ \citenamefont
		{Nori}}]{PhysRevLett.120.140404}%
	\BibitemOpen
	\bibfield  {author} {\bibinfo {author} {\bibfnamefont {A.~F.}\ \bibnamefont
			{Kockum}}, \bibinfo {author} {\bibfnamefont {G.}~\bibnamefont {Johansson}}, \
		and\ \bibinfo {author} {\bibfnamefont {F.}~\bibnamefont {Nori}},\ }\bibfield
	{title} {\enquote {\bibinfo {title} {Decoherence-free interaction between
				giant atoms in waveguide quantum electrodynamics},}\ }\href {\doibase
		10.1103/PhysRevLett.120.140404} {\bibfield  {journal} {\bibinfo  {journal}
			{Phys. Rev. Lett.}\ }\textbf {\bibinfo {volume} {120}},\ \bibinfo {pages}
		{140404} (\bibinfo {year} {2018})}\BibitemShut {NoStop}%
	\bibitem [{\citenamefont {Garc\'{\i}a-Elcano}\ \emph
		{et~al.}(2020)\citenamefont {Garc\'{\i}a-Elcano}, \citenamefont
		{Gonz\'alez-Tudela},\ and\ \citenamefont
		{Bravo-Abad}}]{PhysRevLett.125.163602}%
	\BibitemOpen
	\bibfield  {author} {\bibinfo {author} {\bibfnamefont {I.}~\bibnamefont
			{Garc\'{\i}a-Elcano}}, \bibinfo {author} {\bibfnamefont {A.}~\bibnamefont
			{Gonz\'alez-Tudela}}, \ and\ \bibinfo {author} {\bibfnamefont
			{J.}~\bibnamefont {Bravo-Abad}},\ }\bibfield  {title} {\enquote {\bibinfo
			{title} {Tunable and robust long-range coherent interactions between quantum
				emitters mediated by {W}eyl bound states},}\ }\href {\doibase
		10.1103/PhysRevLett.125.163602} {\bibfield  {journal} {\bibinfo  {journal}
			{Phys. Rev. Lett.}\ }\textbf {\bibinfo {volume} {125}},\ \bibinfo {pages}
		{163602} (\bibinfo {year} {2020})}\BibitemShut {NoStop}%
	\bibitem [{\citenamefont {Wang}\ \emph {et~al.}(2021)\citenamefont {Wang},
		\citenamefont {Liu}, \citenamefont {Kockum}, \citenamefont {Li},\ and\
		\citenamefont {Nori}}]{PhysRevLett.126.043602}%
	\BibitemOpen
	\bibfield  {author} {\bibinfo {author} {\bibfnamefont {X.}~\bibnamefont
			{Wang}}, \bibinfo {author} {\bibfnamefont {T.}~\bibnamefont {Liu}}, \bibinfo
		{author} {\bibfnamefont {A.~F.}\ \bibnamefont {Kockum}}, \bibinfo {author}
		{\bibfnamefont {H.-R.}\ \bibnamefont {Li}}, \ and\ \bibinfo {author}
		{\bibfnamefont {F.}~\bibnamefont {Nori}},\ }\bibfield  {title} {\enquote
		{\bibinfo {title} {Tunable chiral bound states with giant atoms},}\ }\href
	{\doibase 10.1103/PhysRevLett.126.043602} {\bibfield  {journal} {\bibinfo
			{journal} {Phys. Rev. Lett.}\ }\textbf {\bibinfo {volume} {126}},\ \bibinfo
		{pages} {043602} (\bibinfo {year} {2021})}\BibitemShut {NoStop}%
	\bibitem [{\citenamefont {Leonforte}\ \emph {et~al.}(2021)\citenamefont
		{Leonforte}, \citenamefont {Carollo},\ and\ \citenamefont
		{Ciccarello}}]{PhysRevLett.126.063601}%
	\BibitemOpen
	\bibfield  {author} {\bibinfo {author} {\bibfnamefont {L.}~\bibnamefont
			{Leonforte}}, \bibinfo {author} {\bibfnamefont {A.}~\bibnamefont {Carollo}},
		\ and\ \bibinfo {author} {\bibfnamefont {F.}~\bibnamefont {Ciccarello}},\
	}\bibfield  {title} {\enquote {\bibinfo {title} {Vacancy-like dressed states
				in topological waveguide qed},}\ }\href {\doibase
		10.1103/PhysRevLett.126.063601} {\bibfield  {journal} {\bibinfo  {journal}
			{Phys. Rev. Lett.}\ }\textbf {\bibinfo {volume} {126}},\ \bibinfo {pages}
		{063601} (\bibinfo {year} {2021})}\BibitemShut {NoStop}%
	\bibitem [{\citenamefont {De~Bernardis}\ \emph {et~al.}(2021)\citenamefont
		{De~Bernardis}, \citenamefont {Cian}, \citenamefont {Carusotto},
		\citenamefont {Hafezi},\ and\ \citenamefont {Rabl}}]{PhysRevLett.126.103603}%
	\BibitemOpen
	\bibfield  {author} {\bibinfo {author} {\bibfnamefont {D.}~\bibnamefont
			{De~Bernardis}}, \bibinfo {author} {\bibfnamefont {Z.-P.}\ \bibnamefont
			{Cian}}, \bibinfo {author} {\bibfnamefont {I.}~\bibnamefont {Carusotto}},
		\bibinfo {author} {\bibfnamefont {M.}~\bibnamefont {Hafezi}}, \ and\ \bibinfo
		{author} {\bibfnamefont {P.}~\bibnamefont {Rabl}},\ }\bibfield  {title}
	{\enquote {\bibinfo {title} {Light-matter interactions in synthetic magnetic
				fields: {L}andau-photon polaritons},}\ }\href {\doibase
		10.1103/PhysRevLett.126.103603} {\bibfield  {journal} {\bibinfo  {journal}
			{Phys. Rev. Lett.}\ }\textbf {\bibinfo {volume} {126}},\ \bibinfo {pages}
		{103603} (\bibinfo {year} {2021})}\BibitemShut {NoStop}%
	\bibitem [{\citenamefont {Perczel}\ \emph {et~al.}(2017)\citenamefont
		{Perczel}, \citenamefont {Borregaard}, \citenamefont {Chang}, \citenamefont
		{Pichler}, \citenamefont {Yelin}, \citenamefont {Zoller},\ and\ \citenamefont
		{Lukin}}]{PhysRevLett.119.023603}%
	\BibitemOpen
	\bibfield  {author} {\bibinfo {author} {\bibfnamefont {J.}~\bibnamefont
			{Perczel}}, \bibinfo {author} {\bibfnamefont {J.}~\bibnamefont {Borregaard}},
		\bibinfo {author} {\bibfnamefont {D.~E.}\ \bibnamefont {Chang}}, \bibinfo
		{author} {\bibfnamefont {H.}~\bibnamefont {Pichler}}, \bibinfo {author}
		{\bibfnamefont {S.~F.}\ \bibnamefont {Yelin}}, \bibinfo {author}
		{\bibfnamefont {P.}~\bibnamefont {Zoller}}, \ and\ \bibinfo {author}
		{\bibfnamefont {M.~D.}\ \bibnamefont {Lukin}},\ }\bibfield  {title} {\enquote
		{\bibinfo {title} {Topological quantum optics in two-dimensional atomic
				arrays},}\ }\href {\doibase 10.1103/PhysRevLett.119.023603} {\bibfield
		{journal} {\bibinfo  {journal} {Phys. Rev. Lett.}\ }\textbf {\bibinfo
			{volume} {119}},\ \bibinfo {pages} {023603} (\bibinfo {year}
		{2017})}\BibitemShut {NoStop}%
	\bibitem [{\citenamefont {Bettles}\ \emph {et~al.}(2017)\citenamefont
		{Bettles}, \citenamefont {Min\'a\ifmmode~\check{r}\else \v{r}\fi{}},
		\citenamefont {Adams}, \citenamefont {Lesanovsky},\ and\ \citenamefont
		{Olmos}}]{PhysRevA.96.041603}%
	\BibitemOpen
	\bibfield  {author} {\bibinfo {author} {\bibfnamefont {R.~J.}\ \bibnamefont
			{Bettles}}, \bibinfo {author} {\bibfnamefont {J.}~\bibnamefont
			{Min\'a\ifmmode~\check{r}\else \v{r}\fi{}}}, \bibinfo {author} {\bibfnamefont
			{C.~S.}\ \bibnamefont {Adams}}, \bibinfo {author} {\bibfnamefont
			{I.}~\bibnamefont {Lesanovsky}}, \ and\ \bibinfo {author} {\bibfnamefont
			{B.}~\bibnamefont {Olmos}},\ }\bibfield  {title} {\enquote {\bibinfo {title}
			{Topological properties of a dense atomic lattice gas},}\ }\href {\doibase
		10.1103/PhysRevA.96.041603} {\bibfield  {journal} {\bibinfo  {journal} {Phys.
				Rev. A}\ }\textbf {\bibinfo {volume} {96}},\ \bibinfo {pages} {041603}
		(\bibinfo {year} {2017})}\BibitemShut {NoStop}%
	\bibitem [{\citenamefont {Barik}\ \emph {et~al.}(2018)\citenamefont {Barik},
		\citenamefont {Karasahin}, \citenamefont {Flower}, \citenamefont {Cai},
		\citenamefont {Miyake}, \citenamefont {DeGottardi}, \citenamefont {Hafezi},\
		and\ \citenamefont {Waks}}]{Barik2018}%
	\BibitemOpen
	\bibfield  {author} {\bibinfo {author} {\bibfnamefont {S.}~\bibnamefont
			{Barik}}, \bibinfo {author} {\bibfnamefont {A.}~\bibnamefont {Karasahin}},
		\bibinfo {author} {\bibfnamefont {C.}~\bibnamefont {Flower}}, \bibinfo
		{author} {\bibfnamefont {T.}~\bibnamefont {Cai}}, \bibinfo {author}
		{\bibfnamefont {H.}~\bibnamefont {Miyake}}, \bibinfo {author} {\bibfnamefont
			{W.}~\bibnamefont {DeGottardi}}, \bibinfo {author} {\bibfnamefont
			{M.}~\bibnamefont {Hafezi}}, \ and\ \bibinfo {author} {\bibfnamefont
			{E.}~\bibnamefont {Waks}},\ }\bibfield  {title} {\enquote {\bibinfo {title}
			{A topological quantum optics interface},}\ }\href {\doibase
		10.1126/science.aaq0327} {\bibfield  {journal} {\bibinfo  {journal}
			{Science}\ }\textbf {\bibinfo {volume} {359}},\ \bibinfo {pages} {666}
		(\bibinfo {year} {2018})}\BibitemShut {NoStop}%
	\bibitem [{\citenamefont {Bello}\ \emph {et~al.}(2019)\citenamefont {Bello},
		\citenamefont {Platero}, \citenamefont {Cirac},\ and\ \citenamefont
		{Gonz{\'{a}}lez-Tudela}}]{Bello2019}%
	\BibitemOpen
	\bibfield  {author} {\bibinfo {author} {\bibfnamefont {M.}~\bibnamefont
			{Bello}}, \bibinfo {author} {\bibfnamefont {G.}~\bibnamefont {Platero}},
		\bibinfo {author} {\bibfnamefont {J.~I.}\ \bibnamefont {Cirac}}, \ and\
		\bibinfo {author} {\bibfnamefont {A.}~\bibnamefont {Gonz{\'{a}}lez-Tudela}},\
	}\bibfield  {title} {\enquote {\bibinfo {title} {Unconventional quantum
				optics in topological waveguide {QED}},}\ }\href
	{https://doi.org/10.1126/sciadv.aaw0297} {\bibfield  {journal} {\bibinfo
			{journal} {Sci. Adv.}\ }\textbf {\bibinfo {volume} {5}},\ \bibinfo {pages}
		{eaaw0297} (\bibinfo {year} {2019})}\BibitemShut {NoStop}%
	\bibitem [{\citenamefont {Kim}\ \emph {et~al.}(2021)\citenamefont {Kim},
		\citenamefont {Zhang}, \citenamefont {Ferreira}, \citenamefont {Banker},
		\citenamefont {Iverson}, \citenamefont {Sipahigil}, \citenamefont {Bello},
		\citenamefont {Gonz\'alez-Tudela}, \citenamefont {Mirhosseini},\ and\
		\citenamefont {Painter}}]{PhysRevX.11.011015}%
	\BibitemOpen
	\bibfield  {author} {\bibinfo {author} {\bibfnamefont {E.}~\bibnamefont
			{Kim}}, \bibinfo {author} {\bibfnamefont {X.}~\bibnamefont {Zhang}}, \bibinfo
		{author} {\bibfnamefont {V.~S.}\ \bibnamefont {Ferreira}}, \bibinfo {author}
		{\bibfnamefont {J.}~\bibnamefont {Banker}}, \bibinfo {author} {\bibfnamefont
			{J.~K.}\ \bibnamefont {Iverson}}, \bibinfo {author} {\bibfnamefont
			{A.}~\bibnamefont {Sipahigil}}, \bibinfo {author} {\bibfnamefont
			{M.}~\bibnamefont {Bello}}, \bibinfo {author} {\bibfnamefont
			{A.}~\bibnamefont {Gonz\'alez-Tudela}}, \bibinfo {author} {\bibfnamefont
			{M.}~\bibnamefont {Mirhosseini}}, \ and\ \bibinfo {author} {\bibfnamefont
			{O.}~\bibnamefont {Painter}},\ }\bibfield  {title} {\enquote {\bibinfo
			{title} {Quantum electrodynamics in a topological waveguide},}\ }\href
	{\doibase 10.1103/PhysRevX.11.011015} {\bibfield  {journal} {\bibinfo
			{journal} {Phys. Rev. X}\ }\textbf {\bibinfo {volume} {11}},\ \bibinfo
		{pages} {011015} (\bibinfo {year} {2021})}\BibitemShut {NoStop}%
	\bibitem [{\citenamefont {Vega}\ \emph {et~al.}(2021)\citenamefont {Vega},
		\citenamefont {Bello}, \citenamefont {Porras},\ and\ \citenamefont
		{Gonz\'alez-Tudela}}]{PhysRevA.104.053522}%
	\BibitemOpen
	\bibfield  {author} {\bibinfo {author} {\bibfnamefont {C.}~\bibnamefont
			{Vega}}, \bibinfo {author} {\bibfnamefont {M.}~\bibnamefont {Bello}},
		\bibinfo {author} {\bibfnamefont {D.}~\bibnamefont {Porras}}, \ and\ \bibinfo
		{author} {\bibfnamefont {A.}~\bibnamefont {Gonz\'alez-Tudela}},\ }\bibfield
	{title} {\enquote {\bibinfo {title} {Qubit-photon bound states in topological
				waveguides with long-range hoppings},}\ }\href {\doibase
		10.1103/PhysRevA.104.053522} {\bibfield  {journal} {\bibinfo  {journal}
			{Phys. Rev. A}\ }\textbf {\bibinfo {volume} {104}},\ \bibinfo {pages}
		{053522} (\bibinfo {year} {2021})}\BibitemShut {NoStop}%
	\bibitem [{\citenamefont {Vega}\ \emph {et~al.}(2023)\citenamefont {Vega},
		\citenamefont {Porras},\ and\ \citenamefont
		{Gonz\'alez-Tudela}}]{PhysRevResearch.5.023031}%
	\BibitemOpen
	\bibfield  {author} {\bibinfo {author} {\bibfnamefont {C.}~\bibnamefont
			{Vega}}, \bibinfo {author} {\bibfnamefont {D.}~\bibnamefont {Porras}}, \ and\
		\bibinfo {author} {\bibfnamefont {A.}~\bibnamefont {Gonz\'alez-Tudela}},\
	}\bibfield  {title} {\enquote {\bibinfo {title} {Topological multimode
				waveguide {QED}},}\ }\href {\doibase 10.1103/PhysRevResearch.5.023031}
	{\bibfield  {journal} {\bibinfo  {journal} {Phys. Rev. Res.}\ }\textbf
		{\bibinfo {volume} {5}},\ \bibinfo {pages} {023031} (\bibinfo {year}
		{2023})}\BibitemShut {NoStop}%
	\bibitem [{\citenamefont {Bello}\ \emph {et~al.}(2022)\citenamefont {Bello},
		\citenamefont {Platero},\ and\ \citenamefont
		{Gonz\'alez-Tudela}}]{PRXQuantum.3.010336}%
	\BibitemOpen
	\bibfield  {author} {\bibinfo {author} {\bibfnamefont {M.}~\bibnamefont
			{Bello}}, \bibinfo {author} {\bibfnamefont {G.}~\bibnamefont {Platero}}, \
		and\ \bibinfo {author} {\bibfnamefont {A.}~\bibnamefont
			{Gonz\'alez-Tudela}},\ }\bibfield  {title} {\enquote {\bibinfo {title} {Spin
				many-body phases in standard- and topological-waveguide {QED} simulators},}\
	}\href {\doibase 10.1103/PRXQuantum.3.010336} {\bibfield  {journal} {\bibinfo
			{journal} {PRX Quantum}\ }\textbf {\bibinfo {volume} {3}},\ \bibinfo {pages}
		{010336} (\bibinfo {year} {2022})}\BibitemShut {NoStop}%
	\bibitem [{\citenamefont {Nie}\ \emph {et~al.}(2021)\citenamefont {Nie},
		\citenamefont {Antezza}, \citenamefont {Liu},\ and\ \citenamefont
		{Nori}}]{PhysRevLett.127.250402}%
	\BibitemOpen
	\bibfield  {author} {\bibinfo {author} {\bibfnamefont {W.}~\bibnamefont
			{Nie}}, \bibinfo {author} {\bibfnamefont {M.}~\bibnamefont {Antezza}},
		\bibinfo {author} {\bibfnamefont {Y.-x.}\ \bibnamefont {Liu}}, \ and\
		\bibinfo {author} {\bibfnamefont {F.}~\bibnamefont {Nori}},\ }\bibfield
	{title} {\enquote {\bibinfo {title} {Dissipative topological phase transition
				with strong system-environment coupling},}\ }\href {\doibase
		10.1103/PhysRevLett.127.250402} {\bibfield  {journal} {\bibinfo  {journal}
			{Phys. Rev. Lett.}\ }\textbf {\bibinfo {volume} {127}},\ \bibinfo {pages}
		{250402} (\bibinfo {year} {2021})}\BibitemShut {NoStop}%
	\bibitem [{\citenamefont {Chang}\ \emph {et~al.}(2018)\citenamefont {Chang},
		\citenamefont {Douglas}, \citenamefont {Gonz\'alez-Tudela}, \citenamefont
		{Hung},\ and\ \citenamefont {Kimble}}]{RevModPhys.90.031002}%
	\BibitemOpen
	\bibfield  {author} {\bibinfo {author} {\bibfnamefont {D.~E.}\ \bibnamefont
			{Chang}}, \bibinfo {author} {\bibfnamefont {J.~S.}\ \bibnamefont {Douglas}},
		\bibinfo {author} {\bibfnamefont {A.}~\bibnamefont {Gonz\'alez-Tudela}},
		\bibinfo {author} {\bibfnamefont {C.-L.}\ \bibnamefont {Hung}}, \ and\
		\bibinfo {author} {\bibfnamefont {H.~J.}\ \bibnamefont {Kimble}},\ }\bibfield
	{title} {\enquote {\bibinfo {title} {Colloquium: {Q}uantum matter built from
				nanoscopic lattices of atoms and photons},}\ }\href {\doibase
		10.1103/RevModPhys.90.031002} {\bibfield  {journal} {\bibinfo  {journal}
			{Rev. Mod. Phys.}\ }\textbf {\bibinfo {volume} {90}},\ \bibinfo {pages}
		{031002} (\bibinfo {year} {2018})}\BibitemShut {NoStop}%
	\bibitem [{\citenamefont {Tang}\ \emph {et~al.}(2022)\citenamefont {Tang},
		\citenamefont {Nie}, \citenamefont {Tang}, \citenamefont {Chen},
		\citenamefont {Su}, \citenamefont {Lu}, \citenamefont {Nori},\ and\
		\citenamefont {Xia}}]{PhysRevLett.128.203602}%
	\BibitemOpen
	\bibfield  {author} {\bibinfo {author} {\bibfnamefont {J.-S.}\ \bibnamefont
			{Tang}}, \bibinfo {author} {\bibfnamefont {W.}~\bibnamefont {Nie}}, \bibinfo
		{author} {\bibfnamefont {L.}~\bibnamefont {Tang}}, \bibinfo {author}
		{\bibfnamefont {M.}~\bibnamefont {Chen}}, \bibinfo {author} {\bibfnamefont
			{X.}~\bibnamefont {Su}}, \bibinfo {author} {\bibfnamefont {Y.}~\bibnamefont
			{Lu}}, \bibinfo {author} {\bibfnamefont {F.}~\bibnamefont {Nori}}, \ and\
		\bibinfo {author} {\bibfnamefont {K.}~\bibnamefont {Xia}},\ }\bibfield
	{title} {\enquote {\bibinfo {title} {Nonreciprocal single-photon band
				structure},}\ }\href {\doibase 10.1103/PhysRevLett.128.203602} {\bibfield
		{journal} {\bibinfo  {journal} {Phys. Rev. Lett.}\ }\textbf {\bibinfo
			{volume} {128}},\ \bibinfo {pages} {203602} (\bibinfo {year}
		{2022})}\BibitemShut {NoStop}%
	\bibitem [{\citenamefont {Sheremet}\ \emph {et~al.}(2023)\citenamefont
		{Sheremet}, \citenamefont {Petrov}, \citenamefont {Iorsh}, \citenamefont
		{Poshakinskiy},\ and\ \citenamefont {Poddubny}}]{RevModPhys.95.015002}%
	\BibitemOpen
	\bibfield  {author} {\bibinfo {author} {\bibfnamefont {A.~S.}\ \bibnamefont
			{Sheremet}}, \bibinfo {author} {\bibfnamefont {M.~I.}\ \bibnamefont
			{Petrov}}, \bibinfo {author} {\bibfnamefont {I.~V.}\ \bibnamefont {Iorsh}},
		\bibinfo {author} {\bibfnamefont {A.~V.}\ \bibnamefont {Poshakinskiy}}, \
		and\ \bibinfo {author} {\bibfnamefont {A.~N.}\ \bibnamefont {Poddubny}},\
	}\bibfield  {title} {\enquote {\bibinfo {title} {Waveguide quantum
				electrodynamics: Collective radiance and photon-photon correlations},}\
	}\href {\doibase 10.1103/RevModPhys.95.015002} {\bibfield  {journal}
		{\bibinfo  {journal} {Rev. Mod. Phys.}\ }\textbf {\bibinfo {volume} {95}},\
		\bibinfo {pages} {015002} (\bibinfo {year} {2023})}\BibitemShut {NoStop}%
	\bibitem [{\citenamefont {Wang}\ \emph
		{et~al.}(2024{\natexlab{a}})\citenamefont {Wang}, \citenamefont {Zhu},
		\citenamefont {Liu},\ and\ \citenamefont {Nori}}]{PhysRevResearch.6.013279}%
	\BibitemOpen
	\bibfield  {author} {\bibinfo {author} {\bibfnamefont {X.}~\bibnamefont
			{Wang}}, \bibinfo {author} {\bibfnamefont {H.-B.}\ \bibnamefont {Zhu}},
		\bibinfo {author} {\bibfnamefont {T.}~\bibnamefont {Liu}}, \ and\ \bibinfo
		{author} {\bibfnamefont {F.}~\bibnamefont {Nori}},\ }\bibfield  {title}
	{\enquote {\bibinfo {title} {Realizing quantum optics in structured
				environments with giant atoms},}\ }\href {\doibase
		10.1103/PhysRevResearch.6.013279} {\bibfield  {journal} {\bibinfo  {journal}
			{Phys. Rev. Res.}\ }\textbf {\bibinfo {volume} {6}},\ \bibinfo {pages}
		{013279} (\bibinfo {year} {2024}{\natexlab{a}})}\BibitemShut {NoStop}%
	\bibitem [{\citenamefont {Gao}\ \emph {et~al.}(2024)\citenamefont {Gao},
		\citenamefont {Li}, \citenamefont {Wu}, \citenamefont {Liu},\ and\
		\citenamefont {Wang}}]{PhysRevA.110.053706}%
	\BibitemOpen
	\bibfield  {author} {\bibinfo {author} {\bibfnamefont {Z.-M.}\ \bibnamefont
			{Gao}}, \bibinfo {author} {\bibfnamefont {J.-Q.}\ \bibnamefont {Li}},
		\bibinfo {author} {\bibfnamefont {Y.-H.}\ \bibnamefont {Wu}}, \bibinfo
		{author} {\bibfnamefont {W.-X.}\ \bibnamefont {Liu}}, \ and\ \bibinfo
		{author} {\bibfnamefont {X.}~\bibnamefont {Wang}},\ }\bibfield  {title}
	{\enquote {\bibinfo {title} {Harnessing spontaneous emission of correlated
				photon pairs from ladder-type giant atoms},}\ }\href {\doibase
		10.1103/PhysRevA.110.053706} {\bibfield  {journal} {\bibinfo  {journal}
			{Phys. Rev. A}\ }\textbf {\bibinfo {volume} {110}},\ \bibinfo {pages}
		{053706} (\bibinfo {year} {2024})}\BibitemShut {NoStop}%
	\bibitem [{\citenamefont {Lu}\ \emph {et~al.}(2024)\citenamefont {Lu},
		\citenamefont {Tian}, \citenamefont {Lü},\ and\ \citenamefont
		{Shang}}]{arXiv:2405.03675}%
	\BibitemOpen
	\bibfield  {author} {\bibinfo {author} {\bibfnamefont {Z.-G.}\ \bibnamefont
			{Lu}}, \bibinfo {author} {\bibfnamefont {G.}~\bibnamefont {Tian}}, \bibinfo
		{author} {\bibfnamefont {X.-Y.}\ \bibnamefont {Lü}}, \ and\ \bibinfo
		{author} {\bibfnamefont {C.}~\bibnamefont {Shang}},\ }\bibfield  {title}
	{\enquote {\bibinfo {title} {Topological quantum batteries},}\ }\href@noop {}
	{\bibfield  {journal} {\bibinfo  {journal} {arXiv:2405.03675}\ } (\bibinfo
		{year} {2024})}\BibitemShut {NoStop}%
	\bibitem [{\citenamefont {Wang}\ \emph
		{et~al.}(2024{\natexlab{b}})\citenamefont {Wang}, \citenamefont {Li},
		\citenamefont {Liu}, \citenamefont {Miranowicz},\ and\ \citenamefont
		{Nori}}]{PhysRevResearch.6.043226}%
	\BibitemOpen
	\bibfield  {author} {\bibinfo {author} {\bibfnamefont {X.}~\bibnamefont
			{Wang}}, \bibinfo {author} {\bibfnamefont {J.-Q.}\ \bibnamefont {Li}},
		\bibinfo {author} {\bibfnamefont {T.}~\bibnamefont {Liu}}, \bibinfo {author}
		{\bibfnamefont {A.}~\bibnamefont {Miranowicz}}, \ and\ \bibinfo {author}
		{\bibfnamefont {F.}~\bibnamefont {Nori}},\ }\bibfield  {title} {\enquote
		{\bibinfo {title} {Long-range four-body interactions in structured nonlinear
				photonic waveguides},}\ }\href {\doibase 10.1103/PhysRevResearch.6.043226}
	{\bibfield  {journal} {\bibinfo  {journal} {Phys. Rev. Res.}\ }\textbf
		{\bibinfo {volume} {6}},\ \bibinfo {pages} {043226} (\bibinfo {year}
		{2024}{\natexlab{b}})}\BibitemShut {NoStop}%
	\bibitem [{\citenamefont {González-Tudela}\ \emph {et~al.}(2024)\citenamefont
		{González-Tudela}, \citenamefont {Reiserer}, \citenamefont
		{García-Ripoll},\ and\ \citenamefont {García-Vidal}}]{GonzlezTudela2024}%
	\BibitemOpen
	\bibfield  {author} {\bibinfo {author} {\bibfnamefont {A.}~\bibnamefont
			{González-Tudela}}, \bibinfo {author} {\bibfnamefont {A.}~\bibnamefont
			{Reiserer}}, \bibinfo {author} {\bibfnamefont {J.~J.}\ \bibnamefont
			{García-Ripoll}}, \ and\ \bibinfo {author} {\bibfnamefont {F.~J.}\
			\bibnamefont {García-Vidal}},\ }\bibfield  {title} {\enquote {\bibinfo
			{title} {Light–matter interactions in quantum nanophotonic devices},}\
	}\href {\doibase 10.1038/s42254-023-00681-1} {\bibfield  {journal} {\bibinfo
			{journal} {Nat. Rev. Phys.}\ }\textbf {\bibinfo {volume} {6}},\ \bibinfo
		{pages} {166} (\bibinfo {year} {2024})}\BibitemShut {NoStop}%
	\bibitem [{\citenamefont {Wang}\ \emph
		{et~al.}(2024{\natexlab{c}})\citenamefont {Wang}, \citenamefont {Li},
		\citenamefont {Wang}, \citenamefont {Kockum}, \citenamefont {Du},
		\citenamefont {Liu},\ and\ \citenamefont {Nori}}]{arXiv:2404.09829}%
	\BibitemOpen
	\bibfield  {author} {\bibinfo {author} {\bibfnamefont {X.}~\bibnamefont
			{Wang}}, \bibinfo {author} {\bibfnamefont {J.-Q.}\ \bibnamefont {Li}},
		\bibinfo {author} {\bibfnamefont {Z.}~\bibnamefont {Wang}}, \bibinfo {author}
		{\bibfnamefont {A.~F.}\ \bibnamefont {Kockum}}, \bibinfo {author}
		{\bibfnamefont {L.}~\bibnamefont {Du}}, \bibinfo {author} {\bibfnamefont
			{T.}~\bibnamefont {Liu}}, \ and\ \bibinfo {author} {\bibfnamefont
			{F.}~\bibnamefont {Nori}},\ }\bibfield  {title} {\enquote {\bibinfo {title}
			{Nonlinear chiral quantum optics with giant-emitter pairs},}\ }\href@noop {}
	{\bibfield  {journal} {\bibinfo  {journal} {arXiv:2404.09829}\ } (\bibinfo
		{year} {2024}{\natexlab{c}})}\BibitemShut {NoStop}%
	\bibitem [{\citenamefont {Ashida}\ \emph {et~al.}(2020)\citenamefont {Ashida},
		\citenamefont {Gong},\ and\ \citenamefont {Ueda}}]{Ashida2020}%
	\BibitemOpen
	\bibfield  {author} {\bibinfo {author} {\bibfnamefont {Y.}~\bibnamefont
			{Ashida}}, \bibinfo {author} {\bibfnamefont {Z.}~\bibnamefont {Gong}}, \ and\
		\bibinfo {author} {\bibfnamefont {M.}~\bibnamefont {Ueda}},\ }\bibfield
	{title} {\enquote {\bibinfo {title} {Non-{H}ermitian physics},}\ }\href
	{\doibase 10.1080/00018732.2021.1876991} {\bibfield  {journal} {\bibinfo
			{journal} {Adv. Phys.}\ }\textbf {\bibinfo {volume} {69}},\ \bibinfo {pages}
		{249} (\bibinfo {year} {2020})}\BibitemShut {NoStop}%
	\bibitem [{\citenamefont {Gao}\ \emph {et~al.}(2015)\citenamefont {Gao},
		\citenamefont {Estrecho}, \citenamefont {Bliokh}, \citenamefont {Liew},
		\citenamefont {Fraser}, \citenamefont {Brodbeck}, \citenamefont {Kamp},
		\citenamefont {Schneider}, \citenamefont {H\"{o}fling}, \citenamefont
		{Yamamoto}, \citenamefont {Nori}, \citenamefont {Kivshar}, \citenamefont
		{Truscott}, \citenamefont {Dall},\ and\ \citenamefont
		{Ostrovskaya}}]{Gao2015}%
	\BibitemOpen
	\bibfield  {author} {\bibinfo {author} {\bibfnamefont {T.}~\bibnamefont
			{Gao}}, \bibinfo {author} {\bibfnamefont {E.}~\bibnamefont {Estrecho}},
		\bibinfo {author} {\bibfnamefont {K.~Y.}\ \bibnamefont {Bliokh}}, \bibinfo
		{author} {\bibfnamefont {T.~C.~H.}\ \bibnamefont {Liew}}, \bibinfo {author}
		{\bibfnamefont {M.~D.}\ \bibnamefont {Fraser}}, \bibinfo {author}
		{\bibfnamefont {S.}~\bibnamefont {Brodbeck}}, \bibinfo {author}
		{\bibfnamefont {M.}~\bibnamefont {Kamp}}, \bibinfo {author} {\bibfnamefont
			{C.}~\bibnamefont {Schneider}}, \bibinfo {author} {\bibfnamefont
			{S.}~\bibnamefont {H\"{o}fling}}, \bibinfo {author} {\bibfnamefont
			{Y.}~\bibnamefont {Yamamoto}}, \bibinfo {author} {\bibfnamefont
			{F.}~\bibnamefont {Nori}}, \bibinfo {author} {\bibfnamefont {Y.~S.}\
			\bibnamefont {Kivshar}}, \bibinfo {author} {\bibfnamefont {A.~G.}\
			\bibnamefont {Truscott}}, \bibinfo {author} {\bibfnamefont {R.~G.}\
			\bibnamefont {Dall}}, \ and\ \bibinfo {author} {\bibfnamefont {E.~A.}\
			\bibnamefont {Ostrovskaya}},\ }\bibfield  {title} {\enquote {\bibinfo {title}
			{Observation of non-{H}ermitian degeneracies in a chaotic exciton-polariton
				billiard},}\ }\href {http://dx.doi.org/10.1038/nature15522} {\bibfield
		{journal} {\bibinfo  {journal} {Nature}\ }\textbf {\bibinfo {volume} {526}},\
		\bibinfo {pages} {554} (\bibinfo {year} {2015})}\BibitemShut {NoStop}%
	\bibitem [{\citenamefont {Monifi}\ \emph {et~al.}(2016)\citenamefont {Monifi},
		\citenamefont {Zhang}, \citenamefont {\"{O}zdemir}, \citenamefont {Peng},
		\citenamefont {Liu}, \citenamefont {Bo}, \citenamefont {Nori},\ and\
		\citenamefont {Yang}}]{Monifi2016}%
	\BibitemOpen
	\bibfield  {author} {\bibinfo {author} {\bibfnamefont {F.}~\bibnamefont
			{Monifi}}, \bibinfo {author} {\bibfnamefont {J.}~\bibnamefont {Zhang}},
		\bibinfo {author} {\bibfnamefont {Ş.~K.}\ \bibnamefont {\"{O}zdemir}},
		\bibinfo {author} {\bibfnamefont {B.}~\bibnamefont {Peng}}, \bibinfo {author}
		{\bibfnamefont {Y.-x.}\ \bibnamefont {Liu}}, \bibinfo {author} {\bibfnamefont
			{F.}~\bibnamefont {Bo}}, \bibinfo {author} {\bibfnamefont {F.}~\bibnamefont
			{Nori}}, \ and\ \bibinfo {author} {\bibfnamefont {L.}~\bibnamefont {Yang}},\
	}\bibfield  {title} {\enquote {\bibinfo {title} {Optomechanically induced
				stochastic resonance and chaos transfer between optical fields},}\ }\href
	{http://dx.doi.org/10.1038/nphoton.2016.73} {\bibfield  {journal} {\bibinfo
			{journal} {Nat. Photon.}\ }\textbf {\bibinfo {volume} {10}},\ \bibinfo
		{pages} {399} (\bibinfo {year} {2016})}\BibitemShut {NoStop}%
	\bibitem [{\citenamefont {Zhang}\ \emph {et~al.}(2018)\citenamefont {Zhang},
		\citenamefont {Peng}, \citenamefont {\"{O}zdemir}, \citenamefont {Pichler},
		\citenamefont {Krimer}, \citenamefont {Zhao}, \citenamefont {Nori},
		\citenamefont {Liu}, \citenamefont {Rotter},\ and\ \citenamefont
		{Yang}}]{ZhangJ2018}%
	\BibitemOpen
	\bibfield  {author} {\bibinfo {author} {\bibfnamefont {J.}~\bibnamefont
			{Zhang}}, \bibinfo {author} {\bibfnamefont {B.}~\bibnamefont {Peng}},
		\bibinfo {author} {\bibfnamefont {Ş.~K.}\ \bibnamefont {\"{O}zdemir}},
		\bibinfo {author} {\bibfnamefont {K.}~\bibnamefont {Pichler}}, \bibinfo
		{author} {\bibfnamefont {D.~O.}\ \bibnamefont {Krimer}}, \bibinfo {author}
		{\bibfnamefont {G.}~\bibnamefont {Zhao}}, \bibinfo {author} {\bibfnamefont
			{F.}~\bibnamefont {Nori}}, \bibinfo {author} {\bibfnamefont {Y.-x.}\
			\bibnamefont {Liu}}, \bibinfo {author} {\bibfnamefont {S.}~\bibnamefont
			{Rotter}}, \ and\ \bibinfo {author} {\bibfnamefont {L.}~\bibnamefont
			{Yang}},\ }\bibfield  {title} {\enquote {\bibinfo {title} {A phonon laser
				operating at an exceptional point},}\ }\href
	{http://dx.doi.org/10.1038/s41566-018-0213-5} {\bibfield  {journal} {\bibinfo
			{journal} {Nat. Photon.}\ }\textbf {\bibinfo {volume} {12}},\ \bibinfo
		{pages} {479} (\bibinfo {year} {2018})}\BibitemShut {NoStop}%
	\bibitem [{\citenamefont {Lee}(2016)}]{PhysRevLett.116.133903}%
	\BibitemOpen
	\bibfield  {author} {\bibinfo {author} {\bibfnamefont {Tony~E.}\ \bibnamefont
			{Lee}},\ }\bibfield  {title} {\enquote {\bibinfo {title} {Anomalous edge
				state in a non-{H}ermitian lattice},}\ }\href {\doibase
		10.1103/PhysRevLett.116.133903} {\bibfield  {journal} {\bibinfo  {journal}
			{Phys. Rev. Lett.}\ }\textbf {\bibinfo {volume} {116}},\ \bibinfo {pages}
		{133903} (\bibinfo {year} {2016})}\BibitemShut {NoStop}%
	\bibitem [{\citenamefont {Leykam}\ \emph {et~al.}(2017)\citenamefont {Leykam},
		\citenamefont {Bliokh}, \citenamefont {Huang}, \citenamefont {Chong},\ and\
		\citenamefont {Nori}}]{PhysRevLett.118.040401}%
	\BibitemOpen
	\bibfield  {author} {\bibinfo {author} {\bibfnamefont {D.}~\bibnamefont
			{Leykam}}, \bibinfo {author} {\bibfnamefont {K.~Y.}\ \bibnamefont {Bliokh}},
		\bibinfo {author} {\bibfnamefont {C.}~\bibnamefont {Huang}}, \bibinfo
		{author} {\bibfnamefont {Y.~D.}\ \bibnamefont {Chong}}, \ and\ \bibinfo
		{author} {\bibfnamefont {F.}~\bibnamefont {Nori}},\ }\bibfield  {title}
	{\enquote {\bibinfo {title} {Edge modes, degeneracies, and topological
				numbers in non-\uppercase{H}ermitian systems},}\ }\href
	{https://link.aps.org/doi/10.1103/PhysRevLett.118.040401} {\bibfield
		{journal} {\bibinfo  {journal} {Phys. Rev. Lett.}\ }\textbf {\bibinfo
			{volume} {118}},\ \bibinfo {pages} {040401} (\bibinfo {year}
		{2017})}\BibitemShut {NoStop}%
	\bibitem [{\citenamefont {Xu}\ \emph {et~al.}(2017)\citenamefont {Xu},
		\citenamefont {Wang},\ and\ \citenamefont {Duan}}]{PhysRevLett.118.045701}%
	\BibitemOpen
	\bibfield  {author} {\bibinfo {author} {\bibfnamefont {Y.}~\bibnamefont
			{Xu}}, \bibinfo {author} {\bibfnamefont {S.~T.}\ \bibnamefont {Wang}}, \ and\
		\bibinfo {author} {\bibfnamefont {L.~M.}\ \bibnamefont {Duan}},\ }\bibfield
	{title} {\enquote {\bibinfo {title} {Weyl exceptional rings in a
				three-dimensional dissipative cold atomic gas},}\ }\href
	{https://link.aps.org/doi/10.1103/PhysRevLett.118.045701} {\bibfield
		{journal} {\bibinfo  {journal} {Phys. Rev. Lett.}\ }\textbf {\bibinfo
			{volume} {118}},\ \bibinfo {pages} {045701} (\bibinfo {year}
		{2017})}\BibitemShut {NoStop}%
	\bibitem [{\citenamefont {Gong}\ \emph {et~al.}(2018)\citenamefont {Gong},
		\citenamefont {Ashida}, \citenamefont {Kawabata}, \citenamefont {Takasan},
		\citenamefont {Higashikawa},\ and\ \citenamefont {Ueda}}]{arXiv:1802.07964}%
	\BibitemOpen
	\bibfield  {author} {\bibinfo {author} {\bibfnamefont {Z.}~\bibnamefont
			{Gong}}, \bibinfo {author} {\bibfnamefont {Y.}~\bibnamefont {Ashida}},
		\bibinfo {author} {\bibfnamefont {K.}~\bibnamefont {Kawabata}}, \bibinfo
		{author} {\bibfnamefont {K.}~\bibnamefont {Takasan}}, \bibinfo {author}
		{\bibfnamefont {S.}~\bibnamefont {Higashikawa}}, \ and\ \bibinfo {author}
		{\bibfnamefont {M.}~\bibnamefont {Ueda}},\ }\bibfield  {title} {\enquote
		{\bibinfo {title} {Topological phases of non-\uppercase{H}ermitian
				systems},}\ }\href {https://link.aps.org/doi/10.1103/PhysRevX.8.031079}
	{\bibfield  {journal} {\bibinfo  {journal} {Phys. Rev. X}\ }\textbf {\bibinfo
			{volume} {8}},\ \bibinfo {pages} {031079} (\bibinfo {year}
		{2018})}\BibitemShut {NoStop}%
	\bibitem [{\citenamefont {Peng}\ \emph {et~al.}(2014)\citenamefont {Peng},
		\citenamefont {\"{O}zdemir}, \citenamefont {Rotter}, \citenamefont {Yilmaz},
		\citenamefont {Liertzer}, \citenamefont {Monifi}, \citenamefont {Bender},
		\citenamefont {Nori},\ and\ \citenamefont {Yang}}]{Peng2014b}%
	\BibitemOpen
	\bibfield  {author} {\bibinfo {author} {\bibfnamefont {B.}~\bibnamefont
			{Peng}}, \bibinfo {author} {\bibfnamefont {S.~K.}\ \bibnamefont
			{\"{O}zdemir}}, \bibinfo {author} {\bibfnamefont {S.}~\bibnamefont {Rotter}},
		\bibinfo {author} {\bibfnamefont {H.}~\bibnamefont {Yilmaz}}, \bibinfo
		{author} {\bibfnamefont {M.}~\bibnamefont {Liertzer}}, \bibinfo {author}
		{\bibfnamefont {F.}~\bibnamefont {Monifi}}, \bibinfo {author} {\bibfnamefont
			{C.~M.}\ \bibnamefont {Bender}}, \bibinfo {author} {\bibfnamefont
			{F.}~\bibnamefont {Nori}}, \ and\ \bibinfo {author} {\bibfnamefont
			{L.}~\bibnamefont {Yang}},\ }\bibfield  {title} {\enquote {\bibinfo {title}
			{Loss-induced suppression and revival of lasing},}\ }\href {\doibase
		10.1126/science.1258004} {\bibfield  {journal} {\bibinfo  {journal}
			{Science}\ }\textbf {\bibinfo {volume} {346}},\ \bibinfo {pages} {328}
		(\bibinfo {year} {2014})}\BibitemShut {NoStop}%
	\bibitem [{\citenamefont {El-Ganainy}\ \emph {et~al.}(2018)\citenamefont
		{El-Ganainy}, \citenamefont {Makris}, \citenamefont {Khajavikhan},
		\citenamefont {Musslimani}, \citenamefont {Rotter},\ and\ \citenamefont
		{Christodoulides}}]{El-Ganainy2018}%
	\BibitemOpen
	\bibfield  {author} {\bibinfo {author} {\bibfnamefont {R.}~\bibnamefont
			{El-Ganainy}}, \bibinfo {author} {\bibfnamefont {K.~G.}\ \bibnamefont
			{Makris}}, \bibinfo {author} {\bibfnamefont {M.}~\bibnamefont {Khajavikhan}},
		\bibinfo {author} {\bibfnamefont {Z.~H.}\ \bibnamefont {Musslimani}},
		\bibinfo {author} {\bibfnamefont {S.}~\bibnamefont {Rotter}}, \ and\ \bibinfo
		{author} {\bibfnamefont {D.~N.}\ \bibnamefont {Christodoulides}},\ }\bibfield
	{title} {\enquote {\bibinfo {title} {Non-\uppercase{H}ermitian physics and
				\uppercase{PT} symmetry},}\ }\href {http://dx.doi.org/10.1038/nphys4323
		http://10.0.4.14/nphys4323} {\bibfield  {journal} {\bibinfo  {journal} {Nat.
				Phys.}\ }\textbf {\bibinfo {volume} {14}},\ \bibinfo {pages} {11} (\bibinfo
		{year} {2018})}\BibitemShut {NoStop}%
	\bibitem [{\citenamefont {Yao}\ and\ \citenamefont
		{Wang}(2018)}]{ShunyuYao2018}%
	\BibitemOpen
	\bibfield  {author} {\bibinfo {author} {\bibfnamefont {S.}~\bibnamefont
			{Yao}}\ and\ \bibinfo {author} {\bibfnamefont {Z.}~\bibnamefont {Wang}},\
	}\bibfield  {title} {\enquote {\bibinfo {title} {Edge states and topological
				invariants of non-\uppercase{H}ermitian systems},}\ }\href
	{https://link.aps.org/doi/10.1103/PhysRevLett.121.086803} {\bibfield
		{journal} {\bibinfo  {journal} {Phys. Rev. Lett.}\ }\textbf {\bibinfo
			{volume} {121}},\ \bibinfo {pages} {086803} (\bibinfo {year}
		{2018})}\BibitemShut {NoStop}%
	\bibitem [{\citenamefont {Zhang}\ \emph {et~al.}(2020)\citenamefont {Zhang},
		\citenamefont {Yang},\ and\ \citenamefont {Fang}}]{PhysRevLett.125.126402}%
	\BibitemOpen
	\bibfield  {author} {\bibinfo {author} {\bibfnamefont {K.}~\bibnamefont
			{Zhang}}, \bibinfo {author} {\bibfnamefont {Z.}~\bibnamefont {Yang}}, \ and\
		\bibinfo {author} {\bibfnamefont {C.}~\bibnamefont {Fang}},\ }\bibfield
	{title} {\enquote {\bibinfo {title} {Correspondence between winding numbers
				and skin modes in non-{H}ermitian systems},}\ }\href {\doibase
		10.1103/PhysRevLett.125.126402} {\bibfield  {journal} {\bibinfo  {journal}
			{Phys. Rev. Lett.}\ }\textbf {\bibinfo {volume} {125}},\ \bibinfo {pages}
		{126402} (\bibinfo {year} {2020})}\BibitemShut {NoStop}%
	\bibitem [{\citenamefont {Yokomizo}\ and\ \citenamefont
		{Murakami}(2019)}]{PhysRevLett.123.066404}%
	\BibitemOpen
	\bibfield  {author} {\bibinfo {author} {\bibfnamefont {K.}~\bibnamefont
			{Yokomizo}}\ and\ \bibinfo {author} {\bibfnamefont {S.}~\bibnamefont
			{Murakami}},\ }\bibfield  {title} {\enquote {\bibinfo {title} {Non-{B}loch
				band theory of non-{H}ermitian systems},}\ }\href {\doibase
		10.1103/PhysRevLett.123.066404} {\bibfield  {journal} {\bibinfo  {journal}
			{Phys. Rev. Lett.}\ }\textbf {\bibinfo {volume} {123}},\ \bibinfo {pages}
		{066404} (\bibinfo {year} {2019})}\BibitemShut {NoStop}%
	\bibitem [{\citenamefont {Yao}\ \emph {et~al.}(2018)\citenamefont {Yao},
		\citenamefont {Song},\ and\ \citenamefont {Wang}}]{YaoarXiv:1804.04672}%
	\BibitemOpen
	\bibfield  {author} {\bibinfo {author} {\bibfnamefont {S.}~\bibnamefont
			{Yao}}, \bibinfo {author} {\bibfnamefont {F.}~\bibnamefont {Song}}, \ and\
		\bibinfo {author} {\bibfnamefont {Z.}~\bibnamefont {Wang}},\ }\bibfield
	{title} {\enquote {\bibinfo {title} {Non-\uppercase{H}ermitian
				\uppercase{C}hern bands},}\ }\href
	{https://link.aps.org/doi/10.1103/PhysRevLett.121.136802} {\bibfield
		{journal} {\bibinfo  {journal} {Phys. Rev. Lett.}\ }\textbf {\bibinfo
			{volume} {121}},\ \bibinfo {pages} {136802} (\bibinfo {year}
		{2018})}\BibitemShut {NoStop}%
	\bibitem [{\citenamefont {Kunst}\ \emph {et~al.}(2018)\citenamefont {Kunst},
		\citenamefont {Edvardsson}, \citenamefont {Budich},\ and\ \citenamefont
		{Bergholtz}}]{PhysRevLett.121.026808}%
	\BibitemOpen
	\bibfield  {author} {\bibinfo {author} {\bibfnamefont {F.~K.}\ \bibnamefont
			{Kunst}}, \bibinfo {author} {\bibfnamefont {E.}~\bibnamefont {Edvardsson}},
		\bibinfo {author} {\bibfnamefont {J.~C.}\ \bibnamefont {Budich}}, \ and\
		\bibinfo {author} {\bibfnamefont {E.~J.}\ \bibnamefont {Bergholtz}},\
	}\bibfield  {title} {\enquote {\bibinfo {title} {Biorthogonal bulk-boundary
				correspondence in non-{H}ermitian systems},}\ }\href {\doibase
		10.1103/PhysRevLett.121.026808} {\bibfield  {journal} {\bibinfo  {journal}
			{Phys. Rev. Lett.}\ }\textbf {\bibinfo {volume} {121}},\ \bibinfo {pages}
		{026808} (\bibinfo {year} {2018})}\BibitemShut {NoStop}%
	\bibitem [{\citenamefont {Liu}\ \emph {et~al.}(2019)\citenamefont {Liu},
		\citenamefont {Zhang}, \citenamefont {Ai}, \citenamefont {Gong},
		\citenamefont {Kawabata}, \citenamefont {Ueda},\ and\ \citenamefont
		{Nori}}]{PhysRevLett.122.076801}%
	\BibitemOpen
	\bibfield  {author} {\bibinfo {author} {\bibfnamefont {T.}~\bibnamefont
			{Liu}}, \bibinfo {author} {\bibfnamefont {Y.-R.}\ \bibnamefont {Zhang}},
		\bibinfo {author} {\bibfnamefont {Q.}~\bibnamefont {Ai}}, \bibinfo {author}
		{\bibfnamefont {Z.}~\bibnamefont {Gong}}, \bibinfo {author} {\bibfnamefont
			{K.}~\bibnamefont {Kawabata}}, \bibinfo {author} {\bibfnamefont
			{M.}~\bibnamefont {Ueda}}, \ and\ \bibinfo {author} {\bibfnamefont
			{F.}~\bibnamefont {Nori}},\ }\bibfield  {title} {\enquote {\bibinfo {title}
			{Second-order topological phases in non-{H}ermitian systems},}\ }\href
	{\doibase 10.1103/PhysRevLett.122.076801} {\bibfield  {journal} {\bibinfo
			{journal} {Phys. Rev. Lett.}\ }\textbf {\bibinfo {volume} {122}},\ \bibinfo
		{pages} {076801} (\bibinfo {year} {2019})}\BibitemShut {NoStop}%
	\bibitem [{\citenamefont {Song}\ \emph {et~al.}(2019)\citenamefont {Song},
		\citenamefont {Yao},\ and\ \citenamefont {Wang}}]{PhysRevLett.123.170401}%
	\BibitemOpen
	\bibfield  {author} {\bibinfo {author} {\bibfnamefont {F.}~\bibnamefont
			{Song}}, \bibinfo {author} {\bibfnamefont {S.}~\bibnamefont {Yao}}, \ and\
		\bibinfo {author} {\bibfnamefont {Z.}~\bibnamefont {Wang}},\ }\bibfield
	{title} {\enquote {\bibinfo {title} {Non-{H}ermitian skin effect and chiral
				damping in open quantum systems},}\ }\href {\doibase
		10.1103/PhysRevLett.123.170401} {\bibfield  {journal} {\bibinfo  {journal}
			{Phys. Rev. Lett.}\ }\textbf {\bibinfo {volume} {123}},\ \bibinfo {pages}
		{170401} (\bibinfo {year} {2019})}\BibitemShut {NoStop}%
	\bibitem [{\citenamefont {Lee}\ \emph {et~al.}(2019)\citenamefont {Lee},
		\citenamefont {Ahn}, \citenamefont {Zhou},\ and\ \citenamefont
		{Vishwanath}}]{PhysRevLett.123.206404}%
	\BibitemOpen
	\bibfield  {author} {\bibinfo {author} {\bibfnamefont {J.~Y.}\ \bibnamefont
			{Lee}}, \bibinfo {author} {\bibfnamefont {J.}~\bibnamefont {Ahn}}, \bibinfo
		{author} {\bibfnamefont {H.}~\bibnamefont {Zhou}}, \ and\ \bibinfo {author}
		{\bibfnamefont {A.}~\bibnamefont {Vishwanath}},\ }\bibfield  {title}
	{\enquote {\bibinfo {title} {Topological correspondence between {H}ermitian
				and non-{H}ermitian systems: {A}nomalous dynamics},}\ }\href {\doibase
		10.1103/PhysRevLett.123.206404} {\bibfield  {journal} {\bibinfo  {journal}
			{Phys. Rev. Lett.}\ }\textbf {\bibinfo {volume} {123}},\ \bibinfo {pages}
		{206404} (\bibinfo {year} {2019})}\BibitemShut {NoStop}%
	\bibitem [{\citenamefont {Kawabata}\ \emph
		{et~al.}(2019{\natexlab{a}})\citenamefont {Kawabata}, \citenamefont
		{Bessho},\ and\ \citenamefont {Sato}}]{PhysRevLett.123.066405}%
	\BibitemOpen
	\bibfield  {author} {\bibinfo {author} {\bibfnamefont {K.}~\bibnamefont
			{Kawabata}}, \bibinfo {author} {\bibfnamefont {T.}~\bibnamefont {Bessho}}, \
		and\ \bibinfo {author} {\bibfnamefont {M.}~\bibnamefont {Sato}},\ }\bibfield
	{title} {\enquote {\bibinfo {title} {Classification of exceptional points and
				non-{H}ermitian topological semimetals},}\ }\href {\doibase
		10.1103/PhysRevLett.123.066405} {\bibfield  {journal} {\bibinfo  {journal}
			{Phys. Rev. Lett.}\ }\textbf {\bibinfo {volume} {123}},\ \bibinfo {pages}
		{066405} (\bibinfo {year} {2019}{\natexlab{a}})}\BibitemShut {NoStop}%
	\bibitem [{\citenamefont {Nie}\ \emph {et~al.}(2023)\citenamefont {Nie},
		\citenamefont {Shi}, \citenamefont {Liu},\ and\ \citenamefont
		{Nori}}]{PhysRevLett.131.103602}%
	\BibitemOpen
	\bibfield  {author} {\bibinfo {author} {\bibfnamefont {W.}~\bibnamefont
			{Nie}}, \bibinfo {author} {\bibfnamefont {T.}~\bibnamefont {Shi}}, \bibinfo
		{author} {\bibfnamefont {Y.-x.}\ \bibnamefont {Liu}}, \ and\ \bibinfo
		{author} {\bibfnamefont {F.}~\bibnamefont {Nori}},\ }\bibfield  {title}
	{\enquote {\bibinfo {title} {Non-{H}ermitian waveguide cavity {QED} with
				tunable atomic mirrors},}\ }\href {\doibase 10.1103/PhysRevLett.131.103602}
	{\bibfield  {journal} {\bibinfo  {journal} {Phys. Rev. Lett.}\ }\textbf
		{\bibinfo {volume} {131}},\ \bibinfo {pages} {103602} (\bibinfo {year}
		{2023})}\BibitemShut {NoStop}%
	\bibitem [{\citenamefont {Ge}\ \emph {et~al.}(2019)\citenamefont {Ge},
		\citenamefont {Zhang}, \citenamefont {Liu}, \citenamefont {Li}, \citenamefont
		{Fan},\ and\ \citenamefont {Nori}}]{PhysRevB.100.054105}%
	\BibitemOpen
	\bibfield  {author} {\bibinfo {author} {\bibfnamefont {Z.~Y.}\ \bibnamefont
			{Ge}}, \bibinfo {author} {\bibfnamefont {Y.~R.}\ \bibnamefont {Zhang}},
		\bibinfo {author} {\bibfnamefont {T.}~\bibnamefont {Liu}}, \bibinfo {author}
		{\bibfnamefont {S.~W.}\ \bibnamefont {Li}}, \bibinfo {author} {\bibfnamefont
			{H.}~\bibnamefont {Fan}}, \ and\ \bibinfo {author} {\bibfnamefont
			{F.}~\bibnamefont {Nori}},\ }\bibfield  {title} {\enquote {\bibinfo {title}
			{Topological band theory for non-{H}ermitian systems from the {D}irac
				equation},}\ }\href {\doibase 10.1103/PhysRevB.100.054105} {\bibfield
		{journal} {\bibinfo  {journal} {Phys. Rev. B}\ }\textbf {\bibinfo {volume}
			{100}},\ \bibinfo {pages} {054105} (\bibinfo {year} {2019})}\BibitemShut
	{NoStop}%
	\bibitem [{\citenamefont {Zhou}\ and\ \citenamefont
		{Lee}(2019)}]{PhysRevB.99.235112}%
	\BibitemOpen
	\bibfield  {author} {\bibinfo {author} {\bibfnamefont {H.}~\bibnamefont
			{Zhou}}\ and\ \bibinfo {author} {\bibfnamefont {J.~Y.}\ \bibnamefont {Lee}},\
	}\bibfield  {title} {\enquote {\bibinfo {title} {Periodic table for
				topological bands with non-{H}ermitian symmetries},}\ }\href {\doibase
		10.1103/PhysRevB.99.235112} {\bibfield  {journal} {\bibinfo  {journal} {Phys.
				Rev. B}\ }\textbf {\bibinfo {volume} {99}},\ \bibinfo {pages} {235112}
		(\bibinfo {year} {2019})}\BibitemShut {NoStop}%
	\bibitem [{\citenamefont {Zhao}\ \emph {et~al.}(2019)\citenamefont {Zhao},
		\citenamefont {Qiao}, \citenamefont {Wu}, \citenamefont {Midya},
		\citenamefont {Longhi},\ and\ \citenamefont {Feng}}]{Zhao2019}%
	\BibitemOpen
	\bibfield  {author} {\bibinfo {author} {\bibfnamefont {H.}~\bibnamefont
			{Zhao}}, \bibinfo {author} {\bibfnamefont {X.}~\bibnamefont {Qiao}}, \bibinfo
		{author} {\bibfnamefont {T.}~\bibnamefont {Wu}}, \bibinfo {author}
		{\bibfnamefont {B.}~\bibnamefont {Midya}}, \bibinfo {author} {\bibfnamefont
			{S.}~\bibnamefont {Longhi}}, \ and\ \bibinfo {author} {\bibfnamefont
			{L.}~\bibnamefont {Feng}},\ }\bibfield  {title} {\enquote {\bibinfo {title}
			{Non-{H}ermitian topological light steering},}\ }\href {\doibase
		10.1126/science.aay1064} {\bibfield  {journal} {\bibinfo  {journal}
			{Science}\ }\textbf {\bibinfo {volume} {365}},\ \bibinfo {pages} {1163}
		(\bibinfo {year} {2019})}\BibitemShut {NoStop}%
	\bibitem [{\citenamefont {Kawabata}\ \emph
		{et~al.}(2019{\natexlab{b}})\citenamefont {Kawabata}, \citenamefont
		{Shiozaki}, \citenamefont {Ueda},\ and\ \citenamefont
		{Sato}}]{PhysRevX.9.041015}%
	\BibitemOpen
	\bibfield  {author} {\bibinfo {author} {\bibfnamefont {K.}~\bibnamefont
			{Kawabata}}, \bibinfo {author} {\bibfnamefont {K.}~\bibnamefont {Shiozaki}},
		\bibinfo {author} {\bibfnamefont {M.}~\bibnamefont {Ueda}}, \ and\ \bibinfo
		{author} {\bibfnamefont {M.}~\bibnamefont {Sato}},\ }\bibfield  {title}
	{\enquote {\bibinfo {title} {Symmetry and topology in non-{H}ermitian
				physics},}\ }\href {\doibase 10.1103/PhysRevX.9.041015} {\bibfield  {journal}
		{\bibinfo  {journal} {Phys. Rev. X}\ }\textbf {\bibinfo {volume} {9}},\
		\bibinfo {pages} {041015} (\bibinfo {year} {2019}{\natexlab{b}})}\BibitemShut
	{NoStop}%
	\bibitem [{\citenamefont {Borgnia}\ \emph {et~al.}(2020)\citenamefont
		{Borgnia}, \citenamefont {Kruchkov},\ and\ \citenamefont
		{Slager}}]{PhysRevLett.124.056802}%
	\BibitemOpen
	\bibfield  {author} {\bibinfo {author} {\bibfnamefont {D.~S.}\ \bibnamefont
			{Borgnia}}, \bibinfo {author} {\bibfnamefont {A.~J.}\ \bibnamefont
			{Kruchkov}}, \ and\ \bibinfo {author} {\bibfnamefont {R.-J.}\ \bibnamefont
			{Slager}},\ }\bibfield  {title} {\enquote {\bibinfo {title} {Non-{H}ermitian
				boundary modes and topology},}\ }\href {\doibase
		10.1103/PhysRevLett.124.056802} {\bibfield  {journal} {\bibinfo  {journal}
			{Phys. Rev. Lett.}\ }\textbf {\bibinfo {volume} {124}},\ \bibinfo {pages}
		{056802} (\bibinfo {year} {2020})}\BibitemShut {NoStop}%
	\bibitem [{\citenamefont {Liu}\ \emph {et~al.}(2020)\citenamefont {Liu},
		\citenamefont {He}, \citenamefont {Yoshida}, \citenamefont {Xiang},\ and\
		\citenamefont {Nori}}]{PhysRevB.102.235151}%
	\BibitemOpen
	\bibfield  {author} {\bibinfo {author} {\bibfnamefont {T.}~\bibnamefont
			{Liu}}, \bibinfo {author} {\bibfnamefont {J.~J.}\ \bibnamefont {He}},
		\bibinfo {author} {\bibfnamefont {T.}~\bibnamefont {Yoshida}}, \bibinfo
		{author} {\bibfnamefont {Z.-L.}\ \bibnamefont {Xiang}}, \ and\ \bibinfo
		{author} {\bibfnamefont {F.}~\bibnamefont {Nori}},\ }\bibfield  {title}
	{\enquote {\bibinfo {title} {Non-{H}ermitian topological {M}ott insulators in
				one-dimensional fermionic superlattices},}\ }\href {\doibase
		10.1103/PhysRevB.102.235151} {\bibfield  {journal} {\bibinfo  {journal}
			{Phys. Rev. B}\ }\textbf {\bibinfo {volume} {102}},\ \bibinfo {pages}
		{235151} (\bibinfo {year} {2020})}\BibitemShut {NoStop}%
	\bibitem [{\citenamefont {Bliokh}\ \emph {et~al.}(2019)\citenamefont {Bliokh},
		\citenamefont {Leykam}, \citenamefont {Lein},\ and\ \citenamefont
		{Nori}}]{Bliokh2019}%
	\BibitemOpen
	\bibfield  {author} {\bibinfo {author} {\bibfnamefont {K.~Y.}\ \bibnamefont
			{Bliokh}}, \bibinfo {author} {\bibfnamefont {D.}~\bibnamefont {Leykam}},
		\bibinfo {author} {\bibfnamefont {M.}~\bibnamefont {Lein}}, \ and\ \bibinfo
		{author} {\bibfnamefont {F.}~\bibnamefont {Nori}},\ }\bibfield  {title}
	{\enquote {\bibinfo {title} {Topological non-{H}ermitian origin of surface
				{M}axwell waves},}\ }\href {http://dx.doi.org/10.1038/s41467-019-08397-6}
	{\bibfield  {journal} {\bibinfo  {journal} {Nat. Comm.}\ }\textbf {\bibinfo
			{volume} {10}},\ \bibinfo {pages} {580} (\bibinfo {year} {2019})}\BibitemShut
	{NoStop}%
	\bibitem [{\citenamefont {Yokomizo}\ and\ \citenamefont
		{Murakami}(2021)}]{PhysRevB.104.165117}%
	\BibitemOpen
	\bibfield  {author} {\bibinfo {author} {\bibfnamefont {K.}~\bibnamefont
			{Yokomizo}}\ and\ \bibinfo {author} {\bibfnamefont {S.}~\bibnamefont
			{Murakami}},\ }\bibfield  {title} {\enquote {\bibinfo {title} {Scaling rule
				for the critical non-{H}ermitian skin effect},}\ }\href {\doibase
		10.1103/PhysRevB.104.165117} {\bibfield  {journal} {\bibinfo  {journal}
			{Phys. Rev. B}\ }\textbf {\bibinfo {volume} {104}},\ \bibinfo {pages}
		{165117} (\bibinfo {year} {2021})}\BibitemShut {NoStop}%
	\bibitem [{\citenamefont {Okuma}\ \emph {et~al.}(2020)\citenamefont {Okuma},
		\citenamefont {Kawabata}, \citenamefont {Shiozaki},\ and\ \citenamefont
		{Sato}}]{PhysRevLett.124.086801}%
	\BibitemOpen
	\bibfield  {author} {\bibinfo {author} {\bibfnamefont {N.}~\bibnamefont
			{Okuma}}, \bibinfo {author} {\bibfnamefont {K.}~\bibnamefont {Kawabata}},
		\bibinfo {author} {\bibfnamefont {K.}~\bibnamefont {Shiozaki}}, \ and\
		\bibinfo {author} {\bibfnamefont {M.}~\bibnamefont {Sato}},\ }\bibfield
	{title} {\enquote {\bibinfo {title} {Topological origin of non-{H}ermitian
				skin effects},}\ }\href {\doibase 10.1103/PhysRevLett.124.086801} {\bibfield
		{journal} {\bibinfo  {journal} {Phys. Rev. Lett.}\ }\textbf {\bibinfo
			{volume} {124}},\ \bibinfo {pages} {086801} (\bibinfo {year}
		{2020})}\BibitemShut {NoStop}%
	\bibitem [{\citenamefont {Yi}\ and\ \citenamefont
		{Yang}(2020)}]{PhysRevLett.125.186802}%
	\BibitemOpen
	\bibfield  {author} {\bibinfo {author} {\bibfnamefont {Y.}~\bibnamefont
			{Yi}}\ and\ \bibinfo {author} {\bibfnamefont {Z.}~\bibnamefont {Yang}},\
	}\bibfield  {title} {\enquote {\bibinfo {title} {Non-{H}ermitian skin modes
				induced by on-site dissipations and chiral tunneling effect},}\ }\href
	{\doibase 10.1103/PhysRevLett.125.186802} {\bibfield  {journal} {\bibinfo
			{journal} {Phys. Rev. Lett.}\ }\textbf {\bibinfo {volume} {125}},\ \bibinfo
		{pages} {186802} (\bibinfo {year} {2020})}\BibitemShut {NoStop}%
	\bibitem [{\citenamefont {Liu}\ \emph {et~al.}(2021)\citenamefont {Liu},
		\citenamefont {He}, \citenamefont {Yang},\ and\ \citenamefont
		{Nori}}]{PhysRevLett.127.196801}%
	\BibitemOpen
	\bibfield  {author} {\bibinfo {author} {\bibfnamefont {T.}~\bibnamefont
			{Liu}}, \bibinfo {author} {\bibfnamefont {J.~J.}\ \bibnamefont {He}},
		\bibinfo {author} {\bibfnamefont {Z.}~\bibnamefont {Yang}}, \ and\ \bibinfo
		{author} {\bibfnamefont {F.}~\bibnamefont {Nori}},\ }\bibfield  {title}
	{\enquote {\bibinfo {title} {Higher-order {W}eyl-exceptional-ring
				semimetals},}\ }\href {\doibase 10.1103/PhysRevLett.127.196801} {\bibfield
		{journal} {\bibinfo  {journal} {Phys. Rev. Lett.}\ }\textbf {\bibinfo
			{volume} {127}},\ \bibinfo {pages} {196801} (\bibinfo {year}
		{2021})}\BibitemShut {NoStop}%
	\bibitem [{\citenamefont {Bergholtz}\ \emph {et~al.}(2021)\citenamefont
		{Bergholtz}, \citenamefont {Budich},\ and\ \citenamefont
		{Kunst}}]{RevModPhys.93.015005}%
	\BibitemOpen
	\bibfield  {author} {\bibinfo {author} {\bibfnamefont {E.~J.}\ \bibnamefont
			{Bergholtz}}, \bibinfo {author} {\bibfnamefont {J.~C.}\ \bibnamefont
			{Budich}}, \ and\ \bibinfo {author} {\bibfnamefont {F.~K.}\ \bibnamefont
			{Kunst}},\ }\bibfield  {title} {\enquote {\bibinfo {title} {Exceptional
				topology of non-{H}ermitian systems},}\ }\href {\doibase
		10.1103/RevModPhys.93.015005} {\bibfield  {journal} {\bibinfo  {journal}
			{Rev. Mod. Phys.}\ }\textbf {\bibinfo {volume} {93}},\ \bibinfo {pages}
		{015005} (\bibinfo {year} {2021})}\BibitemShut {NoStop}%
	\bibitem [{\citenamefont {Li}\ \emph {et~al.}(2022)\citenamefont {Li},
		\citenamefont {Liang}, \citenamefont {Wang}, \citenamefont {Lu},\ and\
		\citenamefont {Liu}}]{PhysRevLett.128.223903}%
	\BibitemOpen
	\bibfield  {author} {\bibinfo {author} {\bibfnamefont {Y.}~\bibnamefont
			{Li}}, \bibinfo {author} {\bibfnamefont {C.}~\bibnamefont {Liang}}, \bibinfo
		{author} {\bibfnamefont {C.}~\bibnamefont {Wang}}, \bibinfo {author}
		{\bibfnamefont {C.}~\bibnamefont {Lu}}, \ and\ \bibinfo {author}
		{\bibfnamefont {Y.-C.}\ \bibnamefont {Liu}},\ }\bibfield  {title} {\enquote
		{\bibinfo {title} {Gain-loss-induced hybrid skin-topological effect},}\
	}\href {\doibase 10.1103/PhysRevLett.128.223903} {\bibfield  {journal}
		{\bibinfo  {journal} {Phys. Rev. Lett.}\ }\textbf {\bibinfo {volume} {128}},\
		\bibinfo {pages} {223903} (\bibinfo {year} {2022})}\BibitemShut {NoStop}%
	\bibitem [{\citenamefont {Leefmans}\ \emph {et~al.}(2022)\citenamefont
		{Leefmans}, \citenamefont {Dutt}, \citenamefont {Williams}, \citenamefont
		{Yuan}, \citenamefont {Parto}, \citenamefont {Nori}, \citenamefont {Fan},\
		and\ \citenamefont {Marandi}}]{Leefmans2022}%
	\BibitemOpen
	\bibfield  {author} {\bibinfo {author} {\bibfnamefont {C.}~\bibnamefont
			{Leefmans}}, \bibinfo {author} {\bibfnamefont {A.}~\bibnamefont {Dutt}},
		\bibinfo {author} {\bibfnamefont {J.}~\bibnamefont {Williams}}, \bibinfo
		{author} {\bibfnamefont {L.}~\bibnamefont {Yuan}}, \bibinfo {author}
		{\bibfnamefont {M.}~\bibnamefont {Parto}}, \bibinfo {author} {\bibfnamefont
			{F.}~\bibnamefont {Nori}}, \bibinfo {author} {\bibfnamefont {S.}~\bibnamefont
			{Fan}}, \ and\ \bibinfo {author} {\bibfnamefont {A.}~\bibnamefont
			{Marandi}},\ }\bibfield  {title} {\enquote {\bibinfo {title} {Topological
				dissipation in a time-multiplexed photonic resonator network},}\ }\href
	{http://dx.doi.org/10.1038/s41567-021-01492-w} {\bibfield  {journal}
		{\bibinfo  {journal} {Nat. Phys.}\ }\textbf {\bibinfo {volume} {18}},\
		\bibinfo {pages} {442} (\bibinfo {year} {2022})}\BibitemShut {NoStop}%
	\bibitem [{\citenamefont {Zhang}\ \emph {et~al.}(2022)\citenamefont {Zhang},
		\citenamefont {Yang},\ and\ \citenamefont {Fang}}]{Zhang2022}%
	\BibitemOpen
	\bibfield  {author} {\bibinfo {author} {\bibfnamefont {K.}~\bibnamefont
			{Zhang}}, \bibinfo {author} {\bibfnamefont {Z.}~\bibnamefont {Yang}}, \ and\
		\bibinfo {author} {\bibfnamefont {C.}~\bibnamefont {Fang}},\ }\bibfield
	{title} {\enquote {\bibinfo {title} {Universal non-{H}ermitian skin effect in
				two and higher dimensions},}\ }\href
	{https://doi.org/10.1038/s41467-022-30161-6} {\bibfield  {journal} {\bibinfo
			{journal} {Nat. Commun.}\ }\textbf {\bibinfo {volume} {13}},\ \bibinfo
		{pages} {2496} (\bibinfo {year} {2022})}\BibitemShut {NoStop}%
	\bibitem [{\citenamefont {Parto}\ \emph {et~al.}(2023)\citenamefont {Parto},
		\citenamefont {Leefmans}, \citenamefont {Williams}, \citenamefont {Nori},\
		and\ \citenamefont {Marandi}}]{Parto2023}%
	\BibitemOpen
	\bibfield  {author} {\bibinfo {author} {\bibfnamefont {M.}~\bibnamefont
			{Parto}}, \bibinfo {author} {\bibfnamefont {C.}~\bibnamefont {Leefmans}},
		\bibinfo {author} {\bibfnamefont {J.}~\bibnamefont {Williams}}, \bibinfo
		{author} {\bibfnamefont {F.}~\bibnamefont {Nori}}, \ and\ \bibinfo {author}
		{\bibfnamefont {A.}~\bibnamefont {Marandi}},\ }\bibfield  {title} {\enquote
		{\bibinfo {title} {Non-{A}belian effects in dissipative photonic topological
				lattices},}\ }\href {http://dx.doi.org/10.1038/s41467-023-37065-z} {\bibfield
		{journal} {\bibinfo  {journal} {Nat. Comm.}\ }\textbf {\bibinfo {volume}
			{14}},\ \bibinfo {pages} {1440} (\bibinfo {year} {2023})}\BibitemShut
	{NoStop}%
	\bibitem [{\citenamefont {Ren}\ \emph {et~al.}(2022)\citenamefont {Ren},
		\citenamefont {Liu}, \citenamefont {Zhao}, \citenamefont {He}, \citenamefont
		{Pak}, \citenamefont {Li},\ and\ \citenamefont {Jo}}]{Ren2022}%
	\BibitemOpen
	\bibfield  {author} {\bibinfo {author} {\bibfnamefont {Z.}~\bibnamefont
			{Ren}}, \bibinfo {author} {\bibfnamefont {D.}~\bibnamefont {Liu}}, \bibinfo
		{author} {\bibfnamefont {E.}~\bibnamefont {Zhao}}, \bibinfo {author}
		{\bibfnamefont {C.}~\bibnamefont {He}}, \bibinfo {author} {\bibfnamefont
			{K.~K.}\ \bibnamefont {Pak}}, \bibinfo {author} {\bibfnamefont
			{J.}~\bibnamefont {Li}}, \ and\ \bibinfo {author} {\bibfnamefont {G.-B.}\
			\bibnamefont {Jo}},\ }\bibfield  {title} {\enquote {\bibinfo {title} {Chiral
				control of quantum states in non-{H}ermitian spin{\textendash}orbit-coupled
				fermions},}\ }\href {\doibase 10.1038/s41567-021-01491-x} {\bibfield
		{journal} {\bibinfo  {journal} {Nat. Phys.}\ }\textbf {\bibinfo {volume}
			{18}},\ \bibinfo {pages} {385} (\bibinfo {year} {2022})}\BibitemShut
	{NoStop}%
	\bibitem [{\citenamefont {Kawabata}\ \emph {et~al.}(2023)\citenamefont
		{Kawabata}, \citenamefont {Numasawa},\ and\ \citenamefont
		{Ryu}}]{PhysRevX.13.021007}%
	\BibitemOpen
	\bibfield  {author} {\bibinfo {author} {\bibfnamefont {K.}~\bibnamefont
			{Kawabata}}, \bibinfo {author} {\bibfnamefont {T.}~\bibnamefont {Numasawa}},
		\ and\ \bibinfo {author} {\bibfnamefont {S.}~\bibnamefont {Ryu}},\ }\bibfield
	{title} {\enquote {\bibinfo {title} {Entanglement phase transition induced
				by the non-{H}ermitian skin effect},}\ }\href {\doibase
		10.1103/PhysRevX.13.021007} {\bibfield  {journal} {\bibinfo  {journal} {Phys.
				Rev. X}\ }\textbf {\bibinfo {volume} {13}},\ \bibinfo {pages} {021007}
		(\bibinfo {year} {2023})}\BibitemShut {NoStop}%
	\bibitem [{\citenamefont {Zhang}\ \emph {et~al.}(2023)\citenamefont {Zhang},
		\citenamefont {Fang},\ and\ \citenamefont {Yang}}]{PhysRevLett.131.036402}%
	\BibitemOpen
	\bibfield  {author} {\bibinfo {author} {\bibfnamefont {K.}~\bibnamefont
			{Zhang}}, \bibinfo {author} {\bibfnamefont {C.}~\bibnamefont {Fang}}, \ and\
		\bibinfo {author} {\bibfnamefont {Z.}~\bibnamefont {Yang}},\ }\bibfield
	{title} {\enquote {\bibinfo {title} {Dynamical degeneracy splitting and
				directional invisibility in non-{H}ermitian systems},}\ }\href {\doibase
		10.1103/PhysRevLett.131.036402} {\bibfield  {journal} {\bibinfo  {journal}
			{Phys. Rev. Lett.}\ }\textbf {\bibinfo {volume} {131}},\ \bibinfo {pages}
		{036402} (\bibinfo {year} {2023})}\BibitemShut {NoStop}%
	\bibitem [{\citenamefont {Li}\ \emph {et~al.}(2023)\citenamefont {Li},
		\citenamefont {Trauzettel}, \citenamefont {Neupert},\ and\ \citenamefont
		{Zhang}}]{PhysRevLett.131.116601}%
	\BibitemOpen
	\bibfield  {author} {\bibinfo {author} {\bibfnamefont {C.-A.}\ \bibnamefont
			{Li}}, \bibinfo {author} {\bibfnamefont {B.}~\bibnamefont {Trauzettel}},
		\bibinfo {author} {\bibfnamefont {T.}~\bibnamefont {Neupert}}, \ and\
		\bibinfo {author} {\bibfnamefont {S.-B.}\ \bibnamefont {Zhang}},\ }\bibfield
	{title} {\enquote {\bibinfo {title} {Enhancement of second-order
				non-{H}ermitian skin effect by magnetic fields},}\ }\href {\doibase
		10.1103/PhysRevLett.131.116601} {\bibfield  {journal} {\bibinfo  {journal}
			{Phys. Rev. Lett.}\ }\textbf {\bibinfo {volume} {131}},\ \bibinfo {pages}
		{116601} (\bibinfo {year} {2023})}\BibitemShut {NoStop}%
	\bibitem [{\citenamefont {Liu}\ \emph {et~al.}(2023)\citenamefont {Liu},
		\citenamefont {Cai}, \citenamefont {Liu},\ and\ \citenamefont
		{Yang}}]{arXiv:2311.03777}%
	\BibitemOpen
	\bibfield  {author} {\bibinfo {author} {\bibfnamefont {J.}~\bibnamefont
			{Liu}}, \bibinfo {author} {\bibfnamefont {Z.-F.}\ \bibnamefont {Cai}},
		\bibinfo {author} {\bibfnamefont {T.}~\bibnamefont {Liu}}, \ and\ \bibinfo
		{author} {\bibfnamefont {Z.}~\bibnamefont {Yang}},\ }\bibfield  {title}
	{\enquote {\bibinfo {title} {Reentrant non-{H}ermitian skin effect in coupled
				non-{H}ermitian and {H}ermitian chains with correlated disorder},}\ }\href
	{https://doi.org/10.48550/arXiv.2311.03777} {\bibfield  {journal} {\bibinfo
			{journal} {arXiv:2311.03777}\ } (\bibinfo {year} {2023})}\BibitemShut
	{NoStop}%
	\bibitem [{\citenamefont {Cai}\ \emph {et~al.}(2024)\citenamefont {Cai},
		\citenamefont {Liu},\ and\ \citenamefont {Yang}}]{PhysRevA.109.063329}%
	\BibitemOpen
	\bibfield  {author} {\bibinfo {author} {\bibfnamefont {Z.-F.}\ \bibnamefont
			{Cai}}, \bibinfo {author} {\bibfnamefont {T.}~\bibnamefont {Liu}}, \ and\
		\bibinfo {author} {\bibfnamefont {Z.}~\bibnamefont {Yang}},\ }\bibfield
	{title} {\enquote {\bibinfo {title} {Non-{H}ermitian skin effect in
				periodically driven dissipative ultracold atoms},}\ }\href {\doibase
		10.1103/PhysRevA.109.063329} {\bibfield  {journal} {\bibinfo  {journal}
			{Phys. Rev. A}\ }\textbf {\bibinfo {volume} {109}},\ \bibinfo {pages}
		{063329} (\bibinfo {year} {2024})}\BibitemShut {NoStop}%
	\bibitem [{\citenamefont {Li}\ \emph {et~al.}(2024)\citenamefont {Li},
		\citenamefont {Liu},\ and\ \citenamefont {Liu}}]{arXiv:2403.07459}%
	\BibitemOpen
	\bibfield  {author} {\bibinfo {author} {\bibfnamefont {X.}~\bibnamefont
			{Li}}, \bibinfo {author} {\bibfnamefont {J.}~\bibnamefont {Liu}}, \ and\
		\bibinfo {author} {\bibfnamefont {T.}~\bibnamefont {Liu}},\ }\bibfield
	{title} {\enquote {\bibinfo {title} {Localization-delocalization transitions
				in non-{H}ermitian {Aharonov-Bohm} cages},}\ }\href
	{https://doi.org/10.48550/arXiv.2403.07459} {\bibfield  {journal} {\bibinfo
			{journal} {Front. Phys.}\ }\textbf {\bibinfo {volume} {19}},\ \bibinfo
		{pages} {33211} (\bibinfo {year} {2024})}\BibitemShut {NoStop}%
	\bibitem [{\citenamefont {Wang}\ \emph
		{et~al.}(2024{\natexlab{d}})\citenamefont {Wang}, \citenamefont {Song},\ and\
		\citenamefont {Wang}}]{PhysRevX.14.021011}%
	\BibitemOpen
	\bibfield  {author} {\bibinfo {author} {\bibfnamefont {H.-Y.}\ \bibnamefont
			{Wang}}, \bibinfo {author} {\bibfnamefont {F.}~\bibnamefont {Song}}, \ and\
		\bibinfo {author} {\bibfnamefont {Z.}~\bibnamefont {Wang}},\ }\bibfield
	{title} {\enquote {\bibinfo {title} {Amoeba formulation of non-{B}loch band
				theory in arbitrary dimensions},}\ }\href {\doibase
		10.1103/PhysRevX.14.021011} {\bibfield  {journal} {\bibinfo  {journal} {Phys.
				Rev. X}\ }\textbf {\bibinfo {volume} {14}},\ \bibinfo {pages} {021011}
		(\bibinfo {year} {2024}{\natexlab{d}})}\BibitemShut {NoStop}%
	\bibitem [{\citenamefont {Hu}\ \emph {et~al.}(2024)\citenamefont {Hu},
		\citenamefont {Wang}, \citenamefont {Wang},\ and\ \citenamefont
		{Song}}]{PhysRevLett.132.050402}%
	\BibitemOpen
	\bibfield  {author} {\bibinfo {author} {\bibfnamefont {Y.-M.}\ \bibnamefont
			{Hu}}, \bibinfo {author} {\bibfnamefont {H.-Y.}\ \bibnamefont {Wang}},
		\bibinfo {author} {\bibfnamefont {Z.}~\bibnamefont {Wang}}, \ and\ \bibinfo
		{author} {\bibfnamefont {F.}~\bibnamefont {Song}},\ }\bibfield  {title}
	{\enquote {\bibinfo {title} {Geometric origin of non-{B}loch
				$\mathcal{P}\mathcal{T}$ symmetry breaking},}\ }\href {\doibase
		10.1103/PhysRevLett.132.050402} {\bibfield  {journal} {\bibinfo  {journal}
			{Phys. Rev. Lett.}\ }\textbf {\bibinfo {volume} {132}},\ \bibinfo {pages}
		{050402} (\bibinfo {year} {2024})}\BibitemShut {NoStop}%
	\bibitem [{\citenamefont {Leefmans}\ \emph {et~al.}(2024)\citenamefont
		{Leefmans}, \citenamefont {Parto}, \citenamefont {Williams}, \citenamefont
		{Li}, \citenamefont {Dutt}, \citenamefont {Nori},\ and\ \citenamefont
		{Marandi}}]{Leefmans2024}%
	\BibitemOpen
	\bibfield  {author} {\bibinfo {author} {\bibfnamefont {C.~R.}\ \bibnamefont
			{Leefmans}}, \bibinfo {author} {\bibfnamefont {M.}~\bibnamefont {Parto}},
		\bibinfo {author} {\bibfnamefont {J.}~\bibnamefont {Williams}}, \bibinfo
		{author} {\bibfnamefont {G.~H.~Y.}\ \bibnamefont {Li}}, \bibinfo {author}
		{\bibfnamefont {A.}~\bibnamefont {Dutt}}, \bibinfo {author} {\bibfnamefont
			{F.}~\bibnamefont {Nori}}, \ and\ \bibinfo {author} {\bibfnamefont
			{A.}~\bibnamefont {Marandi}},\ }\bibfield  {title} {\enquote {\bibinfo
			{title} {Topological temporally mode-locked laser},}\ }\href
	{http://dx.doi.org/10.1038/s41567-024-02420-4} {\bibfield  {journal}
		{\bibinfo  {journal} {Nat. Phys.}\ }\textbf {\bibinfo {volume} {20}},\
		\bibinfo {pages} {852} (\bibinfo {year} {2024})}\BibitemShut {NoStop}%
	\bibitem [{\citenamefont {Roccati}\ \emph {et~al.}(2022)\citenamefont
		{Roccati}, \citenamefont {Lorenzo}, \citenamefont {Calajò}, \citenamefont
		{Palma}, \citenamefont {Carollo},\ and\ \citenamefont
		{Ciccarello}}]{Roccati2022}%
	\BibitemOpen
	\bibfield  {author} {\bibinfo {author} {\bibfnamefont {F.}~\bibnamefont
			{Roccati}}, \bibinfo {author} {\bibfnamefont {S.}~\bibnamefont {Lorenzo}},
		\bibinfo {author} {\bibfnamefont {G.}~\bibnamefont {Calajò}}, \bibinfo
		{author} {\bibfnamefont {G.~M.}\ \bibnamefont {Palma}}, \bibinfo {author}
		{\bibfnamefont {A.}~\bibnamefont {Carollo}}, \ and\ \bibinfo {author}
		{\bibfnamefont {F.}~\bibnamefont {Ciccarello}},\ }\bibfield  {title}
	{\enquote {\bibinfo {title} {Exotic interactions mediated by a
				non-{H}ermitian photonic bath},}\ }\href {\doibase 10.1364/optica.443955}
	{\bibfield  {journal} {\bibinfo  {journal} {Optica}\ }\textbf {\bibinfo
			{volume} {9}},\ \bibinfo {pages} {565} (\bibinfo {year} {2022})}\BibitemShut
	{NoStop}%
	\bibitem [{\citenamefont {Gong}\ \emph
		{et~al.}(2022{\natexlab{a}})\citenamefont {Gong}, \citenamefont {Bello},
		\citenamefont {Malz},\ and\ \citenamefont {Kunst}}]{PhysRevA.106.053517}%
	\BibitemOpen
	\bibfield  {author} {\bibinfo {author} {\bibfnamefont {Z.}~\bibnamefont
			{Gong}}, \bibinfo {author} {\bibfnamefont {M.}~\bibnamefont {Bello}},
		\bibinfo {author} {\bibfnamefont {D.}~\bibnamefont {Malz}}, \ and\ \bibinfo
		{author} {\bibfnamefont {F.~K.}\ \bibnamefont {Kunst}},\ }\bibfield  {title}
	{\enquote {\bibinfo {title} {Bound states and photon emission in
				non-{H}ermitian nanophotonics},}\ }\href {\doibase
		10.1103/PhysRevA.106.053517} {\bibfield  {journal} {\bibinfo  {journal}
			{Phys. Rev. A}\ }\textbf {\bibinfo {volume} {106}},\ \bibinfo {pages}
		{053517} (\bibinfo {year} {2022}{\natexlab{a}})}\BibitemShut {NoStop}%
	\bibitem [{\citenamefont {Gong}\ \emph
		{et~al.}(2022{\natexlab{b}})\citenamefont {Gong}, \citenamefont {Bello},
		\citenamefont {Malz},\ and\ \citenamefont {Kunst}}]{PhysRevLett.129.223601}%
	\BibitemOpen
	\bibfield  {author} {\bibinfo {author} {\bibfnamefont {Z.}~\bibnamefont
			{Gong}}, \bibinfo {author} {\bibfnamefont {M.}~\bibnamefont {Bello}},
		\bibinfo {author} {\bibfnamefont {D.}~\bibnamefont {Malz}}, \ and\ \bibinfo
		{author} {\bibfnamefont {F.~K.}\ \bibnamefont {Kunst}},\ }\bibfield  {title}
	{\enquote {\bibinfo {title} {Anomalous behaviors of quantum emitters in
				non-{H}ermitian baths},}\ }\href {\doibase 10.1103/PhysRevLett.129.223601}
	{\bibfield  {journal} {\bibinfo  {journal} {Phys. Rev. Lett.}\ }\textbf
		{\bibinfo {volume} {129}},\ \bibinfo {pages} {223601} (\bibinfo {year}
		{2022}{\natexlab{b}})}\BibitemShut {NoStop}%
	\bibitem [{\citenamefont {Du}\ \emph {et~al.}(2023)\citenamefont {Du},
		\citenamefont {Guo}, \citenamefont {Zhang},\ and\ \citenamefont
		{Kockum}}]{PhysRevResearch.5.L042040}%
	\BibitemOpen
	\bibfield  {author} {\bibinfo {author} {\bibfnamefont {L.}~\bibnamefont
			{Du}}, \bibinfo {author} {\bibfnamefont {L.}~\bibnamefont {Guo}}, \bibinfo
		{author} {\bibfnamefont {Y.}~\bibnamefont {Zhang}}, \ and\ \bibinfo {author}
		{\bibfnamefont {A.~F.}\ \bibnamefont {Kockum}},\ }\bibfield  {title}
	{\enquote {\bibinfo {title} {Giant emitters in a structured bath with
				non-{H}ermitian skin effect},}\ }\href {\doibase
		10.1103/PhysRevResearch.5.L042040} {\bibfield  {journal} {\bibinfo  {journal}
			{Phys. Rev. Res.}\ }\textbf {\bibinfo {volume} {5}},\ \bibinfo {pages}
		{L042040} (\bibinfo {year} {2023})}\BibitemShut {NoStop}%
	\bibitem [{\citenamefont {Roccati}\ \emph {et~al.}(2024)\citenamefont
		{Roccati}, \citenamefont {Bello}, \citenamefont {Gong}, \citenamefont {Ueda},
		\citenamefont {Ciccarello}, \citenamefont {Chenu},\ and\ \citenamefont
		{Carollo}}]{Roccati2024}%
	\BibitemOpen
	\bibfield  {author} {\bibinfo {author} {\bibfnamefont {F.}~\bibnamefont
			{Roccati}}, \bibinfo {author} {\bibfnamefont {M.}~\bibnamefont {Bello}},
		\bibinfo {author} {\bibfnamefont {Z.}~\bibnamefont {Gong}}, \bibinfo {author}
		{\bibfnamefont {M.}~\bibnamefont {Ueda}}, \bibinfo {author} {\bibfnamefont
			{F.}~\bibnamefont {Ciccarello}}, \bibinfo {author} {\bibfnamefont
			{A.}~\bibnamefont {Chenu}}, \ and\ \bibinfo {author} {\bibfnamefont
			{A.}~\bibnamefont {Carollo}},\ }\bibfield  {title} {\enquote {\bibinfo
			{title} {Hermitian and non-{H}ermitian topology from photon-mediated
				interactions},}\ }\href {http://dx.doi.org/10.1038/s41467-024-46471-w}
	{\bibfield  {journal} {\bibinfo  {journal} {Nat. Commun.}\ }\textbf {\bibinfo
			{volume} {15}},\ \bibinfo {pages} {2400} (\bibinfo {year}
		{2024})}\BibitemShut {NoStop}%
	\bibitem [{\citenamefont {Lidar}(2020)}]{arXiv:1902.00967}%
	\BibitemOpen
	\bibfield  {author} {\bibinfo {author} {\bibfnamefont {D.~A.}\ \bibnamefont
			{Lidar}},\ }\bibfield  {title} {\enquote {\bibinfo {title} {Lecture notes on
				the theory of open quantum systems},}\ }\href@noop {} {\bibfield  {journal}
		{\bibinfo  {journal} {arXiv:1902.00967}\ } (\bibinfo {year}
		{2020})}\BibitemShut {NoStop}%
	\bibitem [{Non()}]{NonHermitianBathS2023}%
	\BibitemOpen
	\href@noop {} {}\bibinfo {note} {See \uppercase{S}upplemental
		\uppercase{M}aterial for (I) Effective non-Hermitian bath in
		single-excitation subspace, (II) Bulk-boundary correspondence of a
		non-Hermitian SSH bath, (III) Chiral and hidden bound states, (IV) Effects of
		disorder on chiral and extended photon-emitter dressed states, and (V)
		Analytical solution of chiral-extended photon-emitter dressed states, which
		includes Refs.~\cite{Scully1997SM,Breuer2007SM, PhysRevLett.129.070401SM,
			PhysRevLett.118.200401SM, PhysRevLett.127.116801SM}}\BibitemShut {NoStop}%
	\bibitem [{\citenamefont {Zhu}\ \emph {et~al.}(2021)\citenamefont {Zhu},
		\citenamefont {Teo}, \citenamefont {Li},\ and\ \citenamefont
		{Gong}}]{PhysRevB.103.195414}%
	\BibitemOpen
	\bibfield  {author} {\bibinfo {author} {\bibfnamefont {W.}~\bibnamefont
			{Zhu}}, \bibinfo {author} {\bibfnamefont {W.~X.}\ \bibnamefont {Teo}},
		\bibinfo {author} {\bibfnamefont {L.}~\bibnamefont {Li}}, \ and\ \bibinfo
		{author} {\bibfnamefont {J.}~\bibnamefont {Gong}},\ }\bibfield  {title}
	{\enquote {\bibinfo {title} {Delocalization of topological edge states},}\
	}\href {\doibase 10.1103/PhysRevB.103.195414} {\bibfield  {journal} {\bibinfo
			{journal} {Phys. Rev. B}\ }\textbf {\bibinfo {volume} {103}},\ \bibinfo
		{pages} {195414} (\bibinfo {year} {2021})}\BibitemShut {NoStop}%
	\bibitem [{\citenamefont {Wang}\ \emph {et~al.}(2022)\citenamefont {Wang},
		\citenamefont {Wang},\ and\ \citenamefont {Ma}}]{Wang2022}%
	\BibitemOpen
	\bibfield  {author} {\bibinfo {author} {\bibfnamefont {W.}~\bibnamefont
			{Wang}}, \bibinfo {author} {\bibfnamefont {X.}~\bibnamefont {Wang}}, \ and\
		\bibinfo {author} {\bibfnamefont {G.}~\bibnamefont {Ma}},\ }\bibfield
	{title} {\enquote {\bibinfo {title} {Non-{H}ermitian morphing of topological
				modes},}\ }\href {\doibase 10.1038/s41586-022-04929-1} {\bibfield  {journal}
		{\bibinfo  {journal} {Nature}\ }\textbf {\bibinfo {volume} {608}},\ \bibinfo
		{pages} {50} (\bibinfo {year} {2022})}\BibitemShut {NoStop}%
	\bibitem [{\citenamefont {Cohen-Tannoudji}\ \emph {et~al.}(1998)\citenamefont
		{Cohen-Tannoudji}, \citenamefont {Dupont-Roc},\ and\ \citenamefont
		{Grynberg}}]{CCohenTannoudji1Atom}%
	\BibitemOpen
	\bibfield  {author} {\bibinfo {author} {\bibfnamefont {C.}~\bibnamefont
			{Cohen-Tannoudji}}, \bibinfo {author} {\bibfnamefont {J.}~\bibnamefont
			{Dupont-Roc}}, \ and\ \bibinfo {author} {\bibfnamefont {G.}~\bibnamefont
			{Grynberg}},\ }\href@noop {} {\emph {\bibinfo {title} {Atom-Photon
				Interactions: {B}asic Process and Appilcations}}}\ (\bibinfo  {publisher}
	{John Wiley and Sons},\ \bibinfo {year} {1998})\BibitemShut {NoStop}%
	\bibitem [{\citenamefont {Economou}(2006)}]{Economou_2006}%
	\BibitemOpen
	\bibfield  {author} {\bibinfo {author} {\bibfnamefont {E.~N.}\ \bibnamefont
			{Economou}},\ }\href {\doibase 10.1007/3-540-28841-4} {\emph {\bibinfo
			{title} {Green’s Functions in Quantum Physics}}}\ (\bibinfo  {publisher}
	{Springer Berlin Heidelberg},\ \bibinfo {year} {2006})\BibitemShut {NoStop}%
	\bibitem [{\citenamefont {Scully}\ and\ \citenamefont
		{Zubairy}(1997)}]{Scully1997SM}%
	\BibitemOpen
	\bibfield  {author} {\bibinfo {author} {\bibfnamefont {M.~O.}\ \bibnamefont
			{Scully}}\ and\ \bibinfo {author} {\bibfnamefont {M.~S.}\ \bibnamefont
			{Zubairy}},\ }\href
	{https://doi.org/10.1093/acprof:oso/9780199213900.001.0001} {\emph {\bibinfo
			{title} {Quantum Optics}}}\ (\bibinfo  {publisher} {Cambridge University
		Press},\ \bibinfo {year} {1997})\BibitemShut {NoStop}%
	\bibitem [{\citenamefont {Breuer}\ and\ \citenamefont
		{Petruccione}(2007)}]{Breuer2007SM}%
	\BibitemOpen
	\bibfield  {author} {\bibinfo {author} {\bibfnamefont {H.~P.}\ \bibnamefont
			{Breuer}}\ and\ \bibinfo {author} {\bibfnamefont {F.}~\bibnamefont
			{Petruccione}},\ }\href {\doibase 10.1093/acprof:oso/9780199213900.001.0001}
	{\emph {\bibinfo {title} {The Theory of Open Quantum Systems}}}\ (\bibinfo
	{publisher} {Oxford University Press},\ \bibinfo {year} {2007})\BibitemShut
	{NoStop}%
	\bibitem [{\citenamefont {Liang}\ \emph {et~al.}(2022)\citenamefont {Liang},
		\citenamefont {Xie}, \citenamefont {Dong}, \citenamefont {Li}, \citenamefont
		{Li}, \citenamefont {Gadway}, \citenamefont {Yi},\ and\ \citenamefont
		{Yan}}]{PhysRevLett.129.070401SM}%
	\BibitemOpen
	\bibfield  {author} {\bibinfo {author} {\bibfnamefont {Q.}~\bibnamefont
			{Liang}}, \bibinfo {author} {\bibfnamefont {D.}~\bibnamefont {Xie}}, \bibinfo
		{author} {\bibfnamefont {Z.}~\bibnamefont {Dong}}, \bibinfo {author}
		{\bibfnamefont {H.}~\bibnamefont {Li}}, \bibinfo {author} {\bibfnamefont
			{H.}~\bibnamefont {Li}}, \bibinfo {author} {\bibfnamefont {B.}~\bibnamefont
			{Gadway}}, \bibinfo {author} {\bibfnamefont {W.}~\bibnamefont {Yi}}, \ and\
		\bibinfo {author} {\bibfnamefont {B.}~\bibnamefont {Yan}},\ }\bibfield
	{title} {\enquote {\bibinfo {title} {Dynamic signatures of non-{H}ermitian
				skin effect and topology in ultracold atoms},}\ }\href {\doibase
		10.1103/PhysRevLett.129.070401} {\bibfield  {journal} {\bibinfo  {journal}
			{Phys. Rev. Lett.}\ }\textbf {\bibinfo {volume} {129}},\ \bibinfo {pages}
		{070401} (\bibinfo {year} {2022})}\BibitemShut {NoStop}%
	\bibitem [{\citenamefont {Gong}\ \emph {et~al.}(2017)\citenamefont {Gong},
		\citenamefont {Higashikawa},\ and\ \citenamefont
		{Ueda}}]{PhysRevLett.118.200401SM}%
	\BibitemOpen
	\bibfield  {author} {\bibinfo {author} {\bibfnamefont {Z.}~\bibnamefont
			{Gong}}, \bibinfo {author} {\bibfnamefont {S.}~\bibnamefont {Higashikawa}}, \
		and\ \bibinfo {author} {\bibfnamefont {M.}~\bibnamefont {Ueda}},\ }\bibfield
	{title} {\enquote {\bibinfo {title} {Zeno {H}all effect},}\ }\href {\doibase
		10.1103/PhysRevLett.118.200401} {\bibfield  {journal} {\bibinfo  {journal}
			{Phys. Rev. Lett.}\ }\textbf {\bibinfo {volume} {118}},\ \bibinfo {pages}
		{200401} (\bibinfo {year} {2017})}\BibitemShut {NoStop}%
	\bibitem [{\citenamefont {Guo}\ \emph {et~al.}(2021)\citenamefont {Guo},
		\citenamefont {Liu}, \citenamefont {Zhao}, \citenamefont {Liu},\ and\
		\citenamefont {Chen}}]{PhysRevLett.127.116801SM}%
	\BibitemOpen
	\bibfield  {author} {\bibinfo {author} {\bibfnamefont {C.-X.}\ \bibnamefont
			{Guo}}, \bibinfo {author} {\bibfnamefont {C.-H.}\ \bibnamefont {Liu}},
		\bibinfo {author} {\bibfnamefont {X.-M.}\ \bibnamefont {Zhao}}, \bibinfo
		{author} {\bibfnamefont {Y.}~\bibnamefont {Liu}}, \ and\ \bibinfo {author}
		{\bibfnamefont {S.}~\bibnamefont {Chen}},\ }\bibfield  {title} {\enquote
		{\bibinfo {title} {Exact solution of non-{H}ermitian systems with generalized
				boundary conditions: {S}ize-dependent boundary effect and fragility of the
				skin effect},}\ }\href {\doibase 10.1103/PhysRevLett.127.116801} {\bibfield
		{journal} {\bibinfo  {journal} {Phys. Rev. Lett.}\ }\textbf {\bibinfo
			{volume} {127}},\ \bibinfo {pages} {116801} (\bibinfo {year}
		{2021})}\BibitemShut {NoStop}%
\end{thebibliography}

\begin{thebibliography}{18}%
	\makeatletter
	\providecommand \@ifxundefined [1]{%
		\@ifx{#1\undefined}
	}%
	\providecommand \@ifnum [1]{%
		\ifnum #1\expandafter \@firstoftwo
		\else \expandafter \@secondoftwo
		\fi
	}%
	\providecommand \@ifx [1]{%
		\ifx #1\expandafter \@firstoftwo
		\else \expandafter \@secondoftwo
		\fi
	}%
	\providecommand \natexlab [1]{#1}%
	\providecommand \enquote  [1]{``#1''}%
	\providecommand \bibnamefont  [1]{#1}%
	\providecommand \bibfnamefont [1]{#1}%
	\providecommand \citenamefont [1]{#1}%
	\providecommand \href@noop [0]{\@secondoftwo}%
	\providecommand \href [0]{\begingroup \@sanitize@url \@href}%
	\providecommand \@href[1]{\@@startlink{#1}\@@href}%
	\providecommand \@@href[1]{\endgroup#1\@@endlink}%
	\providecommand \@sanitize@url [0]{\catcode `\\12\catcode `\$12\catcode
		`\&12\catcode `\#12\catcode `\^12\catcode `\_12\catcode `\%12\relax}%
	\providecommand \@@startlink[1]{}%
	\providecommand \@@endlink[0]{}%
	\providecommand \url  [0]{\begingroup\@sanitize@url \@url }%
	\providecommand \@url [1]{\endgroup\@href {#1}{\urlprefix }}%
	\providecommand \urlprefix  [0]{URL }%
	\providecommand \Eprint [0]{\href }%
	\providecommand \doibase [0]{http://dx.doi.org/}%
	\providecommand \selectlanguage [0]{\@gobble}%
	\providecommand \bibinfo  [0]{\@secondoftwo}%
	\providecommand \bibfield  [0]{\@secondoftwo}%
	\providecommand \translation [1]{[#1]}%
	\providecommand \BibitemOpen [0]{}%
	\providecommand \bibitemStop [0]{}%
	\providecommand \bibitemNoStop [0]{.\EOS\space}%
	\providecommand \EOS [0]{\spacefactor3000\relax}%
	\providecommand \BibitemShut  [1]{\csname bibitem#1\endcsname}%
	\let\auto@bib@innerbib\@empty
	\bibitem [{\citenamefont {Scully}\ and\ \citenamefont
		{Zubairy}(1997)}]{Scully1997SM}%
	\BibitemOpen
	\bibfield  {author} {\bibinfo {author} {\bibfnamefont {M.~O.}\ \bibnamefont
			{Scully}}\ and\ \bibinfo {author} {\bibfnamefont {M.~S.}\ \bibnamefont
			{Zubairy}},\ }\href
	{https://doi.org/10.1093/acprof:oso/9780199213900.001.0001} {\emph {\bibinfo
			{title} {Quantum Optics}}}\ (\bibinfo  {publisher} {Cambridge University
		Press},\ \bibinfo {year} {1997})\BibitemShut {NoStop}%
	\bibitem [{\citenamefont {Breuer}\ and\ \citenamefont
		{Petruccione}(2007)}]{Breuer2007SM}%
	\BibitemOpen
	\bibfield  {author} {\bibinfo {author} {\bibfnamefont {H.~P.}\ \bibnamefont
			{Breuer}}\ and\ \bibinfo {author} {\bibfnamefont {F.}~\bibnamefont
			{Petruccione}},\ }\href {\doibase 10.1093/acprof:oso/9780199213900.001.0001}
	{\emph {\bibinfo {title} {The Theory of Open Quantum Systems}}}\ (\bibinfo
	{publisher} {Oxford University Press},\ \bibinfo {year} {2007})\BibitemShut
	{NoStop}%
	\bibitem [{\citenamefont {Agarwal}(2012)}]{Agarwal2012SM}%
	\BibitemOpen
	\bibfield  {author} {\bibinfo {author} {\bibfnamefont {G.~S.}\ \bibnamefont
			{Agarwal}},\ }\href {https://doi.org/10.1017/CBO9781139035170} {\emph
		{\bibinfo {title} {Quantum Optics}}}\ (\bibinfo  {publisher} {Cambridge
		University Press},\ \bibinfo {year} {2012})\BibitemShut {NoStop}%
	\bibitem [{\citenamefont {Lidar}(2020)}]{arXiv:1902.00967SM}%
	\BibitemOpen
	\bibfield  {author} {\bibinfo {author} {\bibfnamefont {D.~A.}\ \bibnamefont
			{Lidar}},\ }\bibfield  {title} {\enquote {\bibinfo {title} {Lecture notes on
				the theory of open quantum systems},}\ }\href@noop {} {\bibfield  {journal}
		{\bibinfo  {journal} {arXiv:1902.00967}\ } (\bibinfo {year}
		{2020})}\BibitemShut {NoStop}%
	\bibitem [{\citenamefont {Gong}\ \emph {et~al.}(2018)\citenamefont {Gong},
		\citenamefont {Ashida}, \citenamefont {Kawabata}, \citenamefont {Takasan},
		\citenamefont {Higashikawa},\ and\ \citenamefont {Ueda}}]{Gong2018SM}%
	\BibitemOpen
	\bibfield  {author} {\bibinfo {author} {\bibfnamefont {Z.}~\bibnamefont
			{Gong}}, \bibinfo {author} {\bibfnamefont {Y.}~\bibnamefont {Ashida}},
		\bibinfo {author} {\bibfnamefont {K.}~\bibnamefont {Kawabata}}, \bibinfo
		{author} {\bibfnamefont {K.}~\bibnamefont {Takasan}}, \bibinfo {author}
		{\bibfnamefont {S.}~\bibnamefont {Higashikawa}}, \ and\ \bibinfo {author}
		{\bibfnamefont {M.}~\bibnamefont {Ueda}},\ }\bibfield  {title} {\enquote
		{\bibinfo {title} {Topological phases of non-\uppercase{H}ermitian
				systems},}\ }\href {https://link.aps.org/doi/10.1103/PhysRevX.8.031079}
	{\bibfield  {journal} {\bibinfo  {journal} {Phys. Rev. X}\ }\textbf {\bibinfo
			{volume} {8}},\ \bibinfo {pages} {031079} (\bibinfo {year}
		{2018})}\BibitemShut {NoStop}%
	\bibitem [{\citenamefont {Liu}\ \emph {et~al.}(2019)\citenamefont {Liu},
		\citenamefont {Zhang}, \citenamefont {Ai}, \citenamefont {Gong},
		\citenamefont {Kawabata}, \citenamefont {Ueda},\ and\ \citenamefont
		{Nori}}]{PhysRevLett.122.076801SM}%
	\BibitemOpen
	\bibfield  {author} {\bibinfo {author} {\bibfnamefont {T.}~\bibnamefont
			{Liu}}, \bibinfo {author} {\bibfnamefont {Y.-R.}\ \bibnamefont {Zhang}},
		\bibinfo {author} {\bibfnamefont {Q.}~\bibnamefont {Ai}}, \bibinfo {author}
		{\bibfnamefont {Z.}~\bibnamefont {Gong}}, \bibinfo {author} {\bibfnamefont
			{K.}~\bibnamefont {Kawabata}}, \bibinfo {author} {\bibfnamefont
			{M.}~\bibnamefont {Ueda}}, \ and\ \bibinfo {author} {\bibfnamefont
			{F.}~\bibnamefont {Nori}},\ }\bibfield  {title} {\enquote {\bibinfo {title}
			{Second-order topological phases in non-{H}ermitian systems},}\ }\href
	{\doibase 10.1103/PhysRevLett.122.076801} {\bibfield  {journal} {\bibinfo
			{journal} {Phys. Rev. Lett.}\ }\textbf {\bibinfo {volume} {122}},\ \bibinfo
		{pages} {076801} (\bibinfo {year} {2019})}\BibitemShut {NoStop}%
	\bibitem [{\citenamefont {Sun}\ \emph {et~al.}(2023)\citenamefont {Sun},
		\citenamefont {Shi}, \citenamefont {Liu}, \citenamefont {Zhang},
		\citenamefont {Xiao}, \citenamefont {Jia},\ and\ \citenamefont
		{Hu}}]{PhysRevX.13.031009SM}%
	\BibitemOpen
	\bibfield  {author} {\bibinfo {author} {\bibfnamefont {Y.}~\bibnamefont
			{Sun}}, \bibinfo {author} {\bibfnamefont {T.}~\bibnamefont {Shi}}, \bibinfo
		{author} {\bibfnamefont {Z.}~\bibnamefont {Liu}}, \bibinfo {author}
		{\bibfnamefont {Z.}~\bibnamefont {Zhang}}, \bibinfo {author} {\bibfnamefont
			{L.}~\bibnamefont {Xiao}}, \bibinfo {author} {\bibfnamefont {S.}~\bibnamefont
			{Jia}}, \ and\ \bibinfo {author} {\bibfnamefont {Y.}~\bibnamefont {Hu}},\
	}\bibfield  {title} {\enquote {\bibinfo {title} {Fractional quantum zeno
				effect emerging from non-{H}ermitian physics},}\ }\href {\doibase
		10.1103/PhysRevX.13.031009} {\bibfield  {journal} {\bibinfo  {journal} {Phys.
				Rev. X}\ }\textbf {\bibinfo {volume} {13}},\ \bibinfo {pages} {031009}
		(\bibinfo {year} {2023})}\BibitemShut {NoStop}%
	\bibitem [{\citenamefont {Liang}\ \emph {et~al.}(2022)\citenamefont {Liang},
		\citenamefont {Xie}, \citenamefont {Dong}, \citenamefont {Li}, \citenamefont
		{Li}, \citenamefont {Gadway}, \citenamefont {Yi},\ and\ \citenamefont
		{Yan}}]{PhysRevLett.129.070401SM}%
	\BibitemOpen
	\bibfield  {author} {\bibinfo {author} {\bibfnamefont {Q.}~\bibnamefont
			{Liang}}, \bibinfo {author} {\bibfnamefont {D.}~\bibnamefont {Xie}}, \bibinfo
		{author} {\bibfnamefont {Z.}~\bibnamefont {Dong}}, \bibinfo {author}
		{\bibfnamefont {H.}~\bibnamefont {Li}}, \bibinfo {author} {\bibfnamefont
			{H.}~\bibnamefont {Li}}, \bibinfo {author} {\bibfnamefont {B.}~\bibnamefont
			{Gadway}}, \bibinfo {author} {\bibfnamefont {W.}~\bibnamefont {Yi}}, \ and\
		\bibinfo {author} {\bibfnamefont {B.}~\bibnamefont {Yan}},\ }\bibfield
	{title} {\enquote {\bibinfo {title} {Dynamic signatures of non-{H}ermitian
				skin effect and topology in ultracold atoms},}\ }\href {\doibase
		10.1103/PhysRevLett.129.070401} {\bibfield  {journal} {\bibinfo  {journal}
			{Phys. Rev. Lett.}\ }\textbf {\bibinfo {volume} {129}},\ \bibinfo {pages}
		{070401} (\bibinfo {year} {2022})}\BibitemShut {NoStop}%
	\bibitem [{\citenamefont {Gonz\'alez-Tudela}\ and\ \citenamefont
		{Cirac}(2017)}]{PhysRevA.96.043811SM}%
	\BibitemOpen
	\bibfield  {author} {\bibinfo {author} {\bibfnamefont {A.}~\bibnamefont
			{Gonz\'alez-Tudela}}\ and\ \bibinfo {author} {\bibfnamefont {J.~I.}\
			\bibnamefont {Cirac}},\ }\bibfield  {title} {\enquote {\bibinfo {title}
			{Markovian and non-{M}arkovian dynamics of quantum emitters coupled to
				two-dimensional structured reservoirs},}\ }\href {\doibase
		10.1103/PhysRevA.96.043811} {\bibfield  {journal} {\bibinfo  {journal} {Phys.
				Rev. A}\ }\textbf {\bibinfo {volume} {96}},\ \bibinfo {pages} {043811}
		(\bibinfo {year} {2017})}\BibitemShut {NoStop}%
	\bibitem [{\citenamefont {Gong}\ \emph {et~al.}(2017)\citenamefont {Gong},
		\citenamefont {Higashikawa},\ and\ \citenamefont
		{Ueda}}]{PhysRevLett.118.200401SM}%
	\BibitemOpen
	\bibfield  {author} {\bibinfo {author} {\bibfnamefont {Z.}~\bibnamefont
			{Gong}}, \bibinfo {author} {\bibfnamefont {S.}~\bibnamefont {Higashikawa}}, \
		and\ \bibinfo {author} {\bibfnamefont {M.}~\bibnamefont {Ueda}},\ }\bibfield
	{title} {\enquote {\bibinfo {title} {Zeno {H}all effect},}\ }\href {\doibase
		10.1103/PhysRevLett.118.200401} {\bibfield  {journal} {\bibinfo  {journal}
			{Phys. Rev. Lett.}\ }\textbf {\bibinfo {volume} {118}},\ \bibinfo {pages}
		{200401} (\bibinfo {year} {2017})}\BibitemShut {NoStop}%
	\bibitem [{\citenamefont {Gong}\ \emph
		{et~al.}(2022{\natexlab{a}})\citenamefont {Gong}, \citenamefont {Bello},
		\citenamefont {Malz},\ and\ \citenamefont
		{Kunst}}]{PhysRevLett.129.223601SM}%
	\BibitemOpen
	\bibfield  {author} {\bibinfo {author} {\bibfnamefont {Z.}~\bibnamefont
			{Gong}}, \bibinfo {author} {\bibfnamefont {M.}~\bibnamefont {Bello}},
		\bibinfo {author} {\bibfnamefont {D.}~\bibnamefont {Malz}}, \ and\ \bibinfo
		{author} {\bibfnamefont {F.~K.}\ \bibnamefont {Kunst}},\ }\bibfield  {title}
	{\enquote {\bibinfo {title} {Anomalous behaviors of quantum emitters in
				non-{H}ermitian baths},}\ }\href {\doibase 10.1103/PhysRevLett.129.223601}
	{\bibfield  {journal} {\bibinfo  {journal} {Phys. Rev. Lett.}\ }\textbf
		{\bibinfo {volume} {129}},\ \bibinfo {pages} {223601} (\bibinfo {year}
		{2022}{\natexlab{a}})}\BibitemShut {NoStop}%
	\bibitem [{\citenamefont {Gong}\ \emph
		{et~al.}(2022{\natexlab{b}})\citenamefont {Gong}, \citenamefont {Bello},
		\citenamefont {Malz},\ and\ \citenamefont {Kunst}}]{PhysRevA.106.053517SM}%
	\BibitemOpen
	\bibfield  {author} {\bibinfo {author} {\bibfnamefont {Z.}~\bibnamefont
			{Gong}}, \bibinfo {author} {\bibfnamefont {M.}~\bibnamefont {Bello}},
		\bibinfo {author} {\bibfnamefont {D.}~\bibnamefont {Malz}}, \ and\ \bibinfo
		{author} {\bibfnamefont {F.~K.}\ \bibnamefont {Kunst}},\ }\bibfield  {title}
	{\enquote {\bibinfo {title} {Bound states and photon emission in
				non-{H}ermitian nanophotonics},}\ }\href {\doibase
		10.1103/PhysRevA.106.053517} {\bibfield  {journal} {\bibinfo  {journal}
			{Phys. Rev. A}\ }\textbf {\bibinfo {volume} {106}},\ \bibinfo {pages}
		{053517} (\bibinfo {year} {2022}{\natexlab{b}})}\BibitemShut {NoStop}%
	\bibitem [{\citenamefont {Yao}\ and\ \citenamefont
		{Wang}(2018)}]{ShunyuYao2018SM}%
	\BibitemOpen
	\bibfield  {author} {\bibinfo {author} {\bibfnamefont {S.}~\bibnamefont
			{Yao}}\ and\ \bibinfo {author} {\bibfnamefont {Z.}~\bibnamefont {Wang}},\
	}\bibfield  {title} {\enquote {\bibinfo {title} {Edge states and topological
				invariants of non-\uppercase{H}ermitian systems},}\ }\href
	{https://link.aps.org/doi/10.1103/PhysRevLett.121.086803} {\bibfield
		{journal} {\bibinfo  {journal} {Phys. Rev. Lett.}\ }\textbf {\bibinfo
			{volume} {121}},\ \bibinfo {pages} {086803} (\bibinfo {year}
		{2018})}\BibitemShut {NoStop}%
	\bibitem [{\citenamefont {Yokomizo}\ and\ \citenamefont
		{Murakami}(2019)}]{PhysRevLett.123.066404SM}%
	\BibitemOpen
	\bibfield  {author} {\bibinfo {author} {\bibfnamefont {K.}~\bibnamefont
			{Yokomizo}}\ and\ \bibinfo {author} {\bibfnamefont {S.}~\bibnamefont
			{Murakami}},\ }\bibfield  {title} {\enquote {\bibinfo {title} {Non-{B}loch
				band theory of non-{H}ermitian systems},}\ }\href {\doibase
		10.1103/PhysRevLett.123.066404} {\bibfield  {journal} {\bibinfo  {journal}
			{Phys. Rev. Lett.}\ }\textbf {\bibinfo {volume} {123}},\ \bibinfo {pages}
		{066404} (\bibinfo {year} {2019})}\BibitemShut {NoStop}%
	\bibitem [{\citenamefont {Bello}\ \emph {et~al.}(2019)\citenamefont {Bello},
		\citenamefont {Platero}, \citenamefont {Cirac},\ and\ \citenamefont
		{Gonz{\'{a}}lez-Tudela}}]{Bello2019SM}%
	\BibitemOpen
	\bibfield  {author} {\bibinfo {author} {\bibfnamefont {M.}~\bibnamefont
			{Bello}}, \bibinfo {author} {\bibfnamefont {G.}~\bibnamefont {Platero}},
		\bibinfo {author} {\bibfnamefont {J.~I.}\ \bibnamefont {Cirac}}, \ and\
		\bibinfo {author} {\bibfnamefont {A.}~\bibnamefont {Gonz{\'{a}}lez-Tudela}},\
	}\bibfield  {title} {\enquote {\bibinfo {title} {Unconventional quantum
				optics in topological waveguide {QED}},}\ }\href {\doibase
		10.1126/sciadv.aaw0297} {\bibfield  {journal} {\bibinfo  {journal} {Sci.
				Adv.}\ }\textbf {\bibinfo {volume} {5}} (\bibinfo {year} {2019}),\
		10.1126/sciadv.aaw0297}\BibitemShut {NoStop}%
	\bibitem [{\citenamefont {Guo}\ \emph {et~al.}(2021)\citenamefont {Guo},
		\citenamefont {Liu}, \citenamefont {Zhao}, \citenamefont {Liu},\ and\
		\citenamefont {Chen}}]{PhysRevLett.127.116801SM}%
	\BibitemOpen
	\bibfield  {author} {\bibinfo {author} {\bibfnamefont {C.-X.}\ \bibnamefont
			{Guo}}, \bibinfo {author} {\bibfnamefont {C.-H.}\ \bibnamefont {Liu}},
		\bibinfo {author} {\bibfnamefont {X.-M.}\ \bibnamefont {Zhao}}, \bibinfo
		{author} {\bibfnamefont {Y.}~\bibnamefont {Liu}}, \ and\ \bibinfo {author}
		{\bibfnamefont {S.}~\bibnamefont {Chen}},\ }\bibfield  {title} {\enquote
		{\bibinfo {title} {Exact solution of non-{H}ermitian systems with generalized
				boundary conditions: {S}ize-dependent boundary effect and fragility of the
				skin effect},}\ }\href {\doibase 10.1103/PhysRevLett.127.116801} {\bibfield
		{journal} {\bibinfo  {journal} {Phys. Rev. Lett.}\ }\textbf {\bibinfo
			{volume} {127}},\ \bibinfo {pages} {116801} (\bibinfo {year}
		{2021})}\BibitemShut {NoStop}%
	\bibitem [{\citenamefont {Cohen-Tannoudji}\ \emph {et~al.}(1998)\citenamefont
		{Cohen-Tannoudji}, \citenamefont {Dupont-Roc},\ and\ \citenamefont
		{Grynberg}}]{CCohenTannoudji1AtomSM}%
	\BibitemOpen
	\bibfield  {author} {\bibinfo {author} {\bibfnamefont {C.}~\bibnamefont
			{Cohen-Tannoudji}}, \bibinfo {author} {\bibfnamefont {J.}~\bibnamefont
			{Dupont-Roc}}, \ and\ \bibinfo {author} {\bibfnamefont {G.}~\bibnamefont
			{Grynberg}},\ }\href@noop {} {\emph {\bibinfo {title} {Atom-Photon
				Interactions: {B}asic Process and Appilcations}}}\ (\bibinfo  {publisher}
	{John Wiley and Sons},\ \bibinfo {year} {1998})\BibitemShut {NoStop}%
	\bibitem [{\citenamefont {Economou}(2006)}]{Economou_2006SM}%
	\BibitemOpen
	\bibfield  {author} {\bibinfo {author} {\bibfnamefont {E.~N.}\ \bibnamefont
			{Economou}},\ }\href {\doibase 10.1007/3-540-28841-4} {\emph {\bibinfo
			{title} {Green’s Functions in Quantum Physics}}}\ (\bibinfo  {publisher}
	{Springer Berlin Heidelberg},\ \bibinfo {year} {2006})\BibitemShut {NoStop}%
\end{thebibliography}
%

\end{document}